\documentclass[aps,prx,twocolumn,superscriptaddress,nofootinbib,longbibliography]{revtex4-2}
\usepackage[utf8]{inputenc}
\usepackage[T1]{fontenc}
\usepackage{amsmath,amssymb,amsfonts,mathtools}
\usepackage{bm}
\usepackage{graphicx}
\usepackage{multirow}
\usepackage{xcolor}
\usepackage{hyperref}
\hypersetup{colorlinks=true,linkcolor=blue,citecolor=blue,urlcolor=blue}

\providecommand{\apj}{The Astrophysical Journal}
\providecommand{\apjl}{The Astrophysical Journal Letters}

\providecommand{\aap}{Astronomy \& Astrophysics}
\providecommand{\mnras}{Monthly Notices of the Royal Astronomical Society}
\providecommand{\prd}{Physical Review D}
\providecommand{\prl}{Physical Review Letters}
\providecommand{\cqg}{Classical and Quantum Gravity}
\providecommand{\araa}{Annual Review of Astronomy and Astrophysics}

\newcommand{\Msun}{M_\odot}

\newcommand{\phat}{\hat{p}}
\newcommand{\Ohat}{\hat{\Omega}}

\begin{document}

\title{Gravity Echoes from Supermassive Black Hole Binaries}

\author{Qinyuan Zheng}
\affiliation{Department of Physics, Yale University, New Haven, 06520, CT, USA}

\author{Bence B\'ecsy}
\affiliation{Institute for Gravitational Wave Astronomy and School of Physics and Astronomy, University of Birmingham, Edgbaston, Birmingham B15 2TT, UK}

\author{Chiara M. F. Mingarelli}
\affiliation{Department of Physics, Yale University, New Haven, 06520, CT, USA}
\affiliation{Center for Computational Astrophysics, Flatiron Institute, New York, 10010, NY, USA}

\date{\today}

% ======================================================================
% ======================================================================

\begin{abstract}
\noindent Pulsar timing arrays record gravitational waves from supermassive black hole binaries at two spacetime points: an Earth term, measured when the wave passes the Earth, and a pulsar term, measured when the wave passed each pulsar at an earlier epoch. We show that a future $\mu$Hz-band detection of a nearby massive binary by a mission such as $\mu$Ares would turn PTA pulsar terms into targeted probes of binary evolution. In analogy with supernova light echoes, each pulsar term acts as a gravity echo: a dated snapshot of the binary at an earlier stage of its inspiral. Together, the $\mu$Hz Earth-term measurement and the nHz pulsar-term echoes provide a temporal baseline that neither detector could access alone. For a fiducial equal-mass binary with total mass $10^9\,M_\odot$ at 80~Mpc, we find a combined pulsar timing array echo signal-to-noise ratio of 33, with up to 24 pulsars individually resolving the signal among pulsars with 50-year baselines. The angular dependence of the single-pulsar echo sensitivity alone enables independent sky localization of the source to $\sim$10--100~deg$^2$, and the resolved pulsar-term frequencies directly measure the binary inspiral rate hundreds to thousands of years ago. With sufficient pulsar distance precision, a small set of anchor pulsars could additionally phase-connect the array and trace the post-Newtonian evolution coherently over kpc baselines. The source population required for gravity echoes is drawn from the same massive-end census responsible for the observed nanoHertz stochastic background.
\end{abstract}
\maketitle
% ======================================================================
\section{Introduction}

Detecting an individual supermassive black hole binary (SMBHB) across widely separated gravitational-wave (GW) bands would enable direct observation of binary evolution across decades of GW frequency. Pulsar timing arrays (PTAs) are uniquely suited to this problem since each pulsar term samples the same binary at an earlier epoch than the Earth term, with a look-back time set by the source--pulsar--Earth geometry~\cite{Lee2011,CorbinCornish2010,Mingarelli2012}. A future $\mu$Hz detector such as $\mu$Ares~\cite{Sesana2021} would measure the Earth-term waveform directly, turning pulsar terms into targeted probes of the binary at earlier stages of its inspiral. In analogy with light echoes from supernovae~\cite{Couderc1939, Zwicky1940}, we refer to these delayed pulsar-terms as \emph{gravity echoes}. In this work, we show that gravity echoes define a distinct multiband observable: together, the $\mu$Hz Earth term and the delayed PTA pulsar terms provide access to the inspiral history of a single SMBHB over baselines of hundreds to thousands of years.

The recent nanoHertz GW background (GWB) measurements by PTAs~\cite{NANOGrav15yr,EPTA2023,PPTA2023,CPTA2023,MPTA2025} provide indirect evidence that SMBHBs exist in large numbers, and their amplitude suggests that the most massive systems may be more common than previously inferred from electromagnetic scaling relations~\cite{SatoPolito2024, LiepoldMa2024, Mingarelli_ceiling}. This improves the prospects for a nearby, high-mass binary whose Earth term is observable in the $\mu$Hz band and whose pulsar terms are detectable in PTA data. Such a system would constrain several aspects of fundamental physics that are inaccessible to any single detector. For example, black hole spins encode growth history and can distinguish prolonged accretion from repeated mergers~\cite{Barausse2012,Hughes2003,Berti2008}. Indeed, the inspiral phase evolution is sensitive both to the binary environment, including circumbinary disk torques, and to possible deviations from general relativity at mass scales complementary to LIGO--Virgo--KAGRA~\cite{LVK2025} and LISA~\cite{LISA2017}. 

A $\mu$Ares detection would make this science accessible retroactively. The pulsar-term signals are already present in PTA data, but without an independent Earth-term measurement they must be sought through a blind continuous-wave search over a large parameter space. An Earth-term template constrains the binary masses, spins, sky location, and inclination well enough to convert that problem into a targeted search for echoes in individual pulsars~\cite{Arzoumanian2020targeted,NG15targeted}. In fact, dedicated archival PTA searches can recover echoes independently if a continuous-wave PTA detection supplies the Earth-term template at nHz frequencies~\cite{Mingarelli2012}, though here the GWB introduces additional red noise. The $\mu$Ares program is the cleanest route because the $\mu$Hz--nHz band separation isolates the echoes from PTA confusion noise. Concurrent work by Criswell et al.\ (in prep) explores the complementary post-merger regime, in which the binary has already coalesced.

The main obstacle is the precision with which the retarded pulsar epochs can be determined. Fully exploiting gravity echoes requires pulsar distances accurate enough to preserve phase coherence with the source model over the Earth--pulsar baseline. Only a handful of pulsars currently approach this regime: PSR~J0437$-$4715 has a distance uncertainty of $\delta L_p = 0.11$~pc~\cite{Reardon2024}, and very long baseline interferometry programs such as PSR$\pi$~\cite{Deller2019} and MSPSR$\pi$~\cite{Ding2023} achieve microarcsecond parallaxes for dozens of millisecond pulsars. For most PTA-timed pulsars, however, the look-back time remains too uncertain for coherent phase tracking, and the pulsar terms act as noise. Pulsar distance precision is therefore the central bottleneck of the echo program.

These echoes are loudest for a specific class of SMBHB. Typical LISA-band SMBHBs are not massive enough to produce measurable pulsar-term timing residuals, which remain at femtosecond levels even for nearby systems. By contrast, binaries in the high-mass tail of the $\mu$Ares catalog, $M_{\rm tot} \gtrsim 10^{8.5}\,\Msun$ within a few hundred Mpc, can produce greater than nanosecond-level pulsar terms while remaining observable in the $\mu$Hz band. The $\mu$Hz--nHz combination is therefore the only band pairing in which the Earth term and the delayed pulsar terms of the same SMBHB are both measurable. 

This science program divides into three tiers. Tier~1 is a network detection of the echo signal across the PTA. Tier~2 resolves individual echoes in single pulsars, enabling direct tests of the inspiral rate at distinct look-back times. Tier~3 phase-connects multiple anchor pulsars across the array, permitting precision tracking of the waveform over the full Earth--pulsar baseline. This final tier is the most demanding, requiring pulsar distances known to better than a radiation wavelength~\cite{Mingarelli2012}.

The paper is organized as follows. Section~\ref{sec:signal} defines the echo signal model and its angular dependence. Section~\ref{sec:detectability} maps the detectable parameter space, establishes the astrophysical plausibility of suitable sources, and defines the three detection tiers. Section~\ref{sec:phase} addresses the central technical challenge of maintaining phase coherence across the echo baseline. Section~\ref{sec:skyloc} shows how the angular dependence of the echo signal-to-noise ratio enables sky localization. Section~\ref{sec:science} synthesizes the science return through a worked example of a detected nearby SMBHB, and Sec.~\ref{sec:discussion} discusses opportunities, limitations, and next steps.

% ======================================================================
% ======================================================================

\section{Signal model for gravity echoes}\label{sec:signal}

\begin{figure*}[t]
  \centering
  \includegraphics[width=\textwidth]{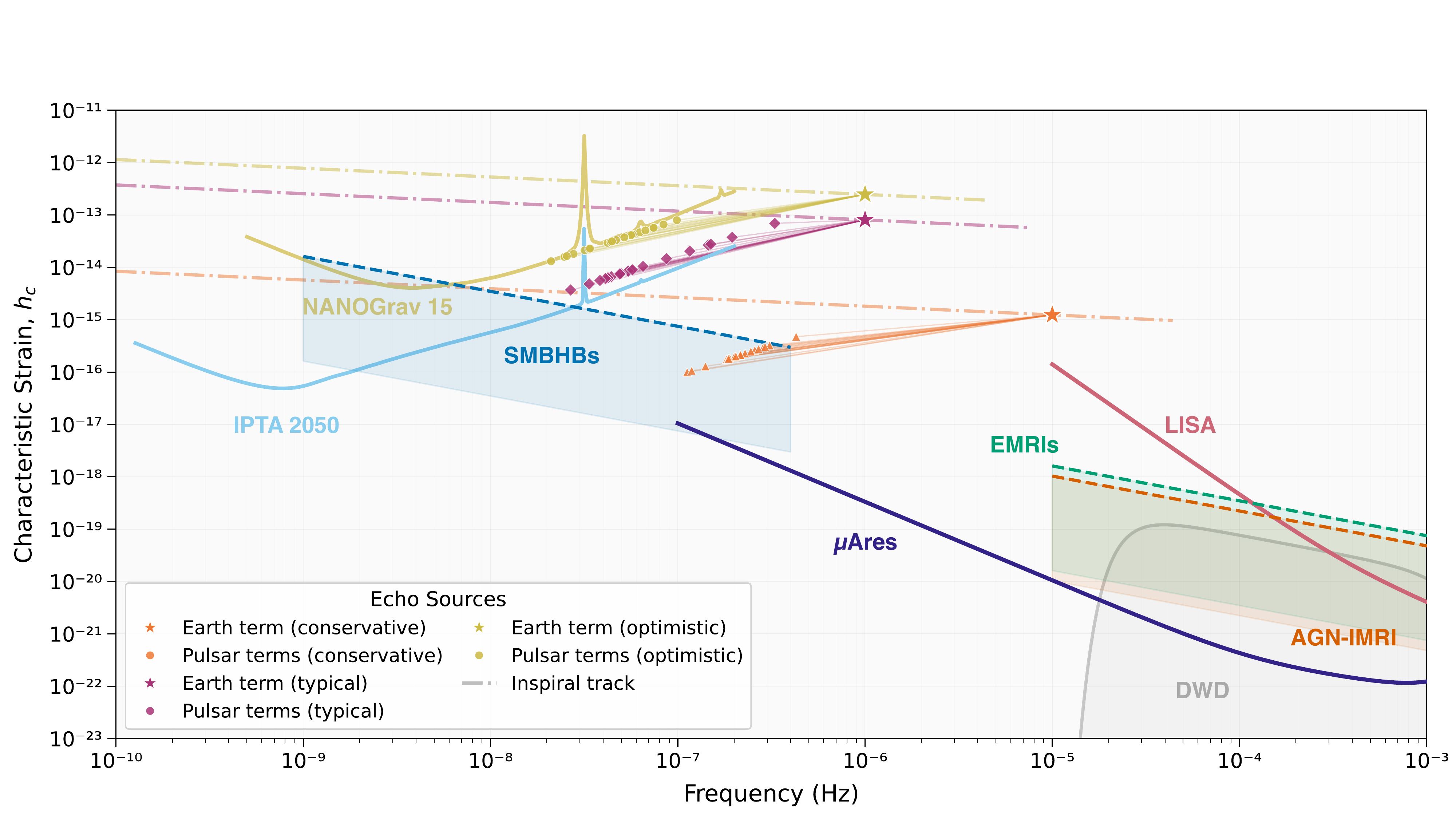}
  \caption{Multiband GW sensitivity landscape from nHz to mHz. Solid curves show detector sensitivity for NANOGrav 15yr~\cite{NG15-continuous}, an IPTA projection to 2050, $\mu$Ares, and LISA. Dashed curves show astrophysical GWB ceiling for SMBHBs~\cite{Mingarelli_ceiling}, AGN-IMRI background~\cite{Mingarelli_AGN_IMRI}, and EMRI background~\cite{BonettiSesana2020}. Three illustrative equal-mass echo sources are shown: Optimistic with $M = 10^9\,M_\odot$ at $D_L = 100$~Mpc, Typical with $M = 6\times10^8\,M_\odot$ at $D_L = 200$~Mpc, and Conservative with $M = 10^8\,M_\odot$ at $D_L = 2$~Gpc. For each source, the dashed-dot line traces the inspiral characteristic strain $h_{c,\rm insp}$, Eq.~\eqref{eq:hc_inspiral}, while the star marks the present-day Earth-term frequency. The smaller markers show the observed pulsar-term strain $h_{c,\rm PTA}$, Eq.~\eqref{eq:hc_pta} at retarded frequencies $f_P < f_E$ for 20 simulated pulsars. These lie below the inspiral track since each pulsar term is only observed for tens of years, which is much shorter than the binary's residence time of tens of millions of years at PTA frequencies.}
  \label{fig:landscape}
\end{figure*}

The observable signature of a gravity echo is a quasi-monochromatic timing residual imprinted on each pulsar by the binary at its retarded pulsar-term frequency ${f_{P,i} < f_E}$, where $f_E$ is the Earth-term frequency. We define the time delay between the Earth term and the pulsar term as
\begin{equation}\label{eq:tau}
  \tau_i = \frac{L_{p,i}(1 + \Ohat \cdot \phat_i)}{c}\, ,
\end{equation}
where $L_{p,i}$ is the distance to pulsar $i$, $\phat_i$ is the unit vector from the Earth to that pulsar, and $\Ohat$ is the GW propagation direction. For PTA pulsars at kpc distances, $\tau_i$ typically spans $10^2$--$10^3$~yr. Since the binary evolves negligibly over the PTA observation time $T_{\rm PTA}$, each pulsar term is well approximated as a single-frequency sinusoid whose amplitude and frequency are set by the source parameters and the retarded epoch. Throughout this section, we model the source as a circular, non-precessing inspiral at leading post-Newtonian order, adding spin corrections in Sec.~\ref{sec:phase} and commenting on residual eccentricity in Sec.~\ref{sec:environment}.

The intrinsic strain amplitude of a circular binary at pulsar-term frequency $f_{P,i}$ is
\begin{equation}\label{eq:strain_amp}
 h_0(f_P) = \frac{4(G\mathcal{M}_c)^{5/3}(\pi f_P)^{2/3}}{c^4 D_L}\,,
\end{equation}
where $\mathcal{M}_c = \eta^{3/5}M_{\rm tot}$ is the chirp mass, $\eta = m_1 m_2/M_{\rm tot}^2$ is the symmetric mass ratio, $M_{\rm tot} = m_1 + m_2$ is the total mass, and $D_L$ is the luminosity distance to the source. After projection through the antenna response functions~\cite{MTW} , the timing residual amplitude in pulsar $i$ is
\begin{equation}\label{eq:residual}
 r_{P,i} = \frac{h_0(f_{P,i})}{2\pi f_{P,i}} \sqrt{ \left(F_+^{(i)} \frac{1+\cos^2\iota}{2}\right)^{\!\!2} + \left(F_\times^{(i)} \cos\iota\right)^{\!\!2} }\,,
\end{equation}
where $\iota$ is the binary inclination, $\psi$ is the polarization angle, and $F_{+,\times}^{(i)}$ are the PTA antenna-pattern functions. %In an all-sky continuous-wave search, the source orientation $(\iota,\psi,\Ohat)$ is unknown. Here, the $\mu$Ares detection fixes those quantities, so the projection becomes known.

For $N_{\rm obs}=T_{\rm PTA}/\Delta t$ times of arrival at cadence $\Delta t$, and assuming white timing noise with variance $\sigma_{\rm TOA}^2$, the optimal matched-filter signal-to-noise ratio in pulsar $i$ is~\cite{JK2012,MooreCB2015,HazbounRS2019}
\begin{equation}\label{eq:snr}
  \rho_i = \frac{r_{P,i}\sqrt{N_{\rm obs}/2}}{\sigma_{\rm TOA}}\,,
\end{equation}
where the factor of $2$ comes from time averaging of the sinusoid. The network SNR adds in quadrature,
\begin{equation}\label{eq:snr_comb}
  \rho_{\rm comb}^2 = \sum_{i=1}^{N_{\rm psr}} \rho_i^2\,,
\end{equation}
where $N_{\rm psr}$ is the number of pulsars in the array. In practice, the network is dominated by a small number of favorably oriented pulsars since both the antenna response and the retarded frequency depend strongly on $(1+\Ohat\cdot\phat_i)$, as we discuss below.

To place the echo signal in the broader multiband context, as in Figure \ref{fig:landscape}, we use two forms of the characteristic strain. The inspiral track over the binary's full lifetime is~\cite{MooreCB2015}
\begin{equation}\label{eq:hc_inspiral}
  h_{c,\rm insp} = h_0\sqrt{\frac{f^2}{\dot{f}}}\,,
\end{equation}
which represents the signal power deposited per logarithmic frequency bin. It is also useful to define the observed pulsar-term characteristic strain, appropriate for a monochromatic source accumulated over the PTA observing time $T_{\rm PTA}$. This is
\begin{equation}\label{eq:hc_pta}
  h_{c,\rm PTA} = h_0\sqrt{f_P\,T_{\rm PTA}}\,.
\end{equation}
The observed $h_{c,\rm PTA}$ lies well below the inspiral track because each pulsar term is measured for only $T_{\rm PTA} \sim 20$--$50$~yr, far shorter than the $\sim 100$~Myr the binary spends at nHz frequencies. Figure~\ref{fig:landscape} shows this multiband landscape from nHz to mHz, including the Earth-term frequency in the $\mu$Ares band and the delayed pulsar terms at lower frequencies in the PTA band.

The PTA sensitivity curves used throughout this work include per-pulsar intrinsic red noise drawn from the NANOGrav 15-year custom-noise analysis~\cite{NG15noise}, injected into the pulsar covariance matrix via \texttt{hasasia}'s \texttt{sim\_pta} routine~\cite{Hazboun2019hasasia}, see Appendix~\ref{app:rednoise} for details. Above $\sim 20$~nHz, this red noise has fallen below the white-noise floor for most NANOGrav millisecond pulsars, and the stochastic GWB steepens from its $f^{-2/3}$ scaling due to source discreteness~\cite{Sesana2008}, with NG15 placing this transition at  $26^{+28}_{-19}$~nHz~\cite{NG15discreteness}. The white-noise matched-filter expression Eq.~\eqref{eq:snr} is therefore adequate for the order-of-magnitude SNR estimates, and we do not add the GWB as a correlated noise source. A full search will require per-pulsar noise modeling, as in e.g. \cite{NG15noise}, but the qualitative conclusions are unchanged. Sensitivity curves and echo source tracks are computed with the interactive visualization tool described in the Software section at the end of this paper.

\subsection{Angular dependence of single-pulsar detectability}\label{sec:snr_angle}

\begin{figure}[t]
  \centering
  \includegraphics[width=\columnwidth]{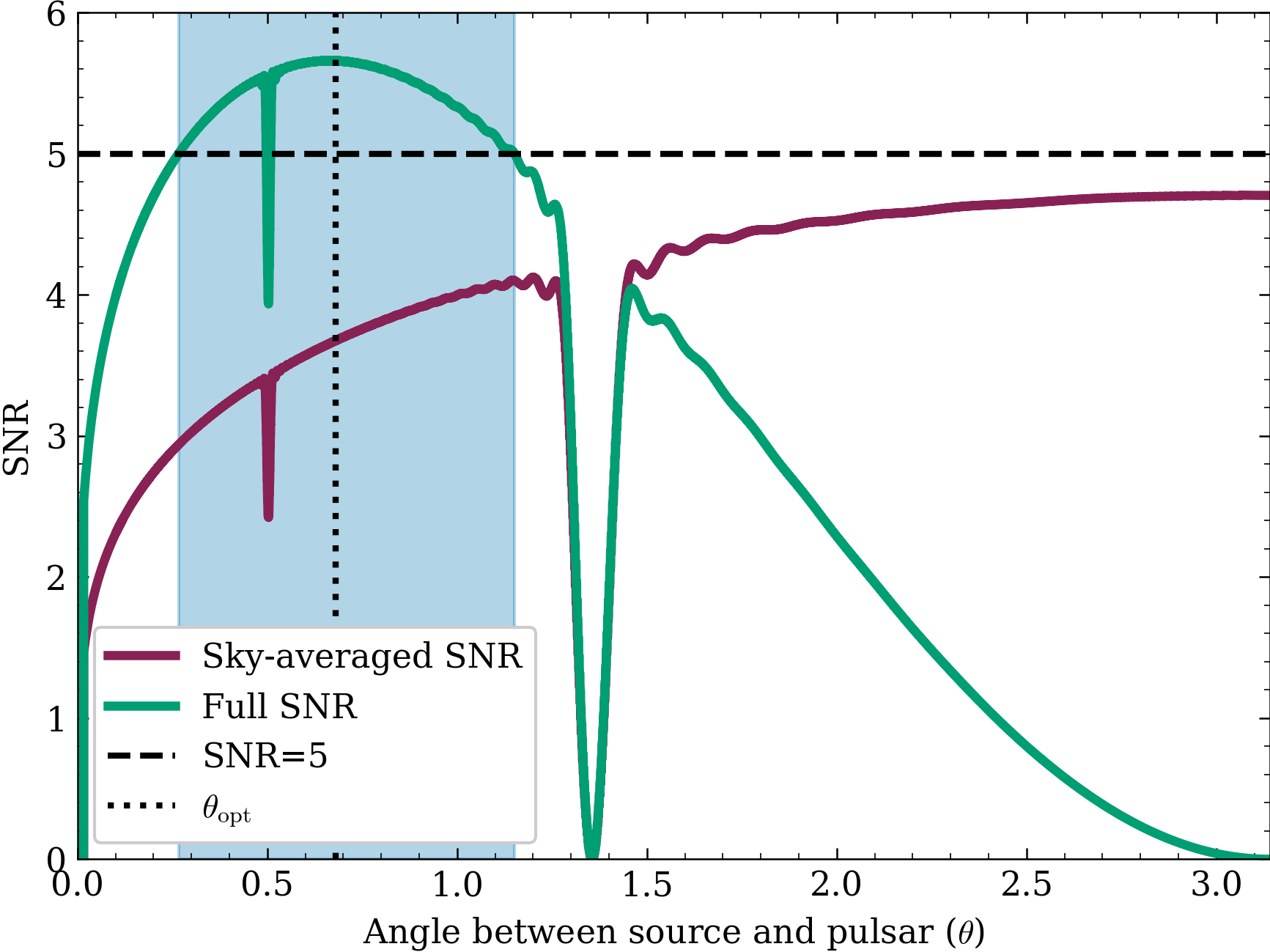}
  \caption{Single-pulsar echo SNR versus source--pulsar angle $\theta$ for the Optimistic binary ($M_{\rm tot}=10^9\;\Msun$, $D_L=100$~Mpc, $L_p \approx 3$~kpc) and a high-precision pulsar with $\sigma_{\rm TOA}=30$ ns. The purple curve shows the enhancement from the frequency shift alone: larger geometric delay lowers $f_P$ and increases the echo amplitude as $r_P \propto f_P^{-1/3}$. The green curve includes the antenna beam pattern, which suppresses the response at large $\theta$ and produces a maximum at $\theta_{\rm opt}\simeq39^\circ$ Eq.~\eqref{eq:theta_opt}. The blue band marks the angular range with $\rho>5$. The adopted $\sigma_{\rm TOA}=30$~ns represents a best-precision target achievable for the most precisely-timed millisecond pulsars, higher than the $\sim100$~ns median assumed for the SKA~Phase~1 and IPTA~2050 arrays in Table~\ref{tab:facilities}.}
  \label{fig:snr_vs_theta}
\end{figure}

Not all pulsars contribute equally to an echo search. The single-pulsar SNR depends strongly on the angular separation $\theta$ between the source and the pulsar, because $\theta$ controls both the geometric delay and the antenna response. Here we analytically solve for the optimal angle between the SMBHB and a pulsar, $\theta_\mathrm{opt}$, Figure \ref{fig:snr_vs_theta}. This angular dependence determines which pulsars dominate the detection and motivates the sky-localization strategy of Sec.~\ref{sec:skyloc}.

Working in the computational frame~\cite{Mingarelli2013}, $\Ohat$ is the GW propagation direction (opposite the source direction), so $\Ohat\cdot\phat = -\cos\theta$, where $\theta$ is the angle between the GW source and the pulsar as seen from the Earth. The factor $1 + \Ohat\cdot\phat$ in Eq.~\eqref{eq:tau} then becomes $1 - \cos\theta$.

The leading-order pulsar-term frequency is
\begin{equation}\label{eq:fP_exact}
 f_P = f_E\!\left(1 + \frac{8\dot{f}\,\tau}{3f_E}\right)^{\!-3/8},
\end{equation}
where $\dot{f}$ is the frequency derivative at $f_E$, expanded to full post-Newtonian order in Eq.~\eqref{eq:fdot}. Larger $\theta$ increases $\tau$ and decreases $f_P$, raising the timing residual amplitude as $r_P \propto f_P^{-1/3}$. The antenna beam patterns $F_{+,\times}^{(i)}$ have a factor of $1/(1-\cos\theta)$ in the denominator, which is offset by a numerator that vanishes when $\theta \to 0$. Averaging over the sky at fixed $\theta$ therefore gives $\langle F^2 \rangle \propto \cos^4(\theta/2)$.

Combining Eqs.~\eqref{eq:fP_exact} and~\eqref{eq:residual} with this antenna-pattern scaling, and writing $\Lambda \equiv 4\dot{f}\,L_p/(3 f_E\,c)$ for the dimensionless chirp accumulated over the light-travel time, the sky-averaged $\rho^2$ takes the form, up to prefactors independent of $\theta$,
\begin{equation}\label{eq:snr_theta}
  \rho^2(\theta) \propto
  \bigl[1 + 4\Lambda\,\sin^2(\theta/2)\bigr]^{1/4}
  \cos^4(\theta/2)\,.
\end{equation}
Setting $\mathrm{d}\rho^2/\mathrm{d}\theta = 0$ to compute the optimal angle, $\theta_\mathrm{opt}$, gives
\begin{equation}\label{eq:theta_opt}
  \sin^2\!\bigl(\theta_{\rm opt}/2\bigr)
 = \frac{\Lambda - 2}{9\Lambda}\,.
\end{equation}

For the sources considered here, $\Lambda$ is large. For example, for $q=1$ and $M_{\rm tot} = 10^9\;\Msun$ at Earth-term frequency $f_E = 1\;\mu$Hz and a pulsar at $L_p = 1$~kpc gives $\Lambda \approx 2000$.  Even a less massive, Typical source with $M_{\rm tot} = 6\times10^8\;\Msun$ yields $\Lambda \approx 850$. In this regime Eq.~\eqref{eq:theta_opt} reduces to $\sin^2(\theta_{\rm opt}/2) = 1/9$, giving $\theta_{\rm opt} = 2\arcsin(1/3) \approx 39^\circ$, consistent with the peak in Fig.~\ref{fig:snr_vs_theta} calculated numerically with the \texttt{gwent}~\cite{Kaiser2021} and \texttt{hasasia}~\cite{HazbounRS2019,Hazboun2019hasasia} software packages. %A peak exists only when $\Lambda > 2$. For sources with minimal frequency evolution, $\Lambda < 2$, the antenna pattern dominates and $\rho$ decreases monotonically with $\theta$.

Counterintuitively, Eq. \eqref{eq:snr_theta} and Figure \ref{fig:snr_vs_theta} show that the most informative pulsars are not those closest to or farthest from the source on the sky, but those at intermediate separations. Each pulsar with $\rho(\theta)$ above threshold constrains the source to lie within an annular ring on the sky centered on that pulsar, while nondetections exclude such rings. This geometry provides the basis for the localization strategy developed in Sec.~\ref{sec:skyloc}.

\section{Detectability}\label{sec:detectability}

Gravity echoes are present in PTA data whether or not the Earth term is ever independently detected. A $\mu$Hz
detection, however, supplies a template that converts each pulsar term into a targeted search, enabling parameter estimation, phase tracking, and post-Newtonian tests. In this section we map the relevant observational regime: the PTA configurations we assume, the pulsar distance precision required for phase coherence, the source masses for which echoes are measurable, and the three detection tiers that define the resulting science program.

The PTA configurations we consider are summarized in Table \ref{tab:facilities}. As a present-day reference we adopt the $\sim131$ canonical IPTA pulsars currently timed across the major regional PTAs~\cite{Mingarelli2026review}, with an effective ensemble timespan of $\sim27$~yr set by the longest-timed EPTA members~\cite{Desvignes2016}. By the $\mu$Ares epoch in $\sim 2050$ or later, SKA Phase~1 is expected to time $\sim200$ millisecond pulsars to $\sim100$~ns precision over at least 20-year baselines~\cite{SKAPTA2025,Keane2025}, while DSA-2000~\cite{DSA2000} should add a further $\sim200$ northern-hemisphere MSPs with comparable precision but monthly cadence. By 2050, the same canonical IPTA pulsars will have accumulated $\sim50$~yr baselines with timing residuals improved to $\sim100$~ns through upgraded backends and the addition of SKA and DSA-2000 data. These long-baseline pulsars are particularly important since the single-pulsar echo SNR scales as $\rho_i \propto \sqrt{N_{\rm obs}}\,/\,\sigma_{\rm TOA}$ Eq.~\eqref{eq:snr}, so baseline length, cadence, and timing precision are all important.

\begin{table}[t]
\caption{Facility configurations adopted in this work. The IPTA~2050 line assumes that upgraded backends, SKA data, and DSA-2000 data improve the median timing precision of the canonical pulsars from $\sim200$~ns to $\sim100$~ns over 50-year baselines. For the NANOGrav 15-yr data, we use the published values from~\cite{NG15-timing}. }\label{tab:facilities}
\centering
\begin{tabular}{lcccc}
\hline\hline
Facility & $N_{\rm psr}$ & $T$ [yr] & $\sigma_{\rm TOA}$ & Cadence \\
\hline
Canonical IPTA & 131 & 27 & 200~ns & biweekly \\
SKA1 & 200 & 20 & 100~ns & biweekly \\
DSA-2000 & 200 & 20 & 100~ns & monthly \\
Canonical IPTA (2050) & 131 & 50 & 100~ns & biweekly \\
\hline\hline
\end{tabular}
\end{table}

%An independent Earth-term measurement from a $\mu$Hz detector turns the PTA continuous-wave search for an individual SMBHB from a blind search into a targeted one. 
Throughout this work we adopt the proposed $\mu$Ares mission as the fiducial Earth-term observatory, using the sensitivity model of Ref.~\cite{Sesana2021}. For the fiducial sources considered below, the Earth-term matched-filter signal-to-noise ratio ranges from $\rho_E\sim2\times10^5$ to $\sim10^7$, so the source masses, spins, sky location, and inclination are effectively known before the PTA echo search begins. Even if low-frequency acceleration noise rises more steeply than the baseline $\mu$Ares design assumes, as in LISA~\cite{LISA2017}, the Earth-term measurement still remains precise enough to support phase-coherent echo tracking.

The limiting PTA requirement is pulsar distance precision. To preserve phase coherence, the uncertainty in the pulsar distance must satisfy $\delta L_p < c/(2\pi f)$, which yields $\approx0.15$~pc at $f=10$~nHz and $\approx0.03$~pc at 50~nHz. 

Since parallax uncertainties scale as $\delta L_p \simeq \sigma_\varpi L_p^2$, the corresponding requirement is
\begin{equation}\label{eq:parallax_req}
\sigma_{\varpi,\rm req} < 1.7\;\mu\text{as}\left(\frac{10\;\text{nHz}}{f}\right)\left(\frac{300\;\text{pc}}{d}\right)^{\!2} \, .
\end{equation}
We define an \emph{anchor pulsar} as one satisfying $\delta L_p < c/(2\pi f)$, so that its echo phase is known to better than one radian and there is no integer-cycle ambiguity. PSR~J0437$-$4715 currently satisfies this threshold, while the next-nearest PTA pulsars remain one to two orders of magnitude away in parallax precision. SKA-era timing and VLBI may plausibly expand the anchor population to $\mathcal{O}(3$--$5)$ at 10~nHz, which is small but not negligible, see Appendix~\ref{app:anchors} for more details. Pulsars with good but sub-threshold distance priors remain useful for detection even when they cannot individually support phase coherence.

\begin{figure}[t]
   \centering
   \includegraphics[width=\columnwidth]{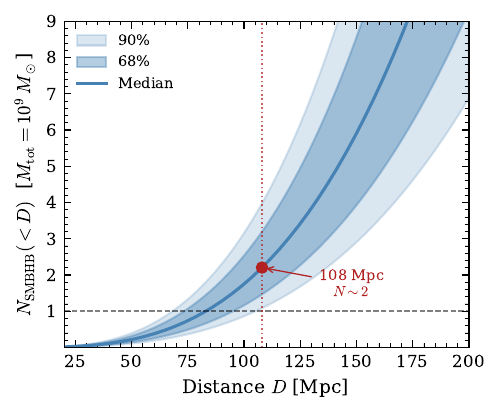}
   \caption{Expected number of SMBHBs with $M_{\rm tot}=10^9\,M_\odot$ and $q>0.9$ as a function of distance. The median (solid), 68\% credible region (dark shading), and 90\% credible region (light shading) are computed from Monte Carlo propagation of uncertainties in the Liepold \& Ma black hole mass function parameters~\cite{LiepoldMa2024} and the Casey-Clyde et al.\ galaxy pairing rate~\cite{CaseyClyde2025}. The dashed line marks $N=1$. The median crosses unity at $D\approx83$~Mpc, comparable to the Tier-3 echo horizon of $\sim80$~Mpc. The red point marks $D=108$~Mpc, with $N\sim2$ expected within the 108~Mpc volume.}
   \label{fig:Nsmbhb}
 \end{figure}

The detectable source population occupies a narrow mass window. LISA-band SMBHBs are too light to produce measurable PTA pulsar terms: even a nearby $10^6\;\Msun$ binary yields only femtosecond-level timing residuals. By contrast, $\mu$Ares-band binaries with $M_{\rm tot}\gtrsim 10^{8.5}\Msun$ can produce nanosecond-level pulsar terms while remaining observable in the $\mu$Hz band. This is the best mass range for echoes, also see Table~\ref{tab:horizon}. Table~\ref{tab:scenarios} summarizes three fiducial cases spanning this regime. The Optimistic source lies within PTA reach, the Typical source lies near the detection boundary, and the Conservative source is effectively inaccessible. The Optimistic distance falls inside the MASSIVE survey volume~\cite{Ma2014}, and population models predict $\sim2$ near-equal-mass PTA-band binaries at this mass within 108~Mpc, see Fig.~\ref{fig:Nsmbhb} and Appendix~\ref{app:population_estimate} for detailed calculations.

\subsection{Detection horizons}\label{sec:horizon}

Given a source in the optimal echo range, the PTA response depends on both the array configuration and the source inclination. We compute the combined network SNR $\rho_{\rm comb}$ by summing Eq.~\eqref{eq:snr} in quadrature over all pulsars in each array configuration listed in Table~\ref{tab:facilities}, evaluating the full antenna pattern at the sky position of each pulsar. We find that the Optimistic source is readily detectable: for a face-on $10^9\;\Msun$ binary at 100~Mpc, the SKA-era array yields $\rho_{\rm comb}\approx17$, the IPTA~2050 pulsars yield $\rho_{\rm comb}\approx20$, and the combined array reaches $\rho_{\rm comb}\approx27$. The horizons listed in Table~\ref{tab:horizon} are for these face-on upper bounds.

\begin{table}[t]
\caption{Echo detectability for a face-on, $10^9\;\Msun$ equal-mass binary at $f_E = 1\;\mu$Hz, assuming white-noise $\sigma_{\rm TOA} = 100$~ns and biweekly cadence. Horizon distances give the maximum $D_L$ at which each tier threshold is met.}
\label{tab:horizon}
\centering
\begin{tabular}{lccccc}
\hline\hline
 & \multicolumn{2}{c}{At 100~Mpc} & \multicolumn{3}{c}{Horizon [Mpc]} \\
\cline{2-3}\cline{4-6}
Array & $\rho_{\rm comb}$ & $N_{\rm det}$ & Tier~1 & Tier~2 & Tier~3 \\
\hline
SKA (200 psr, 20~yr)        & 16.8 &  0 & 336 &  60 &  51 \\
IPTA 2050 (131 psr, 50~yr)  & 20.4 &  2 & 408 &  93 &  80 \\
Combined (331 psr)          & 26.5 &  2 & 529 &  93 &  80 \\
\hline\hline
\end{tabular}
\end{table}

\begin{figure}[t]
  \centering
  \includegraphics[width=\columnwidth]{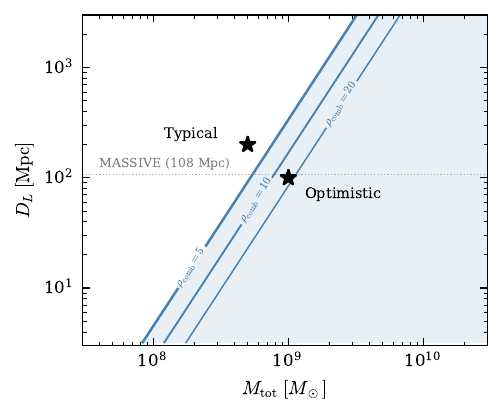}
  \caption{Echo detectability in the $(M_{\rm tot},D_L)$ plane for the SKA-era array parameters in Table~\ref{tab:facilities}, assuming face-on orientation ($\iota = 0$). Contours show the combined PTA echo SNR $\rho_{\rm comb}$ at 5, 10, and 20, computed using the full antenna pattern Eq.~\eqref{eq:residual}. The shaded blue region is detectable ($\rho_{\rm comb} > 5$, Tier~1); Tiers~2 and~3 require progressively closer sources or higher masses. Stars mark the Typical and Optimistic fiducial scenarios from Table~\ref{tab:scenarios}. The dotted horizontal line marks the MASSIVE survey boundary at 108~Mpc~\cite{Ma2014}. The Liepold \& Ma~\cite{LiepoldMa2024} black hole mass function predicts $\sim500$ SMBHs with $M_{\rm BH}>10^9\;\Msun$ in the MASSIVE survey volume.}
  \label{fig:horizon}
\end{figure}

The echo science program divides into three tiers of increasing ambition, each with a distinct horizon in the $(M_{\rm tot},D_L)$ plane. Figure~\ref{fig:horizon} maps the detectability across this parameter space: contours show $\rho_{\rm comb}=5$, 10, and 20 for the SKA-era array, with the shaded region marking the Tier~1 detection zone. The Typical source falls near the detection boundary, while the Optimistic source sits between the $\rho_{\rm comb}=10$ and $\rho_{\rm comb}=20$ contours, Table~\ref{tab:horizon}. The dotted line at 108~Mpc marks the MASSIVE survey volume (not just the MASSIVE footprint).

\textit{Tier~1: Network detection.}---Coherent stacking across the full array yields a combined SNR $\rho_{\rm comb}>5$ even when no individual pulsar resolves the echo. This establishes that the PTA archive contains a deterministic echo associated with the $\mu$Ares source. For face-on sources the combined-array horizon reaches $\sim530$~Mpc for $10^9\;\Msun$.

\textit{Tier~2: Individual echo recovery.}---For more nearby sources, five or more individual pulsars resolve the echo with $\rho_i \ge 3$, providing per-pulsar measurements of $f_{P,i}$ and $h_{P,i}$. The threshold of five comes from the post-Newtonian decomposition described in Section \ref{sec:phase}: separating the Newtonian, 1pN, 1.5pN, and 2pN contributions requires at least four independent baselines, plus one to fix the normalization. For a face-on $10^9\;\Msun$ binary this is reached at $D_L \lesssim 93$~Mpc. The long-baseline IPTA pulsars dominate Tier~2 because individual-pulsar sensitivity scales as $\rho_i \propto \sqrt{T_{\rm obs}}$ Eq.~\eqref{eq:snr}. At this tier the measured echo frequencies at different look-back times already trace the inspiral rate, and the pN frequency shifts in Table~\ref{tab:scenarios} show that 1pN and 1.5pN corrections are individually resolvable.

\textit{Tier~3: Phase-coherent reconstruction.}---The final tier requires multiple anchor pulsars with $\rho_i \ge 3$. The minimum anchor count needed for reliable branch selection has not yet been determined through injection-recovery studies, but the expected population at 10~nHz is $\mathcal{O}(3$--$5)$, see Appendix~\ref{app:anchors}. Once anchor pulsars eliminate the cycle ambiguity, the array traces the waveform coherently over the thousand-year baselines. For a face-on $10^9\;\Msun$ binary the Tier~3 horizon is $\sim80$~Mpc.

Table~\ref{tab:horizon} summarizes these results for each array configuration. The combined array achieves the deepest Tier~1 horizon through the $\rho_{\rm comb}\propto\sqrt{N}$ scaling, but Tiers~2 and~3 are set entirely by the canonical IPTA pulsars, whose longer baselines give higher per-pulsar SNR.

\begin{table*}[t]
\caption{Echo parameters for three SMBHB scenarios with $q=1$, $\chi=0.98$ spins aligned, $\beta_{\rm SO}=7.68$, and $\tau=L_p/c$ ($\theta=90^\circ$). The 1.5pN tail and spin-orbit pieces are given separately since they contain different physics, see Eqs.~\eqref{eq:phase} and~\eqref{eq:fdot}. The 2pN frequency shift is below $0.03$~nHz throughout and is therefore not reported. Frequency shifts are considered significant if they move the signal by one frequency bin, defined by $\delta f_{\rm bin}=1/T_{\rm obs}$. This is $1.6$~nHz at 20~yr and $0.6$~nHz at 50~yr.}
\label{tab:scenarios}
\centering
\setlength{\tabcolsep}{2pt}
\begin{tabular}{lccccccccccccccc}
\hline\hline
 & \multicolumn{2}{c}{Earth term} &
   \multicolumn{4}{c}{Pulsar term} &
   \multicolumn{6}{c}{Gravitational-wave cycles} &
   \multicolumn{3}{c}{$\delta f$ [nHz]} \\
\cline{2-3}\cline{4-7}\cline{8-13}\cline{14-16}
Scenario & $M_{\rm tot}$ & $f_E$ & Pulsar & $\tau$  & $f_P$ & $r_P$ & Newtonian & 1pN & 1.5pN & 1.5pN & 2pN & Total & 1pN & 1.5pN & 1.5pN \\
 & $\Msun$& $\mu$Hz & & yr & nHz & ns & & & tail & SO/$\beta$ & & & & tail & SO \\
\hline
Conservative & $10^8$ & 10 & 1~kpc & 3{,}262 & 189 & <0.01 & 30{,}989 & $+$322 & $-$158 & $+$12.6 & $+$4 & 31{,}253 & $+$1.2 & $-$0.4 & $+$0.3 \\
$D_L\!=\!2$~Gpc & & & J0437 & 512 & 381 & <0.01 & 9{,}713 & $+$157 & $-$95 & $+$7.5 & $+$3 & 9{,}836 & $+$3.8 & $-$1.7 & $+$1.0 \\[3pt]
Typical & $6\!\times\!10^8$ & 1 & 1~kpc & 3{,}262 & 62 & 2.6 & 10{,}030 & $+$157 & $-$91 & $+$7.3 & $+$2 & 10{,}153 & $+$0.6 & $-$0.3 & $+$0.2 \\
$D_L\!=\!200$~Mpc & & & J0437 & 512 & 125 & 2.0 & 3{,}090 & $+$73 & $-$51 & $+$4.1 & $+$2 & 3{,}144 & $+$1.9 & $-$1.0 & $+$0.6 \\[3pt]
Optimistic & $10^9$ & 1 & 1~kpc & 3{,}262 & 45 & 13.3 & 7{,}318 & $+$132 & $-$83 & $+$6.6 & $+$2 & 7{,}420 & $+$0.5 & $-$0.2 & $+$0.1 \\
$D_L\!=\!100$~Mpc & & & J0437 & 512 & 91 & 10.5 & 2{,}273 & $+$62 & $-$48 & $+$3.8 & $+$2 & 2{,}318 & $+$1.6 & $-$0.9 & $+$0.6 \\
\hline\hline
\end{tabular}
\end{table*}

Having identified the mass window $M_{\rm tot}\gtrsim10^{8.5}~M_\odot$ and the Tier-3 horizon of $\sim80$~Mpc, we now ask whether the local Universe contains any such systems. We look to the recent work by Liepold \& Ma~\cite{LiepoldMa2024} for guidance. Following their approach, we integrate the black hole mass function over the MASSIVE survey volume of $2.05\times10^6$~Mpc$^3$ within 108~Mpc~\cite{Ma2014}. We find roughly 500 SMBHs with $M_{\rm BH}>10^9\,M_\odot$, and approximately 1500 with $M_{\rm BH}>5\times10^8\,M_\odot$. Propagating the Schechter-fit parameter uncertainties yields 90\% confidence ranges of 240 to 1010 and 820 to 2650 respectively, consistent with their 90\% band of Figure~4. Folding in the galaxy pairing rate inferred by Casey-Clyde et al.~\cite{CaseyClyde2025}, which is in turned calibrated to the GWB amplitude, gives an expected number of near-equal-mass binaries that crosses unity at $D\approx83$~Mpc. This is comparable to the Tier-3 horizon, with $\sim2$ expected within 108~Mpc, see Fig.~\ref{fig:Nsmbhb} and Appendix~\ref{app:population_estimate} for the full details of this estimate. This same population yields $h_c\approx2.0\times10^{-15}$ through the Phinney formalism~\cite{Phinney2001}, consistent with current PTA measurements. The source population required for gravity echoes is therefore drawn from the same census that produces the observed GWB. If $\mu$Ares identifies such a source anywhere on the sky, the PTA echo search is performed retroactively on the archived timing data, and only one such detection is needed.

% ======================================================================

% ======================================================================
\section{Post-Newtonian phase evolution}\label{sec:phase}

A central technical challenge of the echo program is maintaining phase fidelity over the full Earth--pulsar baseline. Over hundreds to thousands of years of look-back times, the binary accumulates thousands of GW cycles, so even modest post-Newtonian corrections become observationally important. In this section we show why pN terms must be included, why the primary spin measurement must come from the $\mu$Ares Earth term rather than the pulsar network, and how a small set of anchor pulsars can convert the PTA archive into a phase-coherent long-baseline test of the inspiral.

\subsection{The importance of spins}

\begin{figure}[t]
  \centering
  \includegraphics[width=\columnwidth]{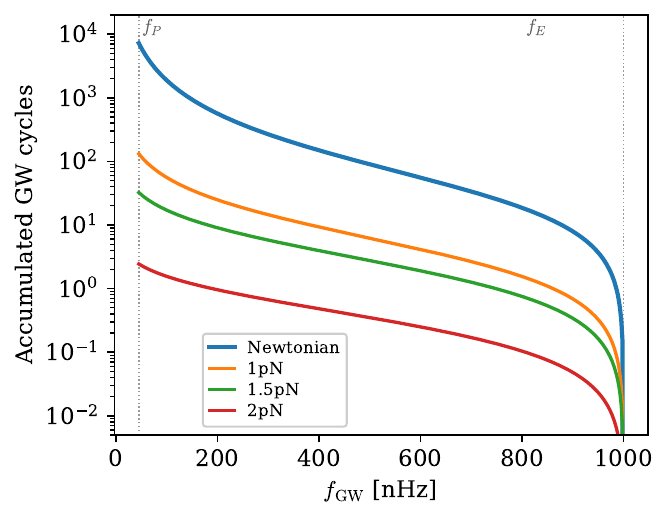}
  \caption{Cumulative GW cycles by post-Newtonian order for the Optimistic binary on the orthogonal 1~kpc fiducial baseline. Parameters and the per-order cycle budget are given in Table~\ref{tab:scenarios}. The vertical dashed lines mark the pulsar-term frequency, $f_\mathrm{P}$, and Earth term frequency, $f_\mathrm{E}$.}
 \label{fig:evolution}
\end{figure}

Spin-orbit coupling enters the GW phase at 1.5pN order, contributing $\sim10$--$100$ cycles over a kpc baseline for $10^9\;\Msun$ binaries~\cite{Mingarelli2012}, see Fig.~\ref{fig:evolution}, comparable to the 1pN mass contribution. Supermassive black holes are expected to carry significant spin, $\chi \sim 0.1$--$0.99$, from mergers and prolonged accretion~\cite{Reynolds2021}. Thus omitting spin mismodels the phase by $\sim10$--$100$ cycles and the matched filter cannot coherently track the echo.

Two post-Newtonian approximants enter our analysis~\cite{Blanchet2006,BIOPS2009}. TaylorT2 gives the GW phase $\Phi(v)$ and time $t(v)$ as closed-form functions of the pN velocity $v=(\pi M_s f)^{1/3}$ with $M_s = GM_{\rm tot}/c^3$. This maps directly onto the echos' pulsar-term frequencies $f_{P,i}$ at known look-back times $\tau_i$. We therefore we use it to compute the total cycles between pulsar and Earth terms. TaylorF2 is the frequency-domain perturbative expansion in which each pN order appears as an additive correction in powers of $v$, which we use to split the total into its per-order contributions in Table~\ref{tab:scenarios} and to write $\dot f$ as an explicit perturbative series in Eq.~\eqref{eq:fdot}.

The TaylorT2 GW phase through 2pN order is
\begin{equation}\label{eq:phase}
  \Phi(v) = \Phi_c - \frac{1}{16\eta}\,v^{-5}
  \!\left[1 + \phi_2\,v^2 + \phi_3\,v^3 + \phi_4\,v^4\right],
\end{equation}
where $\Phi_c$ is the coalescence phase, $\eta = m_1 m_2/M_{\rm tot}^2$ is the symmetric mass ratio, $\phi_2 = \frac{3715}{1008} + \frac{55\eta}{12}$ is the 1pN coefficient, $\phi_3 = -10\pi + \frac{5}{2}\beta_{\rm SO}$ includes the spin-orbit parameter $\beta_{\rm SO} = \frac{1}{12}\sum_i \bigl[113 (m_i/M_{\rm tot})^2 + 75\eta\bigr]\chi_i\cos\kappa_i$, and the 2pN coefficient is $\phi_4 = \frac{15293365}{508032} + \frac{27145\eta}{504} + \frac{3085\eta^2}{72} - 10\,\sigma_{\rm SS}$, with the spin-spin parameter $\sigma_{\rm SS} = \frac{\eta}{48}\bigl[-247\,\hat{s}_1\!\cdot\!\hat{s}_2 + 721\,(\hat{L}\!\cdot\!\hat{s}_1)(\hat{L}\!\cdot\!\hat{s}_2)\bigr]\chi_1\chi_2$~\cite{MBMG2005}.

The number of GW cycles between the pulsar and Earth terms decomposes as
\begin{equation}\label{eq:cycles}
 N_{\rm total} = N_{\rm Newt} + N_{\rm 1pN} + N_{\rm 1.5pN} + N_{\rm 2pN}\,,
\end{equation}
where $N_{\rm total} = \Delta\Phi/(2\pi)$ is computed from Eq.~\eqref{eq:phase}, with $v_P$ obtained from the corresponding TaylorT2 time equation, and the individual pN corrections follow from the TaylorF2 decomposition,
\begin{equation}\label{eq:Nn_pN}
N_n = \frac{3}{256\pi\eta}\,\psi_n\,\Delta[v^{2n-5}]\,,
\end{equation}
where $\psi_n$ are the standard coefficients from Blanchet 2006~\cite{Blanchet2006}. The Newtonian term is defined as the residual $N_{\rm Newt}=N_{\rm total}-\sum_n N_n$. Among these orders, the Newtonian contribution and the 1pN coefficient $\phi_2$ in Eq.~\eqref{eq:phase} depend only on the component masses. The 1.5pN coefficient $\phi_3$ is the sum of two physically distinct pieces, the non-spin tail $-10\pi$ and the spin-orbit contribution $(5/2)\beta_{\rm SO}$, which we report as separate columns in Table~\ref{tab:scenarios}. The 2pN coefficient $\phi_4$ of Eq.~\eqref{eq:phase} is reported as a single column.

For the fiducial aligned-spin case with $\kappa=0$ adopted throughout this work, Thomas precession vanishes and the full spin contribution appears in the 1.5pN spin-orbit column of Table~\ref{tab:scenarios}. For misaligned spins, orbital-plane precession would contribute additional cycles that could help break parameter degeneracies~\cite{Apostolatos1994,Kidder1995}, and will be explored in future work.

The pN terms also introduce frequency shifts in the pulsar terms, as well as additional phase. Whether individual pN frequency shifts are detectable depends on the PTA bin width $\delta f_{\rm bin}=1/T_{\rm obs}$, which is $1.6$~nHz at 20~yr and $0.6$~nHz at 50~yr. 

At J0437 the 1pN shift exceeds the 20-yr bin in the Conservative and Typical scenarios, reaches it in the Optimistic case, and peaks at $+3.8$~nHz for the Conservative source, see Table~\ref{tab:scenarios}. The 1.5pN tail shift at J0437 exceeds the 50-yr bin in all three scenarios and also is resolvable in the 20-yr bin in the Conservative case, while the spin-orbit shift at J0437 reaches the 50-yr bin across all three cases. The tail and spin-orbit pieces carry opposite signs, so the net 1.5pN shift is smaller than either piece individually. The 2pN shift is below $0.01$~nHz throughout and is therefore omitted from Table~\ref{tab:scenarios}. 

With the 1~kpc baseline all pN shifts are at or below the 20-yr bin, and only the Conservative 1pN shift exceeds the 50-yr bin width. The cycle budget carries comparable 1pN and spin-orbit contributions, but $|\Delta f_P|$ itself is mass-dominated, with the 1pN shift exceeding the spin-orbit shift by a factor of a few in every scenario and baseline we explore in Table~\ref{tab:scenarios}. The spin-orbit piece remains the only pulsar-term signature of spin, since the 1pN shift depends only on the component masses. Phase accumulation therefore remains the primary observable, while the measurable 1pN shift and the individually resolved 1.5pN tail and spin-orbit pieces at J0437 provide an independent diagnostic that does not require the more difficult phase connection.

For completeness, we also give the corresponding frequency evolution, which follows from the same pN expansion but is written in terms of the chirp mass $\mathcal{M}_c$ for direct comparison with $\mu$Ares observables:
\begin{equation}\label{eq:fdot}
  \dot{f} = \dot{f}_{\rm N}
  \!\left[1 + \delta_{\rm 1pN} + \delta_{\rm tail} + \delta_{\rm SO} + \delta_{\rm 2pN} + \cdots\right] \, ,
\end{equation}
with the Newtonian rate
\begin{equation}\label{eq:fdotN}
 \dot{f}_{\rm N} = \frac{96}{5}\,\pi^{8/3}\!\left(\frac{G\mathcal{M}_c}{c^3}\right)^{\!5/3}\!f^{11/3}\, ,
 \end{equation}
 and fractional pN corrections $\delta_{\rm 1pN}=-(743/336+11\eta/4)\,v^2$, $\delta_{\rm tail}=4\pi\,v^3$, $\delta_{\rm SO}=-\beta_{\rm SO}\,v^3$, and $\delta_{\rm 2pN}=(34103/18144+13661\eta/2016+59\eta^2/18)\,v^4$, with $\beta_{\rm SO}$ as in Eq.~\eqref{eq:phase}, reducing to $47\chi/6$ for equal masses and aligned spins. The spin-orbit entry $\delta_{\rm SO}$ is the direct $\dot f$ analogue of the $N_{\rm SO}$ cycle count in Table~\ref{tab:scenarios}. For aligned spins, $\delta_{\rm SO}<0$ and decelerates the inspiral. Over a fixed look-back baseline $\tau$, the slower frequency evolution leaves $v_P$ closer to $v_E$ than it would be without spin, so $\Delta f_P = f_P^{\rm spin}-f_P^{\rm nospin}>0$. The total 2pN correction is mass-dominated and same-sign at the few-percent level, and does not alter the sign. The resulting pulsar-term shift $\Delta f_P$ is tabulated in the spin-orbit column of Table~\ref{tab:scenarios}.
 
Figure~\ref{fig:evolution} illustrates the cumulative phase budget for the Optimistic binary over the 1~kpc fiducial baseline. The Newtonian term dominates, but the 1pN and 1.5pN pieces are large enough to be accessed either through frequency-bin shifts at J0437 (Tier~2) or through phase connection across the anchor set (Tier~3).

% ======================================================================
\subsection{The spin-distance degeneracy}\label{sec:spin}

Phase-coherent echo analysis requires the Earth-term measurement, propagated backward to each pulsar's retarded epoch, to track the echo phase to better than roughly one radian. For the Optimistic binary, the spin-orbit contribution over the 1~kpc fiducial baseline is $N_{\rm SO}\approx51$ cycles, see Table~\ref{tab:scenarios}, corresponding to a phase contribution of several hundred radians. Demanding $\delta\Phi_{\rm SO}\lesssim1$ rad therefore requires a fractional spin precision of order $\delta\chi/\chi \lesssim 10^{-3}$.

The $\mu$Ares Earth-term measurement at $f_E \sim 1\;\mu$Hz provides this precision. A single-parameter TaylorF2 Fisher integral over the first year of $\mu$Ares observation of the Optimistic source gives
\begin{equation}\label{eq:sigma_chi_earth}
  \sigma_\chi^{(E)} \;\lesssim\; 10^{-8}\,,
\end{equation}
at $D_L = 100$~Mpc, using the sky-averaged $\mu$Ares noise curve and the 1.5pN spin-orbit phase term. This is a one-sided bound: truncating the integration at one year or at merger, whichever comes first, only reduces the information content, so the true Fisher precision is at least this good. The associated matched-filter SNR is $\rho_E \gtrsim 6\times10^6$ at 100~Mpc, rising to $\rho_E \gtrsim 8\times10^6$ at the golden-binary distance of 80~Mpc. The spin precision exceeds the $10^{-3}$ coherence requirement by roughly five orders of magnitude. A full multiparameter analysis could degrade this precision, but the available margin is so large that the qualitative conclusion would be unchanged. Even if the low-frequency noise floor is an order of magnitude worse than the nominal $\mu$Ares design, the Earth-term spin measurement remains more than sufficient.

By contrast, the pulsar network alone cannot determine the spin, as we show quantitatively in Appendix~\ref{app:spin_degeneracy}. Although each pulsar term accumulates more spin-orbit cycles than the Earth term, its total phase depends on both $\chi$ and $\tau_i$. This produces an approximate spin--distance degeneracy: shifts in $\chi$ can be absorbed by compensating shifts in $\tau_i$, so the marginalized Fisher information on $\chi$ vanishes in the absence of external distance information. Distance priors from VLBI partially break this degeneracy, but even the best current pulsars fall short of the $\delta\chi/\chi \lesssim 10^{-3}$ requirement by more than two orders of magnitude --- see Appendix~\ref{app:spin_degeneracy}. The primary spin information must therefore come from the Earth term, not from the pulsar terms. Improving pulsar distances remains important, but mainly because it reduces cycle ambiguity in the echo phase rather than because it directly constrains $\chi$.

% ======================================================================
\subsection{Phase connection via anchor pulsars}\label{sec:phase_connection}

Once $\chi$ is fixed by the Earth term, the echo phase at each pulsar becomes a probe of its retarded epoch. The role of anchor pulsars is to provide absolute phase reference points at known look-back times. Comparing the backward-propagated $\mu$Ares measurement against the actual echo phases at these fixed anchor points tests whether the inspiral proceeded as predicted over the intervening temporal baseline. Any systematic phase drift, discontinuity, or trend with $\tau_i$ would signal non-vacuum, or unmodeled physics in the binary's evolution.

Current VLBI campaigns such as PSR$\pi$~\cite{Deller2019} and MSPSR$\pi$~\cite{Ding2023} provide $\sim20$ millisecond pulsars with $\lesssim1\%$ distance precision out to $\sim3$~kpc, and this sample should grow with SKA-VLBI baselines. These pulsars are informative, but they are not anchors: their distance uncertainties produce cycle ambiguities
\begin{equation}\label{eq:cycle_ambig}
\Delta N \approx f_P\,\sigma_\tau \, ,
\qquad
\sigma_\tau = (\delta L_p/L_p)\,\tau \, ,
\end{equation}
of order tens to hundreds across the pulsar-term frequency range. For example, a pulsar at $L_p = 1$~kpc with $\delta L_p/L_p = 1\%$ has $\sigma_\tau \approx 33$~yr, giving $\Delta N \approx 50$ at $f_P = 50$~nHz and $\Delta N \approx 100$ at $f_P = 100$~nHz. Optical astrometry is a complementary route: Gaia parallaxes of white-dwarf companions to PTA pulsars already improve a subset of MSP distances by $30$--$50\%$~\cite{Moran2023}, and Roman should extend this approach to fainter systems~\cite{McKinnon2026}.

Only pulsars within $L_{p,\max} \simeq 280$--$390$~pc satisfy $\delta L_p < c/(2\pi f)$ at 10~nHz, the lowest frequency at which we evaluate the anchor criterion and where it is least stringent. The anchor population scales as $N_{\rm anchor}\propto(\sigma_\varpi\,f)^{-3/2}$, Eq.~\eqref{eq:Nanchor_scaling}, falling from $\mathcal{O}(3$--$5)$ at 10~nHz to $\mathcal{O}(1)$ by 30~nHz.

Once $\mu$Ares supplies the Earth-term measurement, each pulsar term becomes a one-dimensional matched filter over $\tau_i$. For the Optimistic source, the combined array reaches $\rho_{\rm comb}\approx27$ at $D_L=100$~Mpc, with hundreds of pulsars contributing at $\rho_i>1$. The anchors fix the absolute phase at known epochs, while the VLBI-calibrated pulsars contribute additional constraints whose likelihoods are periodic in $\tau_i$ with period $1/f_P$, corresponding to neighboring peaks separated by $\delta L_p \sim c/f_P \sim 0.3$~pc at 30~nHz. A realistic phase-connection pipeline will therefore use the anchors to lock the global phase and the VLBI priors to select among residual aliases. Demonstrating this in realistic noise injections is a clear target for future work.

% ======================================================================

\section{Sky localization via echo rings}\label{sec:skyloc}

The angular dependence of the single-pulsar SNR (Sec.~\ref{sec:snr_angle}) implies that each pulsar is sensitive to echoes from sources lying within a characteristic annular ring on the sky, centered on the pulsar at radius $\sim\theta_{\rm opt}$ Eq.~\eqref{eq:theta_opt} with width set by the condition $\rho>\rho_{\rm thr}$. A detection constrains the source to that ring, while a nondetection excludes it. With multiple pulsars, the intersection of detected rings and the complement of nondetected rings rapidly narrows the allowed sky region.

\begin{figure}[t]
  \centering
  \includegraphics[width=\columnwidth]{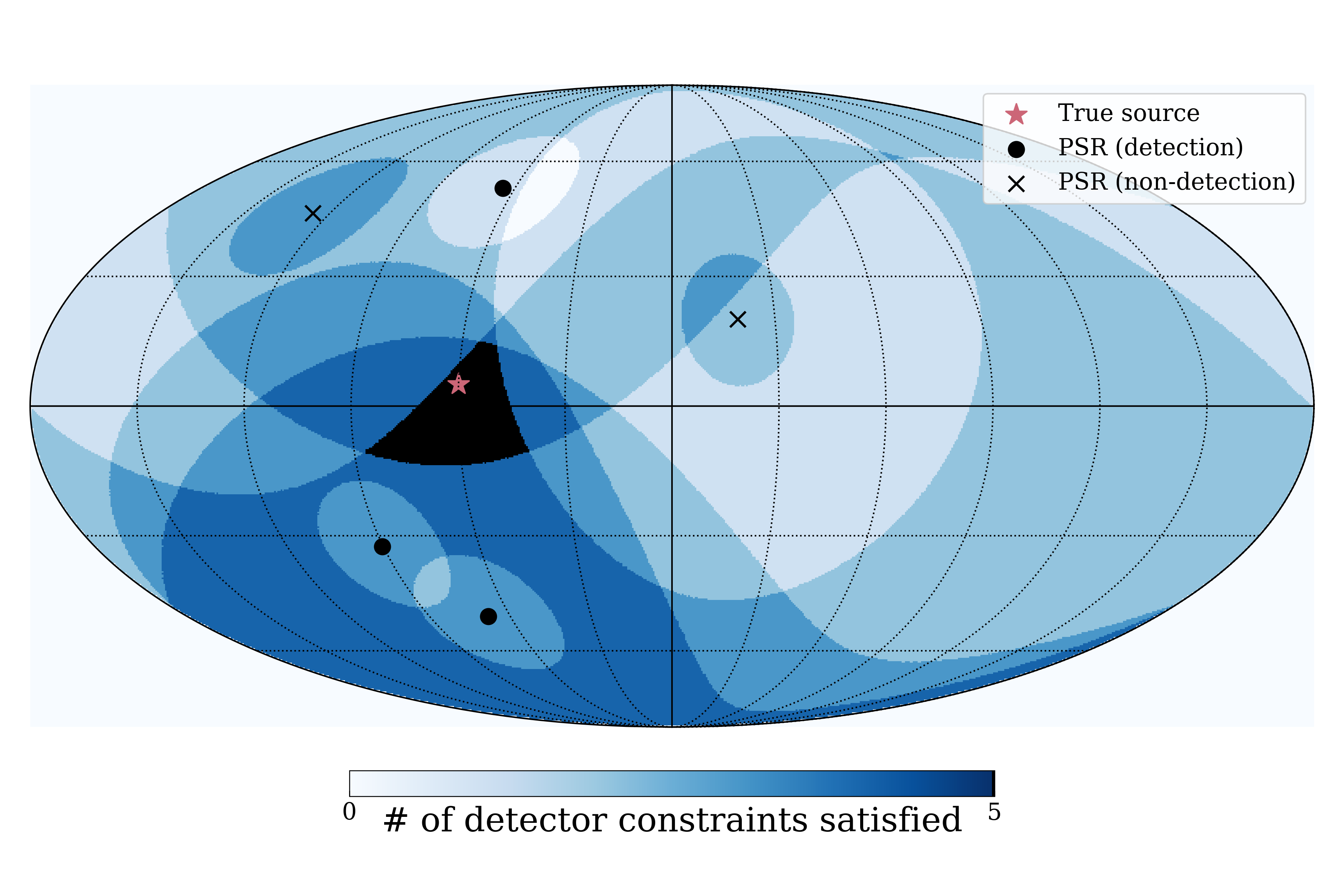}
  \caption{Sky localization of the Optimistic source using five pulsars with $\sigma_{\rm TOA}=30$~ns. Filled circles mark pulsars that detect the echo, while crosses mark nondetections. The color scale shows the number of detection and exclusion constraints satisfied at each sky position. The black patch satisfies all constraints and recovers the true source location (pink star).}
  \label{fig:skyloc}
\end{figure}

Figure~\ref{fig:skyloc} illustrates this for the Optimistic binary (Table~\ref{tab:scenarios}) using five pulsars with $\sigma_{\rm TOA}=30$~ns (a best-precision target, cf.\ Fig.~\ref{fig:snr_vs_theta}) and 30-year baselines. Three pulsars at favorable sky locations detect the echo, while the remaining two contribute through nondetections. The color scale shows the number of constraints satisfied at each sky position, and the black region satisfies all constraints simultaneously. Even this simple configuration, which we chose for illustrative purposes, localizes the source to $\lesssim 1$~sr. This is coarser than the $\mu$Ares sky map, but the echo-ring localization remains valuable: it provides an independent sky constraint from a completely different measurement technique, serves as a consistency check that the detected pulsar terms originate from the same source, and for non-Tier~3 sources where the $\mu$Ares error box is larger, partial overlap between the echo rings and the $\mu$Ares localization could exclude regions of parameter space.

To quantify this behavior statistically, we repeat the exercise over many realizations and different numbers of high-precision pulsars. Table~\ref{tab:skyloc} lists the median, 5th, and 95th percentile sky areas for 5, 10, 20, 50, and 100 pulsars with $\sigma_{\rm TOA}=30$~ns, together with the median number of pulsar terms with SNR$>5$, $N_{\rm det}^{\rm med}$. For small arrays the localization region is typically hundreds to thousands of deg$^2$, consistent with only one or two pulsars detecting the echo\footnote{The detectable range of angles is $16^{\circ}\lesssim\theta\lesssim66^{\circ}$, an annulus covering about 28\% of the sky, so a randomly placed pulsar has a 28\% chance of detecting the pulsar term.}. As more pulsars are added, the localization improves substantially, reaching tens to hundreds of deg$^2$ with 20--50 high-precision pulsars.

\begin{table}[t]
\caption{Statistics of echo-ring localization for the Optimistic binary ($10^9\;\Msun$, $D_L=100$~Mpc) as a function of the number of high-quality pulsars with $\sigma_{\rm TOA}=30$~ns. We report the median number of individually detected pulsar terms with SNR$>5$, $N_{\rm det}^{\rm med}$, together with the median and 5th and 95th percentiles of the reconstructed sky area ($A_{\rm med}$, $A_{\rm 5th}$, $A_{\rm 95th}$).}\label{tab:skyloc}
\centering
\setlength{\tabcolsep}{3pt}
\begin{tabular}{llrrr}
\hline\hline
$N_{\rm psr}$ & $N_{\rm det}^{\rm med}$ & $A_{\rm 5th}$ [deg$^2$] & $A_{\rm med}$ [deg$^2$] & $A_{\rm 95th}$ [deg$^2$] \\
\hline
5   & 1  & 750 & 4,100 & 11,700 \\
10  & 3  & 140 & 1,000 & 3,800 \\
20  & 5  & 27  & 230   & 1,000 \\
50  & 14 & 4   & 36    & 170 \\
100 & 28 & 1   & 9     & 45 \\
\hline\hline
\end{tabular}
\end{table}

These are conservative estimates in the sense that they use only the binary classification of each pulsar as a detection or nondetection within the allowed annulus. In practice, each detected pulsar term also provides an amplitude $r_{P,i}$ Eq.~\eqref{eq:residual}, a frequency $f_{P,i}$ Eq.~\eqref{eq:fP_exact}, and, when phase information is available, an additional timing constraint. Requiring consistency with the Earth-term parameters should therefore further reduce the allowed sky region.

\section{Science return from a golden binary}\label{sec:science}

The detectability analysis in Sec.~\ref{sec:detectability} used the Optimistic scenario at 100~Mpc as a round-number benchmark. We now ask what happens at the boundary where the echo program becomes feasible. Two independent constraints set this boundary. First, the PTA sensitivity analysis gives a Tier-3 horizon of $\sim80$~Mpc for a face-on $10^9\;\Msun$ binary, Table~\ref{tab:horizon}: this is the maximum distance at which two or more anchor pulsars can phase-lock the echo waveform. Second, the population estimate in Appendix~\ref{app:population_estimate} predicts that the expected number of near-equal-mass $10^9\;\Msun$ binaries in the MASSIVE volume crosses unity at $D\approx83$~Mpc, Fig.~\ref{fig:Nsmbhb}. These independent estimates coincide near $\sim80$~Mpc, so the local Universe may contain one source close enough for phase-coherent reconstruction.

We therefore adopt $D_L=80$~Mpc as the fiducial distance for an equal-mass binary with $M_{\rm tot}=10^9\;\Msun$, $f_E=1\;\mu$Hz, and aligned spins $\chi=0.98$. At this distance the source exceeds the Tier~1 and Tier~2 thresholds: $\rho_{\rm comb}\approx33$ and $N(\rho_i\ge3)=24$, with two anchor pulsars detected at $\rho_i\ge3$. We call this a \emph{golden binary}: one system for which the Earth term and the individually resolved pulsar terms are available, and for which the anchor population can potentially phase-lock the waveform. This is the most ambitious of the three tiers, and its feasibility depends on pulsar distances reaching the precision tabulated in Appendix~\ref{app:anchors}. We use this source as a worked example that shows what the echo framework may be able to measure. Table~\ref{tab:science_summary} summarizes the main results.

\begin{table*}[t]
\caption{Demonstrated results and future prospects for a golden binary with $q=1$, $M_\mathrm{tot}=10^9\;\Msun$, $D_L=80$~Mpc, and $\chi=0.98$. The first two blocks list quantities computed in this work. The final block lists prospective physical measurements. Circumbinary disk constraints are estimated quantitatively to illustrate the physics we can probe with echoes.}\label{tab:science_summary}
\centering
\small
\setlength{\tabcolsep}{3pt}
\begin{tabular}{l l l p{5.6cm} r}
\hline\hline
Stage & Measurement & Value & Physical insight & Reference \\
\hline
\multirow{2}{*}{$\mu$Ares Earth term}
& Spin $\chi$ & $\sigma_\chi \lesssim 10^{-8}$ & BH growth history & Eq.~\eqref{eq:sigma_chi_earth} \\
& Masses, sky, inclination & $\rho_E \gtrsim 8\!\times\!10^6$ & Template for echo search & Sec.~\ref{sec:detectability} \\[3pt]
\hline
\multirow{6}{*}{Echo detection}
& Echo frequencies $f_{P,i}$ & 20--170~nHz & Inspiral rate at $100$--$1,000$~yr & Table~\ref{tab:scenarios} \\
& Network echo SNR & $\rho_{\rm comb} \approx 33$ & Confirms binary; horizon $\sim$530~Mpc & Table~\ref{tab:horizon} \\
& J0437 spin bin-shift & $|\Delta f_P| \sim 0.5$--$1.0$~nHz & Spin-induced shift across scenarios, marginal at $T_{\rm obs}=50$~yr & Table~\ref{tab:scenarios} \\
& Sky localization & $\sim$30--230~deg$^2$ & Sky localization with 20--50~psr & Table~\ref{tab:skyloc} \\
& pN consistency test & $\delta\phi_{1\rm pN}\lesssim 0.2\%$, $\delta\phi_{1.5\rm pN}\lesssim 0.7\%$ & Weak-field test of 1pN, 1.5pN dynamics & Sec.~\ref{sec:science} \\
& Pulsar distances & $\delta L_p \sim 0.3$~pc & Cycle-resolved distance at $\sim c/f_P$ & Sec.~\ref{sec:phase_connection} \\[3pt]
\hline\hline
\multirow{5}{*}{Prospective}
& Circumbinary disk & $\delta\Phi \sim -1$ to $-9$~rad & Accretion rate; sensitivity to $\varepsilon \gtrsim 10^{-5}$ & Sec.~\ref{sec:environment} \\
& Boson clouds & --- & Constraints for $\alpha\sim0.1$--$0.3$ & Sec.~\ref{sec:environment} \\
& Dark-matter spike & --- & Dark-matter density profile & Sec.~\ref{sec:environment} \\
& Eccentricity & --- & Circularization history & Sec.~\ref{sec:environment} \\
& Tests of General Relativity & --- & Scalar-tensor gravity; massive graviton & Sec.~\ref{sec:environment} \\
\hline\hline
\end{tabular}
\end{table*}

\textit{Step~1: Earth-term detection.}---$\mu$Ares detects the source at $\rho_E \gtrsim 8\times10^6$, measuring the binary masses, spins, sky location, and inclination. At $f_E=1\;\mu$Hz, the binary merges in $\sim0.8$~yr, so $\mu$Ares follows the final inspiral and merger directly. The PTA search is archival: each pulsar recorded the binary at its own retarded epoch $\tau_i$, when the system was still deep in the adiabatic inspiral. Once the Earth-term waveform is known, the archived pulsar terms become targeted searches. The 2pN truncation used throughout is adequate for the $\delta\phi_{1\rm pN}$ and $\delta\phi_{1.5\rm pN}$ consistency tests reported in Sec.~\ref{sec:science}. A $\delta\phi_{2\rm pN}$ test, or propagation through the strong-field merger, would require 3pN and higher terms.

\textit{Step~2: Echo recovery.}---The matched-filter search over $\tau_i$ then identifies the pulsar terms across the array. For the golden binary, the combined PTA SNR is $\rho_{\rm comb}\approx33$. Among the IPTA~2050 pulsars with 50-year baselines, up to 24 individually resolve the echo at $\rho_i \ge 3$, spanning the pulsar-term frequency range set by the PTA's geometry, Sec.~\ref{sec:horizon}, corresponding to look-back times from hundreds to thousands of years. This is larger than the $N_{\rm det}=2$ reported in Table~\ref{tab:horizon} at 100~Mpc because the per-pulsar $\rho_i$ distribution accumulates just below the threshold detection at that distance, so the 25\% SNR boost from reducing $D_L$ to 80~Mpc boosts tens of pulsars over $\rho_i = 3$. At this stage the PTA confirms the binary's presence through the combined SNR $\rho_{\rm comb}$ and through the consistency of individually recovered echo frequencies. Since each pulsar samples a different retarded epoch, the ensemble of recovered frequencies $f_{P,i}$ maps the binary's past inspiral directly.

\textit{Step~3: Cycle budget.}---The per-order cycle budget at the canonical 1~kpc baseline, $\tau\approx3{,}262$~yr, is read off the Optimistic row of Table~\ref{tab:scenarios}: a Newtonian contribution of order $10^4$ cycles, an $\mathcal{O}(10^2)$ correction at 1pN, and a small net 1.5pN piece arising from a near-cancellation between the tail and spin-orbit terms of Eq.~\eqref{eq:cycles}. The farthest anchor pulsar, at $L_{p,\max}\approx390$~pc and $\tau\approx1{,}272$~yr, Table \ref{tab:anchors}, samples a shorter baseline but has same qualitative structure. 

\textit{Step~4: Phase connection.}---The spin phase uncertainty from $\mu$Ares is negligible, see Sec.~\ref{sec:spin}, so the bottleneck here is pulsar distance. Anchor pulsars provide absolute phase reference points and directly lock the waveform at their respective look-back times. Pulsars with good VLBI distance priors still carry the cycle ambiguities quantified in Sec.~\ref{sec:phase_connection}, but once the anchors fix the global phase they provide additional informative constraints. For the golden binary, two anchors supplemented by $\sim20$ VLBI-calibrated MSPs~\cite{Deller2019,Ding2023} may suffice to constrain the waveform over the full PTA baseline, though a full injection-recovery demonstration with realistic noise would be needed for more concrete projections.

\textit{Step~5: Science}---Once the waveform is phase connected, the echoes become a direct long-baseline test of the binary's secular evolution. A coherent PTA echo data set checks whether the vacuum inspiral predicted by the Earth-term template remains self-consistent over thousands of years. Any failure of coherence that correlates systematically with look-back time signals additional physics beyond the modeled post-Newtonian inspiral. This is the regime in which echoes become a precision probe of binary environments and GW propagation.

\subsection{Consistency tests of the secular inspiral}

With the Earth term detection fixing the waveform, each anchor pulsar independently constrains the accumulated phase over its own look-back baseline with no cycle ambiguity. retarded epochs.

Using the farthest-anchor cycle budget from Step~3 above, in a parameterized post-Newtonian framework where $\phi_n \rightarrow \phi_n(1+\delta\phi_n)$, a sub-radian phase measurement constrains
\begin{equation}\label{eq:dphi_n}
\delta\phi_n \lesssim \frac{1}{2\pi N_n}\, ,
\end{equation}
which implies $\delta\phi_{1\rm pN} \lesssim 0.2\%$ and $\delta\phi_{1.5\rm pN} \lesssim 0.7\%$. This is a weak-field test of SMBHB inspiral dynamics at mass scales and baselines unavailable to either ground-based detectors or LISA.

The longer-baseline VLBI-calibrated pulsars carry the cycle ambiguities discussed in Sec.~\ref{sec:phase_connection}, but once the anchors select the correct branch they extend the tested baseline from $\sim1{,}300$ to $\sim3{,}000$~yr. In that regime the full echo network checks whether the same vacuum inspiral model remains self-consistent across all the independent retarded epochs. %A quantitative injection-recovery study determining the minimum anchor count needed for reliable branch selection has not yet been carried out.

\subsection{Environmental and beyond-GR sensitivity}\label{sec:environment}

The same phase precision that enables a post-Newtonian consistency test also makes the echoes sensitive to any cumulative perturbation to the inspiral: the echo baselines integrate small deviations over thousands of wave cycles, while the anchor pulsars fix the phase well enough that those deviations cannot be absorbed into a biased chirp mass or spin unless they share the same frequency dependence. This makes them sensitive to environmental effects and certain beyond-GR modifications.

Circumbinary disk torques provide the clearest example~\cite{HaimanKocsisMenou2009,Duffell2020,Siwek2024}. In the pulsar-term frequency range of $\sim20$--$170$~nHz, the instantaneous disk torque is subdominant to GW backreaction, but its effect accumulates coherently over the full baseline. For a decoupling frequency $f_{\rm dec}=3$~nHz, representative of gas-pressure-dominated circumbinary disks around $\sim10^9\;\Msun$ binaries~\cite{HaimanKocsisMenou2009}, numerical integration in a constant-torque model gives a dephasing of order $\delta\Phi \approx -1$ rad over 1{,}000~yr, $\approx -7$ rad over the canonical 3{,}262~yr baseline, and $\approx -9$ rad at a J1713-like 3{,}800~yr baseline, setting the $-1$ to $-9$ rad range quoted in Table~\ref{tab:science_summary}. 

Shallower prescriptions that hold the orbital-separation decay rate $\dot a$ constant give $\sim -20$~rad on the same baseline, while steeper prescriptions that hold the orbital-energy decay rate $\dot E$ constant give sub-radian signatures. A useful figure of merit is the disk-to-GW frequency-derivative ratio $\varepsilon \equiv \dot f_{\rm disk}/\dot f_{\rm GW}$: a 1-radian sensitivity over $N_{\rm vac}\sim7{,}300$ cycles at $\tau\approx3{,}262$~yr corresponds to $\varepsilon_{\min} \sim 1/(2\pi N_{\rm vac}) \sim 2\!\times\!10^{-5}$. The negative sign reflects that the disk accelerates the inspiral, so the binary's phase at any past retarded epoch is less advanced than the vacuum-template predicts. 

Even a conservative disk can therefore produce order-radian signatures in the golden-binary regime. Since the induced phase correction scales more steeply with frequency than the vacuum pN terms, it is separable across pulsars sampling different $\tau_i$. A full forecast requires coupling the echo formalism to detailed circumbinary-disk simulations~\cite{SiwekWH2023,Tiede2025}, so we leave that to a dedicated follow-up study.

For completeness, we note that the same logic applies more broadly. Superradiant boson clouds~\cite{Arvanitaki2010,Brito2015,Baumann2019}, whose instability rate peaks at gravitational fine-structure coupling $\alpha \equiv GM\mu_b/(\hbar c)\sim0.1$--$0.3$, dark-matter spikes~\cite{Gondolo1999,AlonsoAlvarez2024}, residual eccentricity~\cite{Peters1964,DOrazio2021,Taylor2016}, dipolar radiation from scalar-tensor gravity~\cite{Yunes2009,Barausse2016}, and massive-graviton dispersion~\cite{Will1998,ZhengLarsenEisenbergMingarelli2026} all imprint phase signatures whose frequency dependence differs from the standard pN hierarchy. Some of these effects are source intrinsic, altering the inspiral rate, while others act during propagation and accumulate with pulsar distance. A massive graviton, for example, produces a pulsar-term phase shift scaling as $\Delta\Phi^{\rm Disp}\propto m_g^2 L_p/\omega$~\cite{ZhengLarsenEisenbergMingarelli2026}, which grows with kpc baseline and cannot be mimicked by a source parameter alone. Dipolar radiation enters at $-1$pN order, where the low velocities probed by the echoes are particularly advantageous.

We do not claim quantitative constraints on these effects here. What we establish is the measurement regime in which such constraints become possible: a $\mu$Ares detection fixes the intrinsic binary parameters, the anchor pulsars control the phase, and the PTA's echoes sample the dephasing across many retarded epochs.

% ======================================================================
\section{Discussion and Conclusions}\label{sec:discussion}

The central result of this work is that a $\mu$Hz detection of a nearby massive SMBHB turns PTA pulsar terms from noise into a usable record of the binary's earlier inspiral. Once the Earth-term waveform is known, the archived pulsar terms become targeted gravity echoes that probe the binary at retarded epochs spanning hundreds to thousands of years. This opportunity is confined to the high-mass tail of the SMBHB population, $M_{\rm tot} \gtrsim 10^{8.5}\;\Msun$ within a few hundred Mpc, but within that window the echoes define a distinct multiband observable. Typical LISA-band sources are not massive enough to generate measurable PTA pulsar terms.

The scientific results have a natural hierarchy. At the lowest useful level, the PTA finds that the Earth-term source left a deterministic retarded imprint across the array. Once individual pulsar terms are resolved, the recovered frequencies $f_{P,i}$ at known look-back times $\tau_i$ already test the secular inspiral rate without requiring full phase connection. In the ideal regime, where multiple anchors remove the cycle ambiguity, the array traces the waveform coherently across thousand-year baselines and enables sub-radian consistency checks of the post-Newtonian phase. This is the regime in which the echoes become sensitive to environmental torques and beyond-GR propagation effects.

The expected number of golden binaries remains uncertain since the sub-parsec binary fraction is poorly constrained. PTA continuous-wave searches~\cite{NG15cw}, improved measurements of the GWB, and better empirical constraints on the massive end of the black hole mass function will sharpen this forecast well before a $\mu$Hz mission flies. The absence of a continuous-wave detection today does not argue against the echo program. A $10^9\;\Msun$ binary that is now in the $\mu$Hz band would have crossed the PTA band hundreds of years before present-day observations began. If the binary is accreting, the $\mu$Ares detection may also be accompanied by an electromagnetic counterpart from the circumbinary disk~\cite{Ennoggi2026}, providing independent confirmation of the source and its host galaxy.

While the full realization of this echo program awaits a $\mu$Hz detector, the conceptual framework and much of the required infrastructure can already be developed with existing PTA data. The first two directions below deliver independent science without a $\mu$Hz input, bounding the local high-mass SMBHB population and determining the achievable Tier-3 anchor set. The third lays the data-analysis groundwork for the eventual Earth-term detection.

\textit{Blind echo searches.}---Current continuous-wave search pipelines~\cite{NG15targeted} already target individual SMBHBs across the nHz band. These searches are sensitive to pulsar terms, though they do not exploit them as echo probes because the source parameters are largely unknown. For a massive binary in the $\mu$Ares detection window that has already evolved through the PTA band, the echo signal from its earlier inspiral persists in the timing residuals. A targeted search toward the most massive nearby galaxies from the MASSIVE catalog~\cite{Ma2014} could set quantitative bounds on the echo-signal amplitude from nearby high-mass binaries, constraining the population independently of the GWB. Even a non-detection would tighten the population models used in Sec.~\ref{sec:detectability}.

\textit{Pulsar distance campaigns.}---The anchor-pulsar population, explored in more detail in Appendix~\ref{app:anchors}, is the long-lead item for phase-coherent science. Continued VLBI campaigns targeting the nearest PTA MSPs, combined with Gaia and Roman astrometry to binary companions~\cite{Moran2023,McKinnon2026}, will determine whether 3--5 anchors are achievable at 10~nHz before $\mu$Ares flies. Every improvement in parallax precision for pulsars within $\sim400$~pc directly expands the Tier~3 capability.

\textit{Pipeline development.}---The Bayesian phase-connection pipeline described in Sec.~\ref{sec:phase_connection}, which must navigate a multimodal posterior from cycle ambiguities across $\sim20$ VLBI-calibrated pulsars, can be developed and tested on simulated data now. Injection-recovery studies with realistic noise models would establish the minimum anchor count and distance precision required for Tier~3 science, and would determine whether the cycle ambiguities of non-anchor pulsars (Sec.~\ref{sec:phase_connection}) can be resolved in practice.

On the PTA side, the same program already needed for continuous-wave discovery also prepares us for echo science: long baselines, low timing residuals, improved cadence, and better pulsar distances. On the space-based side, the missing capability is a detector in the $\mu$Hz band. A mission such as $\mu$Ares would close the frequency gap between LISA and PTAs and provide the Earth-term waveform that the advanced science program rests on. If that capability is realized, decades of archived pulsar timing data become a multiband observatory of SMBHB evolution thousands of years in the past.

\begin{acknowledgments}
C.M.F.M.\ thanks D. Spergel, W. M. Farr, M. Isi, and V. Ozolins for useful conversations. She acknowledges support from the NSF NANOGrav Physics Frontiers Center (award PHY-2020265) and NASA LPS 80NSSC24K0440. During the final stages of preparing this manuscript, the authors became aware of concurrent work by Criswell et al.\ (in prep) on pulsar terms in the post-merger regime.
\end{acknowledgments}

% ======================================================================
\section*{Software}

This work made use of
\texttt{hasasia}~\cite{Hazboun2019hasasia,HazbounRS2019} for PTA sensitivity curve computation and deterministic continuous-wave signal-to-noise ratio estimates,
\texttt{gwent}~\cite{Kaiser2021} for multiband detector noise modeling,
\texttt{NumPy}~\cite{numpy2020},
\texttt{SciPy}~\cite{scipy2020},
\texttt{Matplotlib}~\cite{matplotlib2007},
and \texttt{Streamlit}~\cite{streamlit} for the interactive sensitivity visualization.
The post-Newtonian evolution library and the multiband sensitivity dashboard \texttt{app.py} are available at \url{https://github.com/ChiaraMingarelli/gravity-echoes}. A hosted version of the interactive dashboard is at \url{https://gravityecho.streamlit.app}.
Codes were written and edited with the assistance of Claude Code (Anthropic).

% ======================================================================
\appendix

% ======================================================================

\section{Intrinsic pulsar red noise}\label{app:rednoise}

The PTA sensitivity curves shown in this work include the effect of intrinsic pulsar red noise, which dominates the per-pulsar noise budget below $\sim10$~nHz and therefore sets the practical low-frequency boundary of the pulsar-term frequency range. We model the red noise as a power-law contribution to each pulsar's noise covariance matrix,
\begin{equation}\label{eq:rn_psd}
 P_{\rm RN}(f) = \frac{A^2}{12\pi^2}
 \left(\frac{f}{f_{\rm yr}}\right)^{-\gamma}\mathrm{yr}^3\,,
\end{equation}
where $A$ is the red-noise amplitude and $\gamma$ is the spectral index.

For the IPTA~2050 projections, we draw per-pulsar red-noise parameters from ranges representative of the detected achromatic red noise across the NANOGrav 15-year custom-noise analysis~\cite{NG15noise}: $\log_{10}A$ uniformly distributed in $[-15,-12]$ and $\gamma$ uniformly distributed in $[1,5]$. These intervals span the posterior envelope of per-pulsar red-noise detections across the pulsars favoring Fourier-basis Gaussian-process noise models in that analysis, including both steep spin-noise spectra ($\gamma\sim4$--$5$) and shallower processes ($\gamma\sim1$--$2$).

The red noise is incorporated through the pulsar covariance matrix $\mathbf{N}$ using \texttt{hasasia}'s \texttt{sim\_pta} routine~\cite{Hazboun2019hasasia}, which converts the power-law PSD into a time-domain covariance and adds it to the white-noise diagonal. Deterministic continuous-wave sensitivity is then computed from the full covariance following Hazboun, Romano, \& Smith~\cite{HazbounRS2019}.

Because the pulsar-term frequency range emphasized in the main text lies mostly above $\sim20$~nHz, where intrinsic red noise is subdominant for many PTA pulsars, the order-of-magnitude detectability estimates are reasonably approximated by white noise alone. This approximation breaks down below $\sim10$~nHz, where both pulsar red noise and the stochastic background become important, and is the main reason we treat the lower-frequency extension of the echo program conservatively in the main text.

% ======================================================================

\section{Anchor pulsars at 10~nHz}\label{app:anchors}

Phase-coherent echo analysis requires each pulsar's distance uncertainty to satisfy $\delta L_p < c/(2\pi f)$, so that the pulsar-term phase is known to better than one radian. If the distance is determined by parallax with $\sigma_\varpi \ll \varpi$, the distance uncertainty is $\delta L_p \simeq \sigma_\varpi L_p^2/(1\,{\rm AU})$, with $\sigma_\varpi$ in radians. Combining these gives the maximum distance at which a pulsar can serve as an anchor,
\begin{equation}\label{eq:dmax_anchor}
L_{p,\max}(f,\sigma_\varpi)
= \left[\frac{(1\,{\rm AU})\,c}{2\pi f\,\sigma_\varpi}\right]^{1/2}
\simeq 393~{\rm pc}\;
f_{10}^{-1/2}\,\sigma_{\varpi,1}^{-1/2},
\end{equation}
where $f_{10}\equiv f/(10~{\rm nHz})$ and $\sigma_{\varpi,1}\equiv \sigma_\varpi/(1~\mu{\rm as})$. Even for $\sigma_\varpi=1~\mu$as, strict coherence at 10~nHz is limited to $L_{p,\max}\simeq393$~pc. For a more conservative SKA-era precision of $\sigma_\varpi=1.5$--$2~\mu$as, the horizon shrinks to $L_{p,\max}\simeq278$--$321$~pc.

\subsection{Current census}\label{app:census}

The nearest PTA pulsars and their present distance constraints are summarized in Table~\ref{tab:anchors}. Among currently timed PTA millisecond pulsars, only PSR~J0437$-$4715 satisfies the 10~nHz coherence threshold. Its distance, $156.96\pm0.11$~pc, is obtained from the orbital-period derivative via the Shklovskii effect~\cite{Reardon2024}, not from parallax alone. The timing parallax by itself, $L_p=155.4\pm0.9$~pc, would not meet the anchor requirement. This makes J0437 exceptional: its short orbital period, bright companion, and long timing baseline enable a level of astrometric precision that is not generally available for the rest of the PTA.

The next-best candidate is PSR~J0030$+$0451 at $L_p\simeq330$~pc~\cite{Ding2023}, which remains tens of times above the parallax precision required for strict phase coherence at 10~nHz. It is therefore important to distinguish between true anchors, which satisfy $\delta L_p<c/(2\pi f)$ and can lock the echo phase absolutely, and pulsars with merely useful distance priors, which still constrain the search volume and help select among cycle branches.

\subsection{Population estimate}\label{app:population}

To estimate the size of the future anchor population, we write
\begin{equation}\label{eq:Nanchor_def}
N_{\rm anchor}(f) = \eta_{\rm PTA}\,N_{\rm loc}(<L_{p,\max}),
\end{equation}
where $N_{\rm loc}$ is the number of observable millisecond pulsars within $L_{p,\max}$ and $\eta_{\rm PTA}$ is the fraction that are PTA-grade, meaning bright, rotationally stable, non-eclipsing, and astrometrically clean.

For the local MSP density we adopt the observed surface density $\Sigma_{\rm MSP}\simeq157\pm85~{\rm kpc}^{-2}$ above $0.3$~mJy~kpc$^2$~\cite{Lyne1998}, together with a vertical scale height $z_0\simeq0.5$~kpc~\cite{Levin2013}. Within a few hundred parsecs the scale height is effectively constant, so the accessible local population scales approximately as volume, $N_{\rm loc}(<L_{p,\max})\propto L_{p,\max}^3$. Combining this with Eq.~\eqref{eq:dmax_anchor} gives
\begin{equation}\label{eq:Nanchor_scaling}
N_{\rm anchor}(f)\propto(\sigma_\varpi\,f)^{-3/2}.
\end{equation}
This steep dependence is the main reason the anchor population is expected to be only $\mathcal{O}(3$--$5)$ at 10~nHz and to drop to $\mathcal{O}(1)$ by 30~nHz.

This estimate should be interpreted cautiously. The uncertainty is dominated not by the local MSP count itself, but by $\eta_{\rm PTA}$: only a subset of nearby MSPs will be suitable for PTA-quality timing and for high-precision astrometry.

\begin{table}[t]
\caption{Nearest PTA millisecond pulsars and their distance constraints. The column $\sigma_{\varpi,\rm req}$ gives the parallax precision needed for phase coherence at 10~nHz, and the Factor column is the ratio of the current parallax uncertainty to this requirement. Only J0437$-$4715 currently satisfies $\delta L_p < 0.15$~pc.}\label{tab:anchors}
\centering
\begin{tabular}{lccccll}
\hline\hline
Pulsar & $L_p$ & $\delta L_p$ & $\sigma_{\varpi,\rm req}$ & Factor & Status & Ref. \\
       & pc    & pc           & $\mu$as                    &        &        &      \\
\hline
J0437$-$4715  & 157 & 0.11 & 6.3 & --- & anchor & \cite{Reardon2024} \\
J0636$-$3044  & 230 & 37 & 2.9 & $240\times$ & SKA target & \cite{Shamohammadi2024} \\
J0030$+$0451  & 329 & 5.4 & 1.4 & $35\times$ & SKA target & \cite{Ding2023} \\
J1744$-$1134  & 395 & 4.7 & 1.0 & $30\times$ & beyond horizon & \cite{Deller2019} \\
J2124$-$3358  & 410 & 17 & 0.9 & $110\times$ & beyond horizon & \cite{Deller2019} \\
\hline\hline
\end{tabular}
\end{table}
% ======================================================================

\section{Binary population estimate}\label{app:population_estimate}

In this appendix we estimate the expected number of nearby SMBHBs currently inspiralling between $f_{\rm GW}=1$~nHz and merger. These are the sources whose Earth- and pulsar-term GW emission is accessible to the echo program. The calculation has four ingredients: (i) the local black hole mass function from Liepold \& Ma~\cite{LiepoldMa2024}, (ii) the mass-specific binary occupation fraction derived from the pairing rate of Casey-Clyde et al.~\cite{CaseyClyde2025}, (iii) the near-equal-mass weighting appropriate to the fiducial binaries considered in the main text, and (iv) propagation of the associated parameter uncertainties.

\textit{Step~1: Black-hole host census.}---The Liepold \& Ma~\cite{LiepoldMa2024} $z=0$ black-hole mass function, derived from the MASSIVE survey~\cite{Ma2014} via the $M_{\rm BH}$--$M_*$ relation of McConnell \& Ma~\cite{McConnellMa2013}, is well described by a modified Schechter form,
\begin{equation}\label{eq:bhmf}
\frac{dn}{d\ln M_{\rm BH}}
=
\phi_*
\left(\frac{M_{\rm BH}}{M_s}\right)^{\alpha+1}
\exp\!\left[-\left(\frac{M_{\rm BH}}{M_s}\right)^{\beta}\right],
\end{equation}
with parameters given by Liepold \& Ma~\cite{LiepoldMa2024}. Because this is a \emph{single-black-hole} mass function, it must be evaluated at the component mass rather than the total binary mass. For the fiducial equal-mass binary with $M_{\rm tot}=10^9\,M_\odot$, each component has $M_{\rm BH}=5\times10^8\,M_\odot$. Evaluated in a 0.2-dex mass bin at this mass, the Liepold \& Ma mass function predicts $\sim1000$ galaxies hosting a black hole at this mass within the MASSIVE survey volume ($2.05\times10^6$~Mpc$^3$, $D<108$~Mpc).

\textit{Step~2: Mass-specific binary occupation.}---The Casey-Clyde et al.~\cite{CaseyClyde2025} mass-averaged occupation fraction $\mathcal{F}_{\rm BHB}\approx2.6\%$ cannot be applied directly, because the PTA-band residence time depends strongly on mass. For a circular GW-driven binary, the time to coalescence from $f_{\rm GW}=1$~nHz is
\begin{equation}\label{eq:tc}
t_c = \frac{5}{256}\,
\frac{c^5}{(G\mathcal{M}_c)^{5/3}(\pi f_{\rm GW})^{8/3}},
\end{equation}
where $\mathcal{M}_c=M_{\rm tot}[q/(1+q)^2]^{3/5}$ is the chirp mass. For $M_{\rm tot}=10^9\,M_\odot$ and $q=1$, this gives $t_c\approx82$~Myr. $t_c$ is the residence time in the inspiral window $1\,{\rm nHz}\le f_{\rm GW}\le f_{\rm ISCO}$, and therefore sets the steady-state occupation of that window by binaries at mass $M_{\rm tot}$. Using the Casey-Clyde et al.\ local galaxy pairing rate $\dot f_0=0.04\pm0.01~{\rm Gyr}^{-1}$, the mass-specific occupation fraction is then
\begin{equation}\label{eq:Pactive}
P_{\rm active} = \dot f_0\,t_c \approx 0.04\times0.082 = 3.3\times10^{-3},
\end{equation}
or $0.33\%$. This is nearly an order of magnitude smaller than the mass-averaged value because the latter is dominated by more numerous low-mass SMBHs with much longer PTA-band residence times.

\textit{Step~3: Mass-ratio weighting.}---Following Liepold \& Ma~\cite{LiepoldMa2024} Eq.~(6), we adopt a mass-ratio prior $p(q)\propto q^2$ normalized on $q\in[0.01,1]$. The fraction of binaries with $q>0.9$ is then
\[
f_q = 27.1\% .
\]

\textit{Step~4: Expected count.}---The number of PTA-band SMBHBs with $M_{\rm tot}\approx10^9\,M_\odot$ and $q>0.9$ within distance $D$ is
\begin{equation}\label{eq:Npta}
N_{\rm SMBHB}(<D)
=
\frac{dn}{d\ln M}\,
\Delta\ln M\,
\frac{4\pi}{3}D^3\,
P_{\rm active}\,
f_q.
\end{equation}
Eq.~\eqref{eq:Npta} counts binaries at any inspiral frequency above 1~nHz, not only those currently in the $\mu$Ares band. We propagate uncertainties in the black hole mass function parameters $(\phi_*,\alpha,\beta,M_s)$ and in $\dot f_0$ using Monte Carlo sampling. The median expectation crosses $N=1$ at $D\approx83$~Mpc, comparable to the Tier-3 echo horizon of $\sim80$~Mpc derived from the strain sensitivity in the main text. Within the local $108$~Mpc sphere, we obtain
\[
N_{\rm SMBHB} = 2.2^{+1.0}_{-0.7}\;(68\%) ,
\qquad
N_{\rm SMBHB} = 2.2^{+1.8}_{-1.1}\;(90\%).
\]
The dominant uncertainty sources are the black hole mass function cutoff mass $M_s$ and the pairing-rate normalization $\dot f_0$, which contribute comparable spread.

\textit{Frequency distribution.}---The GW-driven evolution $df/dt\propto f^{11/3}$~\cite{Peters1964} piles binaries up at the lowest PTA frequencies, so the predicted population is strongly concentrated within the first few nHz. The echo program does \emph{not} require a local source to be in the $\mu$Ares band simultaneously with its PTA-band echoes. $\mu$Ares detects the Earth-term signal at the epoch when the source happens to pass through the $\mu$Hz band, while the PTA pulsar terms record the system's much earlier evolution.

\textit{Self-consistency with the GWB.}---The same Liepold \& Ma-based census yields a GWB amplitude $h_c(f=1\,{\rm yr}^{-1})\approx2.0\times10^{-15}$ through the Phinney formalism~\cite{Phinney2001}, using the mass-ratio and redshift moments quoted by Sato-Polito et al.~\cite{SatoPolito2024}. This is consistent with current PTA measurements. The local source count inferred here is therefore self-consistent with the observed nHz background.

% ======================================================================

\section{Spin-distance degeneracy in the pulsar network}\label{app:spin_degeneracy}

One might expect the pulsar terms themselves to constrain the binary spin, since each pulsar term accumulates $N_{\rm SO}^{(P)}\sim30$--$100$ spin-orbit cycles depending on the fiducial scenario and $\tau_i$, substantially more than the $\sim2$--$3$ spin-orbit cycles accumulated in the $\mu$Ares Earth term. However, the pulsar network suffers from an exact degeneracy: the phase of pulsar $i$ depends on both the spin $\chi$ and the look-back time $\tau_i$, so a shift in $\chi$ can always be absorbed by a compensating shift in $\tau_i$ if the distance is unconstrained.

With no prior on $\tau_i$, the marginalized Fisher information on $\chi$ therefore vanishes. Adding a Gaussian prior of width $\sigma_{\tau,i}$ on each look-back time shifts $F_{\tau_i\tau_i}\rightarrow F_{\tau_i\tau_i}+\sigma_{\tau,i}^{-2}$, giving
\begin{equation}\label{eq:fisher_prior}
  \sigma_\chi^{-2}
  = \sum_{i=1}^{N_p}
  \rho_i^2
  \left(\frac{\partial\Phi_i}{\partial\chi}\right)^2
  w_i,
  \qquad
  w_i = \frac{1}{1 + F_{\tau_i\tau_i}\sigma_{\tau,i}^2}\,.
\end{equation}
This interpolates smoothly between perfect distance knowledge ($w_i=1$) and no knowledge ($w_i=0$). Since $F_{\tau_i\tau_i}=\rho_i^2(2\pi f_{P,i})^2$ is very large, realistic weights are small. We retain only the 1.5pN spin-orbit contribution to $\partial\Phi/\partial\chi$, which is the leading-$\chi$ piece; the 2pN spin-spin term enters at order $\chi^2$ and is subdominant for this estimate. In the limit $w_i\ll1$, the per-pulsar contribution reduces to
\begin{equation}\label{eq:sigma_chi_prior}
  \sigma_\chi^{(i)}
  \approx
  \frac{\chi\,f_{P,i}\,\sigma_{\tau,i}}{N_{{\rm SO},i}},
\end{equation}
where $\sigma_{\tau,i}=(\delta L_{p,i}/L_{p,i})\tau_i$. The signal-to-noise ratio cancels: the bottleneck is the distance prior, not the SNR.

Evaluating Eq.~\eqref{eq:sigma_chi_prior} for current high-quality pulsars reveals an instructive asymmetry. PSR~J0437$-$4715, with its exceptional distance precision $(\delta L_p/L_p=0.07\%)$ but modest baseline $(\tau=512~{\rm yr})$, yields $\sigma_\chi\approx0.03$. PSR~J1713$+$0747, despite accumulating the largest spin-orbit cycle budget in the array $(N_{\rm SO}\approx54$ over the canonical $\tau=L_p/c\approx3{,}836~{\rm yr}$), gives only $\sigma_\chi\approx0.9$, because its fractional distance uncertainty $(\delta L_p/L_p=0.94\%)$ is amplified by the long baseline through $\sigma_{\tau,i}\propto(\delta L_p/L_p)\tau_i$. The pulsar with the richest post-Newtonian content is therefore not the one that best constrains spin.

The same asymmetry appears in the frequency-domain bin-shift test. The spin-induced pulsar-term shift $|\Delta f_P|$ is below the 20- and 30-yr PTA bin widths for every scenario and canonical baseline, and only at J0437 does the shift reach the 50-yr bin width of $0.63$~nHz, with $|\Delta f_P|\approx0.6$--$1.0$~nHz in the Conservative and Typical scenarios.

The situation does not improve enough by combining many pulsars. A Fisher sum over the $\sim20$ VLBI-calibrated MSPs with $\delta L_p/L_p\lesssim1\%$ distributed over 0.5--3~kpc~\cite{Deller2019,Ding2023} gives $\sigma_\chi\sim0.2$, more than two orders of magnitude above the $\delta\chi/\chi\lesssim10^{-3}$ requirement for phase-coherent echo analysis. Even a factor-of-ten improvement in VLBI precision would only bring this down to $\sigma_\chi\sim0.02$. The pulsar network alone cannot determine the spin well enough for coherent tracking, and the primary spin measurement must come from the $\mu$Ares Earth term.

\bibliography{references}

\begin{thebibliography}{81}%
\makeatletter
\providecommand \@ifxundefined [1]{%
 \@ifx{#1\undefined}
}%
\providecommand \@ifnum [1]{%
 \ifnum #1\expandafter \@firstoftwo
 \else \expandafter \@secondoftwo
 \fi
}%
\providecommand \@ifx [1]{%
 \ifx #1\expandafter \@firstoftwo
 \else \expandafter \@secondoftwo
 \fi
}%
\providecommand \natexlab [1]{#1}%
\providecommand \enquote  [1]{``#1''}%
\providecommand \bibnamefont  [1]{#1}%
\providecommand \bibfnamefont [1]{#1}%
\providecommand \citenamefont [1]{#1}%
\providecommand \href@noop [0]{\@secondoftwo}%
\providecommand \href [0]{\begingroup \@sanitize@url \@href}%
\providecommand \@href[1]{\@@startlink{#1}\@@href}%
\providecommand \@@href[1]{\endgroup#1\@@endlink}%
\providecommand \@sanitize@url [0]{\catcode `\\12\catcode `\$12\catcode
  `\&12\catcode `\#12\catcode `\^12\catcode `\_12\catcode `\%12\relax}%
\providecommand \@@startlink[1]{}%
\providecommand \@@endlink[0]{}%
\providecommand \url  [0]{\begingroup\@sanitize@url \@url }%
\providecommand \@url [1]{\endgroup\@href {#1}{\urlprefix }}%
\providecommand \urlprefix  [0]{URL }%
\providecommand \Eprint [0]{\href }%
\providecommand \doibase [0]{https://doi.org/}%
\providecommand \selectlanguage [0]{\@gobble}%
\providecommand \bibinfo  [0]{\@secondoftwo}%
\providecommand \bibfield  [0]{\@secondoftwo}%
\providecommand \translation [1]{[#1]}%
\providecommand \BibitemOpen [0]{}%
\providecommand \bibitemStop [0]{}%
\providecommand \bibitemNoStop [0]{.\EOS\space}%
\providecommand \EOS [0]{\spacefactor3000\relax}%
\providecommand \BibitemShut  [1]{\csname bibitem#1\endcsname}%
\let\auto@bib@innerbib\@empty
%</preamble>
\bibitem [{\citenamefont {{Lee}}\ \emph {et~al.}(2011)\citenamefont {{Lee}},
  \citenamefont {{Wex}}, \citenamefont {{Kramer}}, \citenamefont {{Stappers}},
  \citenamefont {{Bassa}}, \citenamefont {{Janssen}}, \citenamefont
  {{Karuppusamy}},\ and\ \citenamefont {{Smits}}}]{Lee2011}%
  \BibitemOpen
  \bibfield  {author} {\bibinfo {author} {\bibfnamefont {K.~J.}\ \bibnamefont
  {{Lee}}}, \bibinfo {author} {\bibfnamefont {N.}~\bibnamefont {{Wex}}},
  \bibinfo {author} {\bibfnamefont {M.}~\bibnamefont {{Kramer}}}, \bibinfo
  {author} {\bibfnamefont {B.~W.}\ \bibnamefont {{Stappers}}}, \bibinfo
  {author} {\bibfnamefont {C.~G.}\ \bibnamefont {{Bassa}}}, \bibinfo {author}
  {\bibfnamefont {G.~H.}\ \bibnamefont {{Janssen}}}, \bibinfo {author}
  {\bibfnamefont {R.}~\bibnamefont {{Karuppusamy}}},\ and\ \bibinfo {author}
  {\bibfnamefont {R.}~\bibnamefont {{Smits}}},\ }\bibfield  {title} {\bibinfo
  {title} {{Gravitational wave astronomy of single sources with a pulsar timing
  array}},\ }\href {https://doi.org/10.1111/j.1365-2966.2011.18622.x}
  {\bibfield  {journal} {\bibinfo  {journal} {\mnras}\ }\textbf {\bibinfo
  {volume} {414}},\ \bibinfo {pages} {3251} (\bibinfo {year} {2011})},\ \Eprint
  {https://arxiv.org/abs/1103.0115} {arXiv:1103.0115 [astro-ph.HE]}
  \BibitemShut {NoStop}%
\bibitem [{\citenamefont {{Corbin}}\ and\ \citenamefont
  {{Cornish}}(2010)}]{CorbinCornish2010}%
  \BibitemOpen
  \bibfield  {author} {\bibinfo {author} {\bibfnamefont {V.}~\bibnamefont
  {{Corbin}}}\ and\ \bibinfo {author} {\bibfnamefont {N.~J.}\ \bibnamefont
  {{Cornish}}},\ }\bibfield  {title} {\bibinfo {title} {{Pulsar Timing Array
  Observations of Massive Black Hole Binaries}},\ }\href
  {https://doi.org/10.48550/arXiv.1008.1782} {\bibfield  {journal} {\bibinfo
  {journal} {arXiv e-prints}\ ,\ \bibinfo {eid} {arXiv:1008.1782}} (\bibinfo
  {year} {2010})},\ \Eprint {https://arxiv.org/abs/1008.1782} {arXiv:1008.1782
  [astro-ph.HE]} \BibitemShut {NoStop}%
\bibitem [{\citenamefont {{Mingarelli}}\ \emph {et~al.}(2012)\citenamefont
  {{Mingarelli}}, \citenamefont {{Grover}}, \citenamefont {{Sidery}},
  \citenamefont {{Smith}},\ and\ \citenamefont {{Vecchio}}}]{Mingarelli2012}%
  \BibitemOpen
  \bibfield  {author} {\bibinfo {author} {\bibfnamefont {C.~M.~F.}\
  \bibnamefont {{Mingarelli}}}, \bibinfo {author} {\bibfnamefont
  {K.}~\bibnamefont {{Grover}}}, \bibinfo {author} {\bibfnamefont
  {T.}~\bibnamefont {{Sidery}}}, \bibinfo {author} {\bibfnamefont {R.~J.~E.}\
  \bibnamefont {{Smith}}},\ and\ \bibinfo {author} {\bibfnamefont
  {A.}~\bibnamefont {{Vecchio}}},\ }\bibfield  {title} {\bibinfo {title}
  {{Observing the Dynamics of Supermassive Black Hole Binaries with Pulsar
  Timing Arrays}},\ }\href {https://doi.org/10.1103/PhysRevLett.109.081104}
  {\bibfield  {journal} {\bibinfo  {journal} {\prl}\ }\textbf {\bibinfo
  {volume} {109}},\ \bibinfo {eid} {081104} (\bibinfo {year} {2012})},\ \Eprint
  {https://arxiv.org/abs/1207.5645} {arXiv:1207.5645 [astro-ph.HE]}
  \BibitemShut {NoStop}%
\bibitem [{\citenamefont {{Sesana}}\ \emph {et~al.}(2021)\citenamefont
  {{Sesana}}, \citenamefont {{Korsakova}}, \citenamefont {{Arca Sedda}},
  \citenamefont {{Baibhav}}, \citenamefont {{Barausse}}, \citenamefont
  {{Barke}}, \citenamefont {{Berti}}, \citenamefont {{Bonetti}}, \citenamefont
  {{Capelo}}, \citenamefont {{Caprini}}, \citenamefont {{Garcia-Bellido}},
  \citenamefont {{Haiman}}, \citenamefont {{Jani}}, \citenamefont {{Jennrich}},
  \citenamefont {{Johansson}}, \citenamefont {{Khan}}, \citenamefont {{Korol}},
  \citenamefont {{Lamberts}}, \citenamefont {{Lupi}}, \citenamefont
  {{Mangiagli}}, \citenamefont {{Mayer}}, \citenamefont {{Nardini}},
  \citenamefont {{Pacucci}}, \citenamefont {{Petiteau}}, \citenamefont
  {{Raccanelli}}, \citenamefont {{Rajendran}}, \citenamefont {{Regan}},
  \citenamefont {{Shao}}, \citenamefont {{Spallicci}}, \citenamefont
  {{Tamanini}}, \citenamefont {{Volonteri}}, \citenamefont {{Warburton}},
  \citenamefont {{Wong}},\ and\ \citenamefont {{Zumalacarregui}}}]{Sesana2021}%
  \BibitemOpen
  \bibfield  {author} {\bibinfo {author} {\bibfnamefont {A.}~\bibnamefont
  {{Sesana}}}, \bibinfo {author} {\bibfnamefont {N.}~\bibnamefont
  {{Korsakova}}}, \bibinfo {author} {\bibfnamefont {M.}~\bibnamefont {{Arca
  Sedda}}}, \bibinfo {author} {\bibfnamefont {V.}~\bibnamefont {{Baibhav}}},
  \bibinfo {author} {\bibfnamefont {E.}~\bibnamefont {{Barausse}}}, \bibinfo
  {author} {\bibfnamefont {S.}~\bibnamefont {{Barke}}}, \bibinfo {author}
  {\bibfnamefont {E.}~\bibnamefont {{Berti}}}, \bibinfo {author} {\bibfnamefont
  {M.}~\bibnamefont {{Bonetti}}}, \bibinfo {author} {\bibfnamefont {P.~R.}\
  \bibnamefont {{Capelo}}}, \bibinfo {author} {\bibfnamefont {C.}~\bibnamefont
  {{Caprini}}}, \bibinfo {author} {\bibfnamefont {J.}~\bibnamefont
  {{Garcia-Bellido}}}, \bibinfo {author} {\bibfnamefont {Z.}~\bibnamefont
  {{Haiman}}}, \bibinfo {author} {\bibfnamefont {K.}~\bibnamefont {{Jani}}},
  \bibinfo {author} {\bibfnamefont {O.}~\bibnamefont {{Jennrich}}}, \bibinfo
  {author} {\bibfnamefont {P.~H.}\ \bibnamefont {{Johansson}}}, \bibinfo
  {author} {\bibfnamefont {F.~M.}\ \bibnamefont {{Khan}}}, \bibinfo {author}
  {\bibfnamefont {V.}~\bibnamefont {{Korol}}}, \bibinfo {author} {\bibfnamefont
  {A.}~\bibnamefont {{Lamberts}}}, \bibinfo {author} {\bibfnamefont
  {A.}~\bibnamefont {{Lupi}}}, \bibinfo {author} {\bibfnamefont
  {A.}~\bibnamefont {{Mangiagli}}}, \bibinfo {author} {\bibfnamefont
  {L.}~\bibnamefont {{Mayer}}}, \bibinfo {author} {\bibfnamefont
  {G.}~\bibnamefont {{Nardini}}}, \bibinfo {author} {\bibfnamefont
  {F.}~\bibnamefont {{Pacucci}}}, \bibinfo {author} {\bibfnamefont
  {A.}~\bibnamefont {{Petiteau}}}, \bibinfo {author} {\bibfnamefont
  {A.}~\bibnamefont {{Raccanelli}}}, \bibinfo {author} {\bibfnamefont
  {S.}~\bibnamefont {{Rajendran}}}, \bibinfo {author} {\bibfnamefont
  {J.}~\bibnamefont {{Regan}}}, \bibinfo {author} {\bibfnamefont
  {L.}~\bibnamefont {{Shao}}}, \bibinfo {author} {\bibfnamefont
  {A.}~\bibnamefont {{Spallicci}}}, \bibinfo {author} {\bibfnamefont
  {N.}~\bibnamefont {{Tamanini}}}, \bibinfo {author} {\bibfnamefont
  {M.}~\bibnamefont {{Volonteri}}}, \bibinfo {author} {\bibfnamefont
  {N.}~\bibnamefont {{Warburton}}}, \bibinfo {author} {\bibfnamefont
  {K.}~\bibnamefont {{Wong}}},\ and\ \bibinfo {author} {\bibfnamefont
  {M.}~\bibnamefont {{Zumalacarregui}}},\ }\bibfield  {title} {\bibinfo {title}
  {{Unveiling the gravitational universe at {\ensuremath{\mu}}-Hz
  frequencies}},\ }\href {https://doi.org/10.1007/s10686-021-09709-9}
  {\bibfield  {journal} {\bibinfo  {journal} {Experimental Astronomy}\ }\textbf
  {\bibinfo {volume} {51}},\ \bibinfo {pages} {1333} (\bibinfo {year}
  {2021})},\ \Eprint {https://arxiv.org/abs/1908.11391} {arXiv:1908.11391
  [astro-ph.IM]} \BibitemShut {NoStop}%
\bibitem [{\citenamefont {{Couderc}}(1939)}]{Couderc1939}%
  \BibitemOpen
  \bibfield  {author} {\bibinfo {author} {\bibfnamefont {P.}~\bibnamefont
  {{Couderc}}},\ }\bibfield  {title} {\bibinfo {title} {{Les aur{\'e}oles
  lumineuses des Nov{\ae}}},\ }\href@noop {} {\bibfield  {journal} {\bibinfo
  {journal} {Annales d'Astrophysique}\ }\textbf {\bibinfo {volume} {2}},\
  \bibinfo {pages} {271} (\bibinfo {year} {1939})}\BibitemShut {NoStop}%
\bibitem [{\citenamefont {{Zwicky}}(1940)}]{Zwicky1940}%
  \BibitemOpen
  \bibfield  {author} {\bibinfo {author} {\bibfnamefont {F.}~\bibnamefont
  {{Zwicky}}},\ }\bibfield  {title} {\bibinfo {title} {{Types of Novae}},\
  }\href {https://doi.org/10.1103/RevModPhys.12.66} {\bibfield  {journal}
  {\bibinfo  {journal} {Reviews of Modern Physics}\ }\textbf {\bibinfo {volume}
  {12}},\ \bibinfo {pages} {66} (\bibinfo {year} {1940})}\BibitemShut {NoStop}%
\bibitem [{\citenamefont {{Agazie}}\ \emph
  {et~al.}(2023{\natexlab{a}})\citenamefont {{Agazie}}, \citenamefont
  {{Anumarlapudi}}, \citenamefont {{Archibald}}, \citenamefont {{Arzoumanian}},
  \citenamefont {{Baker}}, \citenamefont {{B{\'e}csy}}, \citenamefont
  {{Blecha}}, \citenamefont {{Brazier}}, \citenamefont {{Brook}}, \citenamefont
  {{Burke-Spolaor}}, \citenamefont {{Burnette}}, \citenamefont {{Case}},
  \citenamefont {{Charisi}}, \citenamefont {{Chatterjee}}, \citenamefont
  {{Chatziioannou}}, \citenamefont {{Cheeseboro}}, \citenamefont {{Chen}},
  \citenamefont {{Cohen}}, \citenamefont {{Cordes}}, \citenamefont {{Cornish}},
  \citenamefont {{Crawford}}, \citenamefont {{Cromartie}}, \citenamefont
  {{Crowter}}, \citenamefont {{Cutler}}, \citenamefont {{Decesar}},
  \citenamefont {{Degan}}, \citenamefont {{Demorest}}, \citenamefont {{Deng}},
  \citenamefont {{Dolch}}, \citenamefont {{Drachler}}, \citenamefont {{Ellis}},
  \citenamefont {{Ferrara}}, \citenamefont {{Fiore}}, \citenamefont
  {{Fonseca}}, \citenamefont {{Freedman}}, \citenamefont {{Garver-Daniels}},
  \citenamefont {{Gentile}}, \citenamefont {{Gersbach}}, \citenamefont
  {{Glaser}}, \citenamefont {{Good}}, \citenamefont {{G{\"u}ltekin}},
  \citenamefont {{Hazboun}}, \citenamefont {{Hourihane}}, \citenamefont
  {{Islo}}, \citenamefont {{Jennings}}, \citenamefont {{Johnson}},
  \citenamefont {{Jones}}, \citenamefont {{Kaiser}}, \citenamefont {{Kaplan}},
  \citenamefont {{Kelley}}, \citenamefont {{Kerr}}, \citenamefont {{Key}},
  \citenamefont {{Klein}}, \citenamefont {{Laal}}, \citenamefont {{Lam}},
  \citenamefont {{Lamb}}, \citenamefont {{Lazio}}, \citenamefont
  {{Lewandowska}}, \citenamefont {{Littenberg}}, \citenamefont {{Liu}},
  \citenamefont {{Lommen}}, \citenamefont {{Lorimer}}, \citenamefont {{Luo}},
  \citenamefont {{Lynch}}, \citenamefont {{Ma}}, \citenamefont {{Madison}},
  \citenamefont {{Mattson}}, \citenamefont {{McEwen}}, \citenamefont {{McKee}},
  \citenamefont {{McLaughlin}}, \citenamefont {{McMann}}, \citenamefont
  {{Meyers}}, \citenamefont {{Meyers}}, \citenamefont {{Mingarelli}},
  \citenamefont {{Mitridate}}, \citenamefont {{Natarajan}}, \citenamefont
  {{Ng}}, \citenamefont {{Nice}}, \citenamefont {{Ocker}}, \citenamefont
  {{Olum}}, \citenamefont {{Pennucci}}, \citenamefont {{Perera}}, \citenamefont
  {{Petrov}}, \citenamefont {{Pol}}, \citenamefont {{Radovan}}, \citenamefont
  {{Ransom}}, \citenamefont {{Ray}}, \citenamefont {{Romano}}, \citenamefont
  {{Sardesai}}, \citenamefont {{Schmiedekamp}}, \citenamefont {{Schmiedekamp}},
  \citenamefont {{Schmitz}}, \citenamefont {{Schult}}, \citenamefont
  {{Shapiro-Albert}}, \citenamefont {{Siemens}}, \citenamefont {{Simon}},
  \citenamefont {{Siwek}}, \citenamefont {{Stairs}}, \citenamefont
  {{Stinebring}}, \citenamefont {{Stovall}}, \citenamefont {{Sun}},
  \citenamefont {{Susobhanan}}, \citenamefont {{Swiggum}}, \citenamefont
  {{Taylor}}, \citenamefont {{Taylor}}, \citenamefont {{Turner}}, \citenamefont
  {{Unal}}, \citenamefont {{Vallisneri}}, \citenamefont {{van Haasteren}},
  \citenamefont {{Vigeland}}, \citenamefont {{Wahl}}, \citenamefont {{Wang}},
  \citenamefont {{Witt}}, \citenamefont {{Young}},\ and\ \citenamefont
  {{Nanograv Collaboration}}}]{NANOGrav15yr}%
  \BibitemOpen
  \bibfield  {author} {\bibinfo {author} {\bibfnamefont {G.}~\bibnamefont
  {{Agazie}}}, \bibinfo {author} {\bibfnamefont {A.}~\bibnamefont
  {{Anumarlapudi}}}, \bibinfo {author} {\bibfnamefont {A.~M.}\ \bibnamefont
  {{Archibald}}}, \bibinfo {author} {\bibfnamefont {Z.}~\bibnamefont
  {{Arzoumanian}}}, \bibinfo {author} {\bibfnamefont {P.~T.}\ \bibnamefont
  {{Baker}}}, \bibinfo {author} {\bibfnamefont {B.}~\bibnamefont
  {{B{\'e}csy}}}, \bibinfo {author} {\bibfnamefont {L.}~\bibnamefont
  {{Blecha}}}, \bibinfo {author} {\bibfnamefont {A.}~\bibnamefont {{Brazier}}},
  \bibinfo {author} {\bibfnamefont {P.~R.}\ \bibnamefont {{Brook}}}, \bibinfo
  {author} {\bibfnamefont {S.}~\bibnamefont {{Burke-Spolaor}}}, \bibinfo
  {author} {\bibfnamefont {R.}~\bibnamefont {{Burnette}}}, \bibinfo {author}
  {\bibfnamefont {R.}~\bibnamefont {{Case}}}, \bibinfo {author} {\bibfnamefont
  {M.}~\bibnamefont {{Charisi}}}, \bibinfo {author} {\bibfnamefont
  {S.}~\bibnamefont {{Chatterjee}}}, \bibinfo {author} {\bibfnamefont
  {K.}~\bibnamefont {{Chatziioannou}}}, \bibinfo {author} {\bibfnamefont
  {B.~D.}\ \bibnamefont {{Cheeseboro}}}, \bibinfo {author} {\bibfnamefont
  {S.}~\bibnamefont {{Chen}}}, \bibinfo {author} {\bibfnamefont
  {T.}~\bibnamefont {{Cohen}}}, \bibinfo {author} {\bibfnamefont {J.~M.}\
  \bibnamefont {{Cordes}}}, \bibinfo {author} {\bibfnamefont {N.~J.}\
  \bibnamefont {{Cornish}}}, \bibinfo {author} {\bibfnamefont {F.}~\bibnamefont
  {{Crawford}}}, \bibinfo {author} {\bibfnamefont {H.~T.}\ \bibnamefont
  {{Cromartie}}}, \bibinfo {author} {\bibfnamefont {K.}~\bibnamefont
  {{Crowter}}}, \bibinfo {author} {\bibfnamefont {C.~J.}\ \bibnamefont
  {{Cutler}}}, \bibinfo {author} {\bibfnamefont {M.~E.}\ \bibnamefont
  {{Decesar}}}, \bibinfo {author} {\bibfnamefont {D.}~\bibnamefont {{Degan}}},
  \bibinfo {author} {\bibfnamefont {P.~B.}\ \bibnamefont {{Demorest}}},
  \bibinfo {author} {\bibfnamefont {H.}~\bibnamefont {{Deng}}}, \bibinfo
  {author} {\bibfnamefont {T.}~\bibnamefont {{Dolch}}}, \bibinfo {author}
  {\bibfnamefont {B.}~\bibnamefont {{Drachler}}}, \bibinfo {author}
  {\bibfnamefont {J.~A.}\ \bibnamefont {{Ellis}}}, \bibinfo {author}
  {\bibfnamefont {E.~C.}\ \bibnamefont {{Ferrara}}}, \bibinfo {author}
  {\bibfnamefont {W.}~\bibnamefont {{Fiore}}}, \bibinfo {author} {\bibfnamefont
  {E.}~\bibnamefont {{Fonseca}}}, \bibinfo {author} {\bibfnamefont {G.~E.}\
  \bibnamefont {{Freedman}}}, \bibinfo {author} {\bibfnamefont
  {N.}~\bibnamefont {{Garver-Daniels}}}, \bibinfo {author} {\bibfnamefont
  {P.~A.}\ \bibnamefont {{Gentile}}}, \bibinfo {author} {\bibfnamefont {K.~A.}\
  \bibnamefont {{Gersbach}}}, \bibinfo {author} {\bibfnamefont
  {J.}~\bibnamefont {{Glaser}}}, \bibinfo {author} {\bibfnamefont {D.~C.}\
  \bibnamefont {{Good}}}, \bibinfo {author} {\bibfnamefont {K.}~\bibnamefont
  {{G{\"u}ltekin}}}, \bibinfo {author} {\bibfnamefont {J.~S.}\ \bibnamefont
  {{Hazboun}}}, \bibinfo {author} {\bibfnamefont {S.}~\bibnamefont
  {{Hourihane}}}, \bibinfo {author} {\bibfnamefont {K.}~\bibnamefont {{Islo}}},
  \bibinfo {author} {\bibfnamefont {R.~J.}\ \bibnamefont {{Jennings}}},
  \bibinfo {author} {\bibfnamefont {A.~D.}\ \bibnamefont {{Johnson}}}, \bibinfo
  {author} {\bibfnamefont {M.~L.}\ \bibnamefont {{Jones}}}, \bibinfo {author}
  {\bibfnamefont {A.~R.}\ \bibnamefont {{Kaiser}}}, \bibinfo {author}
  {\bibfnamefont {D.~L.}\ \bibnamefont {{Kaplan}}}, \bibinfo {author}
  {\bibfnamefont {L.~Z.}\ \bibnamefont {{Kelley}}}, \bibinfo {author}
  {\bibfnamefont {M.}~\bibnamefont {{Kerr}}}, \bibinfo {author} {\bibfnamefont
  {J.~S.}\ \bibnamefont {{Key}}}, \bibinfo {author} {\bibfnamefont {T.~C.}\
  \bibnamefont {{Klein}}}, \bibinfo {author} {\bibfnamefont {N.}~\bibnamefont
  {{Laal}}}, \bibinfo {author} {\bibfnamefont {M.~T.}\ \bibnamefont {{Lam}}},
  \bibinfo {author} {\bibfnamefont {W.~G.}\ \bibnamefont {{Lamb}}}, \bibinfo
  {author} {\bibfnamefont {T.~J.~W.}\ \bibnamefont {{Lazio}}}, \bibinfo
  {author} {\bibfnamefont {N.}~\bibnamefont {{Lewandowska}}}, \bibinfo {author}
  {\bibfnamefont {T.~B.}\ \bibnamefont {{Littenberg}}}, \bibinfo {author}
  {\bibfnamefont {T.}~\bibnamefont {{Liu}}}, \bibinfo {author} {\bibfnamefont
  {A.}~\bibnamefont {{Lommen}}}, \bibinfo {author} {\bibfnamefont {D.~R.}\
  \bibnamefont {{Lorimer}}}, \bibinfo {author} {\bibfnamefont {J.}~\bibnamefont
  {{Luo}}}, \bibinfo {author} {\bibfnamefont {R.~S.}\ \bibnamefont {{Lynch}}},
  \bibinfo {author} {\bibfnamefont {C.-P.}\ \bibnamefont {{Ma}}}, \bibinfo
  {author} {\bibfnamefont {D.~R.}\ \bibnamefont {{Madison}}}, \bibinfo {author}
  {\bibfnamefont {M.~A.}\ \bibnamefont {{Mattson}}}, \bibinfo {author}
  {\bibfnamefont {A.}~\bibnamefont {{McEwen}}}, \bibinfo {author}
  {\bibfnamefont {J.~W.}\ \bibnamefont {{McKee}}}, \bibinfo {author}
  {\bibfnamefont {M.~A.}\ \bibnamefont {{McLaughlin}}}, \bibinfo {author}
  {\bibfnamefont {N.}~\bibnamefont {{McMann}}}, \bibinfo {author}
  {\bibfnamefont {B.~W.}\ \bibnamefont {{Meyers}}}, \bibinfo {author}
  {\bibfnamefont {P.~M.}\ \bibnamefont {{Meyers}}}, \bibinfo {author}
  {\bibfnamefont {C.~M.~F.}\ \bibnamefont {{Mingarelli}}}, \bibinfo {author}
  {\bibfnamefont {A.}~\bibnamefont {{Mitridate}}}, \bibinfo {author}
  {\bibfnamefont {P.}~\bibnamefont {{Natarajan}}}, \bibinfo {author}
  {\bibfnamefont {C.}~\bibnamefont {{Ng}}}, \bibinfo {author} {\bibfnamefont
  {D.~J.}\ \bibnamefont {{Nice}}}, \bibinfo {author} {\bibfnamefont {S.~K.}\
  \bibnamefont {{Ocker}}}, \bibinfo {author} {\bibfnamefont {K.~D.}\
  \bibnamefont {{Olum}}}, \bibinfo {author} {\bibfnamefont {T.~T.}\
  \bibnamefont {{Pennucci}}}, \bibinfo {author} {\bibfnamefont {B.~B.~P.}\
  \bibnamefont {{Perera}}}, \bibinfo {author} {\bibfnamefont {P.}~\bibnamefont
  {{Petrov}}}, \bibinfo {author} {\bibfnamefont {N.~S.}\ \bibnamefont {{Pol}}},
  \bibinfo {author} {\bibfnamefont {H.~A.}\ \bibnamefont {{Radovan}}}, \bibinfo
  {author} {\bibfnamefont {S.~M.}\ \bibnamefont {{Ransom}}}, \bibinfo {author}
  {\bibfnamefont {P.~S.}\ \bibnamefont {{Ray}}}, \bibinfo {author}
  {\bibfnamefont {J.~D.}\ \bibnamefont {{Romano}}}, \bibinfo {author}
  {\bibfnamefont {S.~C.}\ \bibnamefont {{Sardesai}}}, \bibinfo {author}
  {\bibfnamefont {A.}~\bibnamefont {{Schmiedekamp}}}, \bibinfo {author}
  {\bibfnamefont {C.}~\bibnamefont {{Schmiedekamp}}}, \bibinfo {author}
  {\bibfnamefont {K.}~\bibnamefont {{Schmitz}}}, \bibinfo {author}
  {\bibfnamefont {L.}~\bibnamefont {{Schult}}}, \bibinfo {author}
  {\bibfnamefont {B.~J.}\ \bibnamefont {{Shapiro-Albert}}}, \bibinfo {author}
  {\bibfnamefont {X.}~\bibnamefont {{Siemens}}}, \bibinfo {author}
  {\bibfnamefont {J.}~\bibnamefont {{Simon}}}, \bibinfo {author} {\bibfnamefont
  {M.~S.}\ \bibnamefont {{Siwek}}}, \bibinfo {author} {\bibfnamefont {I.~H.}\
  \bibnamefont {{Stairs}}}, \bibinfo {author} {\bibfnamefont {D.~R.}\
  \bibnamefont {{Stinebring}}}, \bibinfo {author} {\bibfnamefont
  {K.}~\bibnamefont {{Stovall}}}, \bibinfo {author} {\bibfnamefont {J.~P.}\
  \bibnamefont {{Sun}}}, \bibinfo {author} {\bibfnamefont {A.}~\bibnamefont
  {{Susobhanan}}}, \bibinfo {author} {\bibfnamefont {J.~K.}\ \bibnamefont
  {{Swiggum}}}, \bibinfo {author} {\bibfnamefont {J.}~\bibnamefont {{Taylor}}},
  \bibinfo {author} {\bibfnamefont {S.~R.}\ \bibnamefont {{Taylor}}}, \bibinfo
  {author} {\bibfnamefont {J.~E.}\ \bibnamefont {{Turner}}}, \bibinfo {author}
  {\bibfnamefont {C.}~\bibnamefont {{Unal}}}, \bibinfo {author} {\bibfnamefont
  {M.}~\bibnamefont {{Vallisneri}}}, \bibinfo {author} {\bibfnamefont
  {R.}~\bibnamefont {{van Haasteren}}}, \bibinfo {author} {\bibfnamefont
  {S.~J.}\ \bibnamefont {{Vigeland}}}, \bibinfo {author} {\bibfnamefont
  {H.~M.}\ \bibnamefont {{Wahl}}}, \bibinfo {author} {\bibfnamefont
  {Q.}~\bibnamefont {{Wang}}}, \bibinfo {author} {\bibfnamefont {C.~A.}\
  \bibnamefont {{Witt}}}, \bibinfo {author} {\bibfnamefont {O.}~\bibnamefont
  {{Young}}},\ and\ \bibinfo {author} {\bibnamefont {{Nanograv
  Collaboration}}},\ }\bibfield  {title} {\bibinfo {title} {{The NANOGrav 15 yr
  Data Set: Evidence for a Gravitational-wave Background}},\ }\href
  {https://doi.org/10.3847/2041-8213/acdac6} {\bibfield  {journal} {\bibinfo
  {journal} {\apjl}\ }\textbf {\bibinfo {volume} {951}},\ \bibinfo {eid} {L8}
  (\bibinfo {year} {2023}{\natexlab{a}})},\ \Eprint
  {https://arxiv.org/abs/2306.16213} {arXiv:2306.16213 [astro-ph.HE]}
  \BibitemShut {NoStop}%
\bibitem [{\citenamefont {{EPTA Collaboration}}\ \emph
  {et~al.}(2023)\citenamefont {{EPTA Collaboration}}, \citenamefont {{InPTA
  Collaboration}}, \citenamefont {{Antoniadis}}, \citenamefont {{Arumugam}},
  \citenamefont {{Arumugam}}, \citenamefont {{Babak}}, \citenamefont
  {{Bagchi}}, \citenamefont {{Bak Nielsen}}, \citenamefont {{Bassa}},
  \citenamefont {{Bathula}}, \citenamefont {{Berthereau}}, \citenamefont
  {{Bonetti}}, \citenamefont {{Bortolas}}, \citenamefont {{Brook}},
  \citenamefont {{Burgay}}, \citenamefont {{Caballero}}, \citenamefont
  {{Chalumeau}}, \citenamefont {{Champion}}, \citenamefont {{Chanlaridis}},
  \citenamefont {{Chen}}, \citenamefont {{Cognard}}, \citenamefont
  {{Dandapat}}, \citenamefont {{Deb}}, \citenamefont {{Desai}}, \citenamefont
  {{Desvignes}}, \citenamefont {{Dhanda-Batra}}, \citenamefont {{Dwivedi}},
  \citenamefont {{Falxa}}, \citenamefont {{Ferdman}}, \citenamefont
  {{Franchini}}, \citenamefont {{Gair}}, \citenamefont {{Goncharov}},
  \citenamefont {{Gopakumar}}, \citenamefont {{Graikou}}, \citenamefont
  {{Grie{\ss}meier}}, \citenamefont {{Guillemot}}, \citenamefont {{Guo}},
  \citenamefont {{Gupta}}, \citenamefont {{Hisano}}, \citenamefont {{Hu}},
  \citenamefont {{Iraci}}, \citenamefont {{Izquierdo-Villalba}}, \citenamefont
  {{Jang}}, \citenamefont {{Jawor}}, \citenamefont {{Janssen}}, \citenamefont
  {{Jessner}}, \citenamefont {{Joshi}}, \citenamefont {{Kareem}}, \citenamefont
  {{Karuppusamy}}, \citenamefont {{Keane}}, \citenamefont {{Keith}},
  \citenamefont {{Kharbanda}}, \citenamefont {{Kikunaga}}, \citenamefont
  {{Kolhe}}, \citenamefont {{Kramer}}, \citenamefont {{Krishnakumar}},
  \citenamefont {{Lackeos}}, \citenamefont {{Lee}}, \citenamefont {{Liu}},
  \citenamefont {{Liu}}, \citenamefont {{Lyne}}, \citenamefont {{McKee}},
  \citenamefont {{Maan}}, \citenamefont {{Main}}, \citenamefont {{Mickaliger}},
  \citenamefont {{Ni{\c{t}}u}}, \citenamefont {{Nobleson}}, \citenamefont
  {{Paladi}}, \citenamefont {{Parthasarathy}}, \citenamefont {{Perera}},
  \citenamefont {{Perrodin}}, \citenamefont {{Petiteau}}, \citenamefont
  {{Porayko}}, \citenamefont {{Possenti}}, \citenamefont {{Prabu}},
  \citenamefont {{Quelquejay Leclere}}, \citenamefont {{Rana}}, \citenamefont
  {{Samajdar}}, \citenamefont {{Sanidas}}, \citenamefont {{Sesana}},
  \citenamefont {{Shaifullah}}, \citenamefont {{Singha}}, \citenamefont
  {{Speri}}, \citenamefont {{Spiewak}}, \citenamefont {{Srivastava}},
  \citenamefont {{Stappers}}, \citenamefont {{Surnis}}, \citenamefont
  {{Susarla}}, \citenamefont {{Susobhanan}}, \citenamefont {{Takahashi}},
  \citenamefont {{Tarafdar}}, \citenamefont {{Theureau}}, \citenamefont
  {{Tiburzi}}, \citenamefont {{van der Wateren}}, \citenamefont {{Vecchio}},
  \citenamefont {{Venkatraman Krishnan}}, \citenamefont {{Verbiest}},
  \citenamefont {{Wang}}, \citenamefont {{Wang}},\ and\ \citenamefont
  {{Wu}}}]{EPTA2023}%
  \BibitemOpen
  \bibfield  {author} {\bibinfo {author} {\bibnamefont {{EPTA Collaboration}}},
  \bibinfo {author} {\bibnamefont {{InPTA Collaboration}}}, \bibinfo {author}
  {\bibfnamefont {J.}~\bibnamefont {{Antoniadis}}}, \bibinfo {author}
  {\bibfnamefont {P.}~\bibnamefont {{Arumugam}}}, \bibinfo {author}
  {\bibfnamefont {S.}~\bibnamefont {{Arumugam}}}, \bibinfo {author}
  {\bibfnamefont {S.}~\bibnamefont {{Babak}}}, \bibinfo {author} {\bibfnamefont
  {M.}~\bibnamefont {{Bagchi}}}, \bibinfo {author} {\bibfnamefont {A.-S.}\
  \bibnamefont {{Bak Nielsen}}}, \bibinfo {author} {\bibfnamefont {C.~G.}\
  \bibnamefont {{Bassa}}}, \bibinfo {author} {\bibfnamefont {A.}~\bibnamefont
  {{Bathula}}}, \bibinfo {author} {\bibfnamefont {A.}~\bibnamefont
  {{Berthereau}}}, \bibinfo {author} {\bibfnamefont {M.}~\bibnamefont
  {{Bonetti}}}, \bibinfo {author} {\bibfnamefont {E.}~\bibnamefont
  {{Bortolas}}}, \bibinfo {author} {\bibfnamefont {P.~R.}\ \bibnamefont
  {{Brook}}}, \bibinfo {author} {\bibfnamefont {M.}~\bibnamefont {{Burgay}}},
  \bibinfo {author} {\bibfnamefont {R.~N.}\ \bibnamefont {{Caballero}}},
  \bibinfo {author} {\bibfnamefont {A.}~\bibnamefont {{Chalumeau}}}, \bibinfo
  {author} {\bibfnamefont {D.~J.}\ \bibnamefont {{Champion}}}, \bibinfo
  {author} {\bibfnamefont {S.}~\bibnamefont {{Chanlaridis}}}, \bibinfo {author}
  {\bibfnamefont {S.}~\bibnamefont {{Chen}}}, \bibinfo {author} {\bibfnamefont
  {I.}~\bibnamefont {{Cognard}}}, \bibinfo {author} {\bibfnamefont
  {S.}~\bibnamefont {{Dandapat}}}, \bibinfo {author} {\bibfnamefont
  {D.}~\bibnamefont {{Deb}}}, \bibinfo {author} {\bibfnamefont
  {S.}~\bibnamefont {{Desai}}}, \bibinfo {author} {\bibfnamefont
  {G.}~\bibnamefont {{Desvignes}}}, \bibinfo {author} {\bibfnamefont
  {N.}~\bibnamefont {{Dhanda-Batra}}}, \bibinfo {author} {\bibfnamefont
  {C.}~\bibnamefont {{Dwivedi}}}, \bibinfo {author} {\bibfnamefont
  {M.}~\bibnamefont {{Falxa}}}, \bibinfo {author} {\bibfnamefont {R.~D.}\
  \bibnamefont {{Ferdman}}}, \bibinfo {author} {\bibfnamefont {A.}~\bibnamefont
  {{Franchini}}}, \bibinfo {author} {\bibfnamefont {J.~R.}\ \bibnamefont
  {{Gair}}}, \bibinfo {author} {\bibfnamefont {B.}~\bibnamefont {{Goncharov}}},
  \bibinfo {author} {\bibfnamefont {A.}~\bibnamefont {{Gopakumar}}}, \bibinfo
  {author} {\bibfnamefont {E.}~\bibnamefont {{Graikou}}}, \bibinfo {author}
  {\bibfnamefont {J.-M.}\ \bibnamefont {{Grie{\ss}meier}}}, \bibinfo {author}
  {\bibfnamefont {L.}~\bibnamefont {{Guillemot}}}, \bibinfo {author}
  {\bibfnamefont {Y.~J.}\ \bibnamefont {{Guo}}}, \bibinfo {author}
  {\bibfnamefont {Y.}~\bibnamefont {{Gupta}}}, \bibinfo {author} {\bibfnamefont
  {S.}~\bibnamefont {{Hisano}}}, \bibinfo {author} {\bibfnamefont
  {H.}~\bibnamefont {{Hu}}}, \bibinfo {author} {\bibfnamefont {F.}~\bibnamefont
  {{Iraci}}}, \bibinfo {author} {\bibfnamefont {D.}~\bibnamefont
  {{Izquierdo-Villalba}}}, \bibinfo {author} {\bibfnamefont {J.}~\bibnamefont
  {{Jang}}}, \bibinfo {author} {\bibfnamefont {J.}~\bibnamefont {{Jawor}}},
  \bibinfo {author} {\bibfnamefont {G.~H.}\ \bibnamefont {{Janssen}}}, \bibinfo
  {author} {\bibfnamefont {A.}~\bibnamefont {{Jessner}}}, \bibinfo {author}
  {\bibfnamefont {B.~C.}\ \bibnamefont {{Joshi}}}, \bibinfo {author}
  {\bibfnamefont {F.}~\bibnamefont {{Kareem}}}, \bibinfo {author}
  {\bibfnamefont {R.}~\bibnamefont {{Karuppusamy}}}, \bibinfo {author}
  {\bibfnamefont {E.~F.}\ \bibnamefont {{Keane}}}, \bibinfo {author}
  {\bibfnamefont {M.~J.}\ \bibnamefont {{Keith}}}, \bibinfo {author}
  {\bibfnamefont {D.}~\bibnamefont {{Kharbanda}}}, \bibinfo {author}
  {\bibfnamefont {T.}~\bibnamefont {{Kikunaga}}}, \bibinfo {author}
  {\bibfnamefont {N.}~\bibnamefont {{Kolhe}}}, \bibinfo {author} {\bibfnamefont
  {M.}~\bibnamefont {{Kramer}}}, \bibinfo {author} {\bibfnamefont {M.~A.}\
  \bibnamefont {{Krishnakumar}}}, \bibinfo {author} {\bibfnamefont
  {K.}~\bibnamefont {{Lackeos}}}, \bibinfo {author} {\bibfnamefont {K.~J.}\
  \bibnamefont {{Lee}}}, \bibinfo {author} {\bibfnamefont {K.}~\bibnamefont
  {{Liu}}}, \bibinfo {author} {\bibfnamefont {Y.}~\bibnamefont {{Liu}}},
  \bibinfo {author} {\bibfnamefont {A.~G.}\ \bibnamefont {{Lyne}}}, \bibinfo
  {author} {\bibfnamefont {J.~W.}\ \bibnamefont {{McKee}}}, \bibinfo {author}
  {\bibfnamefont {Y.}~\bibnamefont {{Maan}}}, \bibinfo {author} {\bibfnamefont
  {R.~A.}\ \bibnamefont {{Main}}}, \bibinfo {author} {\bibfnamefont {M.~B.}\
  \bibnamefont {{Mickaliger}}}, \bibinfo {author} {\bibfnamefont {I.~C.}\
  \bibnamefont {{Ni{\c{t}}u}}}, \bibinfo {author} {\bibfnamefont
  {K.}~\bibnamefont {{Nobleson}}}, \bibinfo {author} {\bibfnamefont {A.~K.}\
  \bibnamefont {{Paladi}}}, \bibinfo {author} {\bibfnamefont {A.}~\bibnamefont
  {{Parthasarathy}}}, \bibinfo {author} {\bibfnamefont {B.~B.~P.}\ \bibnamefont
  {{Perera}}}, \bibinfo {author} {\bibfnamefont {D.}~\bibnamefont
  {{Perrodin}}}, \bibinfo {author} {\bibfnamefont {A.}~\bibnamefont
  {{Petiteau}}}, \bibinfo {author} {\bibfnamefont {N.~K.}\ \bibnamefont
  {{Porayko}}}, \bibinfo {author} {\bibfnamefont {A.}~\bibnamefont
  {{Possenti}}}, \bibinfo {author} {\bibfnamefont {T.}~\bibnamefont {{Prabu}}},
  \bibinfo {author} {\bibfnamefont {H.}~\bibnamefont {{Quelquejay Leclere}}},
  \bibinfo {author} {\bibfnamefont {P.}~\bibnamefont {{Rana}}}, \bibinfo
  {author} {\bibfnamefont {A.}~\bibnamefont {{Samajdar}}}, \bibinfo {author}
  {\bibfnamefont {S.~A.}\ \bibnamefont {{Sanidas}}}, \bibinfo {author}
  {\bibfnamefont {A.}~\bibnamefont {{Sesana}}}, \bibinfo {author}
  {\bibfnamefont {G.}~\bibnamefont {{Shaifullah}}}, \bibinfo {author}
  {\bibfnamefont {J.}~\bibnamefont {{Singha}}}, \bibinfo {author}
  {\bibfnamefont {L.}~\bibnamefont {{Speri}}}, \bibinfo {author} {\bibfnamefont
  {R.}~\bibnamefont {{Spiewak}}}, \bibinfo {author} {\bibfnamefont
  {A.}~\bibnamefont {{Srivastava}}}, \bibinfo {author} {\bibfnamefont {B.~W.}\
  \bibnamefont {{Stappers}}}, \bibinfo {author} {\bibfnamefont
  {M.}~\bibnamefont {{Surnis}}}, \bibinfo {author} {\bibfnamefont {S.~C.}\
  \bibnamefont {{Susarla}}}, \bibinfo {author} {\bibfnamefont {A.}~\bibnamefont
  {{Susobhanan}}}, \bibinfo {author} {\bibfnamefont {K.}~\bibnamefont
  {{Takahashi}}}, \bibinfo {author} {\bibfnamefont {P.}~\bibnamefont
  {{Tarafdar}}}, \bibinfo {author} {\bibfnamefont {G.}~\bibnamefont
  {{Theureau}}}, \bibinfo {author} {\bibfnamefont {C.}~\bibnamefont
  {{Tiburzi}}}, \bibinfo {author} {\bibfnamefont {E.}~\bibnamefont {{van der
  Wateren}}}, \bibinfo {author} {\bibfnamefont {A.}~\bibnamefont {{Vecchio}}},
  \bibinfo {author} {\bibfnamefont {V.}~\bibnamefont {{Venkatraman Krishnan}}},
  \bibinfo {author} {\bibfnamefont {J.~P.~W.}\ \bibnamefont {{Verbiest}}},
  \bibinfo {author} {\bibfnamefont {J.}~\bibnamefont {{Wang}}}, \bibinfo
  {author} {\bibfnamefont {L.}~\bibnamefont {{Wang}}},\ and\ \bibinfo {author}
  {\bibfnamefont {Z.}~\bibnamefont {{Wu}}},\ }\bibfield  {title} {\bibinfo
  {title} {{The second data release from the European Pulsar Timing Array. III.
  Search for gravitational wave signals}},\ }\href
  {https://doi.org/10.1051/0004-6361/202346844} {\bibfield  {journal} {\bibinfo
   {journal} {\aap}\ }\textbf {\bibinfo {volume} {678}},\ \bibinfo {eid} {A50}
  (\bibinfo {year} {2023})},\ \Eprint {https://arxiv.org/abs/2306.16214}
  {arXiv:2306.16214 [astro-ph.HE]} \BibitemShut {NoStop}%
\bibitem [{\citenamefont {{Reardon}}\ \emph {et~al.}(2023)\citenamefont
  {{Reardon}}, \citenamefont {{Zic}}, \citenamefont {{Shannon}}, \citenamefont
  {{Hobbs}}, \citenamefont {{Bailes}}, \citenamefont {{Di Marco}},
  \citenamefont {{Kapur}}, \citenamefont {{Rogers}}, \citenamefont {{Thrane}},
  \citenamefont {{Askew}}, \citenamefont {{Bhat}}, \citenamefont {{Cameron}},
  \citenamefont {{Cury{\l}o}}, \citenamefont {{Coles}}, \citenamefont {{Dai}},
  \citenamefont {{Goncharov}}, \citenamefont {{Kerr}}, \citenamefont
  {{Kulkarni}}, \citenamefont {{Levin}}, \citenamefont {{Lower}}, \citenamefont
  {{Manchester}}, \citenamefont {{Mandow}}, \citenamefont {{Miles}},
  \citenamefont {{Nathan}}, \citenamefont {{Os{\l}owski}}, \citenamefont
  {{Russell}}, \citenamefont {{Spiewak}}, \citenamefont {{Zhang}},\ and\
  \citenamefont {{Zhu}}}]{PPTA2023}%
  \BibitemOpen
  \bibfield  {author} {\bibinfo {author} {\bibfnamefont {D.~J.}\ \bibnamefont
  {{Reardon}}}, \bibinfo {author} {\bibfnamefont {A.}~\bibnamefont {{Zic}}},
  \bibinfo {author} {\bibfnamefont {R.~M.}\ \bibnamefont {{Shannon}}}, \bibinfo
  {author} {\bibfnamefont {G.~B.}\ \bibnamefont {{Hobbs}}}, \bibinfo {author}
  {\bibfnamefont {M.}~\bibnamefont {{Bailes}}}, \bibinfo {author}
  {\bibfnamefont {V.}~\bibnamefont {{Di Marco}}}, \bibinfo {author}
  {\bibfnamefont {A.}~\bibnamefont {{Kapur}}}, \bibinfo {author} {\bibfnamefont
  {A.~F.}\ \bibnamefont {{Rogers}}}, \bibinfo {author} {\bibfnamefont
  {E.}~\bibnamefont {{Thrane}}}, \bibinfo {author} {\bibfnamefont
  {J.}~\bibnamefont {{Askew}}}, \bibinfo {author} {\bibfnamefont {N.~D.~R.}\
  \bibnamefont {{Bhat}}}, \bibinfo {author} {\bibfnamefont {A.}~\bibnamefont
  {{Cameron}}}, \bibinfo {author} {\bibfnamefont {M.}~\bibnamefont
  {{Cury{\l}o}}}, \bibinfo {author} {\bibfnamefont {W.~A.}\ \bibnamefont
  {{Coles}}}, \bibinfo {author} {\bibfnamefont {S.}~\bibnamefont {{Dai}}},
  \bibinfo {author} {\bibfnamefont {B.}~\bibnamefont {{Goncharov}}}, \bibinfo
  {author} {\bibfnamefont {M.}~\bibnamefont {{Kerr}}}, \bibinfo {author}
  {\bibfnamefont {A.}~\bibnamefont {{Kulkarni}}}, \bibinfo {author}
  {\bibfnamefont {Y.}~\bibnamefont {{Levin}}}, \bibinfo {author} {\bibfnamefont
  {M.~E.}\ \bibnamefont {{Lower}}}, \bibinfo {author} {\bibfnamefont {R.~N.}\
  \bibnamefont {{Manchester}}}, \bibinfo {author} {\bibfnamefont
  {R.}~\bibnamefont {{Mandow}}}, \bibinfo {author} {\bibfnamefont {M.~T.}\
  \bibnamefont {{Miles}}}, \bibinfo {author} {\bibfnamefont {R.~S.}\
  \bibnamefont {{Nathan}}}, \bibinfo {author} {\bibfnamefont {S.}~\bibnamefont
  {{Os{\l}owski}}}, \bibinfo {author} {\bibfnamefont {C.~J.}\ \bibnamefont
  {{Russell}}}, \bibinfo {author} {\bibfnamefont {R.}~\bibnamefont
  {{Spiewak}}}, \bibinfo {author} {\bibfnamefont {S.}~\bibnamefont {{Zhang}}},\
  and\ \bibinfo {author} {\bibfnamefont {X.-J.}\ \bibnamefont {{Zhu}}},\
  }\bibfield  {title} {\bibinfo {title} {{Search for an Isotropic
  Gravitational-wave Background with the Parkes Pulsar Timing Array}},\ }\href
  {https://doi.org/10.3847/2041-8213/acdd02} {\bibfield  {journal} {\bibinfo
  {journal} {\apjl}\ }\textbf {\bibinfo {volume} {951}},\ \bibinfo {eid} {L6}
  (\bibinfo {year} {2023})},\ \Eprint {https://arxiv.org/abs/2306.16215}
  {arXiv:2306.16215 [astro-ph.HE]} \BibitemShut {NoStop}%
\bibitem [{\citenamefont {{Xu}}\ \emph {et~al.}(2023)\citenamefont {{Xu}},
  \citenamefont {{Chen}}, \citenamefont {{Guo}}, \citenamefont {{Jiang}},
  \citenamefont {{Wang}}, \citenamefont {{Xu}}, \citenamefont {{Xue}},
  \citenamefont {{Caballero}}, \citenamefont {{Yuan}}, \citenamefont {{Xu}},
  \citenamefont {{Wang}}, \citenamefont {{Hao}}, \citenamefont {{Luo}},
  \citenamefont {{Lee}}, \citenamefont {{Han}}, \citenamefont {{Jiang}},
  \citenamefont {{Shen}}, \citenamefont {{Wang}}, \citenamefont {{Wang}},
  \citenamefont {{Xu}}, \citenamefont {{Wu}}, \citenamefont {{Manchester}},
  \citenamefont {{Qian}}, \citenamefont {{Guan}}, \citenamefont {{Huang}},
  \citenamefont {{Sun}},\ and\ \citenamefont {{Zhu}}}]{CPTA2023}%
  \BibitemOpen
  \bibfield  {author} {\bibinfo {author} {\bibfnamefont {H.}~\bibnamefont
  {{Xu}}}, \bibinfo {author} {\bibfnamefont {S.}~\bibnamefont {{Chen}}},
  \bibinfo {author} {\bibfnamefont {Y.}~\bibnamefont {{Guo}}}, \bibinfo
  {author} {\bibfnamefont {J.}~\bibnamefont {{Jiang}}}, \bibinfo {author}
  {\bibfnamefont {B.}~\bibnamefont {{Wang}}}, \bibinfo {author} {\bibfnamefont
  {J.}~\bibnamefont {{Xu}}}, \bibinfo {author} {\bibfnamefont {Z.}~\bibnamefont
  {{Xue}}}, \bibinfo {author} {\bibfnamefont {R.~N.}\ \bibnamefont
  {{Caballero}}}, \bibinfo {author} {\bibfnamefont {J.}~\bibnamefont {{Yuan}}},
  \bibinfo {author} {\bibfnamefont {Y.}~\bibnamefont {{Xu}}}, \bibinfo {author}
  {\bibfnamefont {J.}~\bibnamefont {{Wang}}}, \bibinfo {author} {\bibfnamefont
  {L.}~\bibnamefont {{Hao}}}, \bibinfo {author} {\bibfnamefont
  {J.}~\bibnamefont {{Luo}}}, \bibinfo {author} {\bibfnamefont
  {K.}~\bibnamefont {{Lee}}}, \bibinfo {author} {\bibfnamefont
  {J.}~\bibnamefont {{Han}}}, \bibinfo {author} {\bibfnamefont
  {P.}~\bibnamefont {{Jiang}}}, \bibinfo {author} {\bibfnamefont
  {Z.}~\bibnamefont {{Shen}}}, \bibinfo {author} {\bibfnamefont
  {M.}~\bibnamefont {{Wang}}}, \bibinfo {author} {\bibfnamefont
  {N.}~\bibnamefont {{Wang}}}, \bibinfo {author} {\bibfnamefont
  {R.}~\bibnamefont {{Xu}}}, \bibinfo {author} {\bibfnamefont {X.}~\bibnamefont
  {{Wu}}}, \bibinfo {author} {\bibfnamefont {R.}~\bibnamefont {{Manchester}}},
  \bibinfo {author} {\bibfnamefont {L.}~\bibnamefont {{Qian}}}, \bibinfo
  {author} {\bibfnamefont {X.}~\bibnamefont {{Guan}}}, \bibinfo {author}
  {\bibfnamefont {M.}~\bibnamefont {{Huang}}}, \bibinfo {author} {\bibfnamefont
  {C.}~\bibnamefont {{Sun}}},\ and\ \bibinfo {author} {\bibfnamefont
  {Y.}~\bibnamefont {{Zhu}}},\ }\bibfield  {title} {\bibinfo {title}
  {{Searching for the Nano-Hertz Stochastic Gravitational Wave Background with
  the Chinese Pulsar Timing Array Data Release I}},\ }\href
  {https://doi.org/10.1088/1674-4527/acdfa5} {\bibfield  {journal} {\bibinfo
  {journal} {Research in Astronomy and Astrophysics}\ }\textbf {\bibinfo
  {volume} {23}},\ \bibinfo {eid} {075024} (\bibinfo {year} {2023})},\ \Eprint
  {https://arxiv.org/abs/2306.16216} {arXiv:2306.16216 [astro-ph.HE]}
  \BibitemShut {NoStop}%
\bibitem [{\citenamefont {{Miles}}\ \emph {et~al.}(2025)\citenamefont
  {{Miles}}, \citenamefont {{Shannon}}, \citenamefont {{Reardon}},
  \citenamefont {{Bailes}}, \citenamefont {{Champion}}, \citenamefont
  {{Geyer}}, \citenamefont {{Gitika}}, \citenamefont {{Grunthal}},
  \citenamefont {{Keith}}, \citenamefont {{Kramer}}, \citenamefont
  {{Kulkarni}}, \citenamefont {{Nathan}}, \citenamefont {{Parthasarathy}},
  \citenamefont {{Singha}}, \citenamefont {{Theureau}}, \citenamefont
  {{Thrane}}, \citenamefont {{Abbate}}, \citenamefont {{Buchner}},
  \citenamefont {{Cameron}}, \citenamefont {{Camilo}}, \citenamefont
  {{Moreschi}}, \citenamefont {{Shaifullah}}, \citenamefont {{Shamohammadi}},
  \citenamefont {{Possenti}},\ and\ \citenamefont {{Krishnan}}}]{MPTA2025}%
  \BibitemOpen
  \bibfield  {author} {\bibinfo {author} {\bibfnamefont {M.~T.}\ \bibnamefont
  {{Miles}}}, \bibinfo {author} {\bibfnamefont {R.~M.}\ \bibnamefont
  {{Shannon}}}, \bibinfo {author} {\bibfnamefont {D.~J.}\ \bibnamefont
  {{Reardon}}}, \bibinfo {author} {\bibfnamefont {M.}~\bibnamefont {{Bailes}}},
  \bibinfo {author} {\bibfnamefont {D.~J.}\ \bibnamefont {{Champion}}},
  \bibinfo {author} {\bibfnamefont {M.}~\bibnamefont {{Geyer}}}, \bibinfo
  {author} {\bibfnamefont {P.}~\bibnamefont {{Gitika}}}, \bibinfo {author}
  {\bibfnamefont {K.}~\bibnamefont {{Grunthal}}}, \bibinfo {author}
  {\bibfnamefont {M.~J.}\ \bibnamefont {{Keith}}}, \bibinfo {author}
  {\bibfnamefont {M.}~\bibnamefont {{Kramer}}}, \bibinfo {author}
  {\bibfnamefont {A.~D.}\ \bibnamefont {{Kulkarni}}}, \bibinfo {author}
  {\bibfnamefont {R.~S.}\ \bibnamefont {{Nathan}}}, \bibinfo {author}
  {\bibfnamefont {A.}~\bibnamefont {{Parthasarathy}}}, \bibinfo {author}
  {\bibfnamefont {J.}~\bibnamefont {{Singha}}}, \bibinfo {author}
  {\bibfnamefont {G.}~\bibnamefont {{Theureau}}}, \bibinfo {author}
  {\bibfnamefont {E.}~\bibnamefont {{Thrane}}}, \bibinfo {author}
  {\bibfnamefont {F.}~\bibnamefont {{Abbate}}}, \bibinfo {author}
  {\bibfnamefont {S.}~\bibnamefont {{Buchner}}}, \bibinfo {author}
  {\bibfnamefont {A.~D.}\ \bibnamefont {{Cameron}}}, \bibinfo {author}
  {\bibfnamefont {F.}~\bibnamefont {{Camilo}}}, \bibinfo {author}
  {\bibfnamefont {B.~E.}\ \bibnamefont {{Moreschi}}}, \bibinfo {author}
  {\bibfnamefont {G.}~\bibnamefont {{Shaifullah}}}, \bibinfo {author}
  {\bibfnamefont {M.}~\bibnamefont {{Shamohammadi}}}, \bibinfo {author}
  {\bibfnamefont {A.}~\bibnamefont {{Possenti}}},\ and\ \bibinfo {author}
  {\bibfnamefont {V.~V.}\ \bibnamefont {{Krishnan}}},\ }\bibfield  {title}
  {\bibinfo {title} {{The MeerKAT Pulsar Timing Array: the first search for
  gravitational waves with the MeerKAT radio telescope}},\ }\href
  {https://doi.org/10.1093/mnras/stae2571} {\bibfield  {journal} {\bibinfo
  {journal} {\mnras}\ }\textbf {\bibinfo {volume} {536}},\ \bibinfo {pages}
  {1489} (\bibinfo {year} {2025})},\ \Eprint {https://arxiv.org/abs/2412.01153}
  {arXiv:2412.01153 [astro-ph.HE]} \BibitemShut {NoStop}%
\bibitem [{\citenamefont {{Sato-Polito}}\ \emph {et~al.}(2024)\citenamefont
  {{Sato-Polito}}, \citenamefont {{Zaldarriaga}},\ and\ \citenamefont
  {{Quataert}}}]{SatoPolito2024}%
  \BibitemOpen
  \bibfield  {author} {\bibinfo {author} {\bibfnamefont {G.}~\bibnamefont
  {{Sato-Polito}}}, \bibinfo {author} {\bibfnamefont {M.}~\bibnamefont
  {{Zaldarriaga}}},\ and\ \bibinfo {author} {\bibfnamefont {E.}~\bibnamefont
  {{Quataert}}},\ }\bibfield  {title} {\bibinfo {title} {{Where are the
  supermassive black holes measured by PTAs?}},\ }\href
  {https://doi.org/10.1103/PhysRevD.110.063020} {\bibfield  {journal} {\bibinfo
   {journal} {Phys. Rev. D}\ }\textbf {\bibinfo {volume} {110}},\ \bibinfo
  {pages} {063020} (\bibinfo {year} {2024})}\BibitemShut {NoStop}%
\bibitem [{\citenamefont {{Liepold}}\ and\ \citenamefont
  {{Ma}}(2024)}]{LiepoldMa2024}%
  \BibitemOpen
  \bibfield  {author} {\bibinfo {author} {\bibfnamefont {E.~R.}\ \bibnamefont
  {{Liepold}}}\ and\ \bibinfo {author} {\bibfnamefont {C.-P.}\ \bibnamefont
  {{Ma}}},\ }\bibfield  {title} {\bibinfo {title} {{Big Galaxies and Big Black
  Holes: The Massive Ends of the Local Stellar and Black Hole Mass Functions
  and the Implications for Nanohertz Gravitational Waves}},\ }\href
  {https://doi.org/10.3847/2041-8213/ad66b8} {\bibfield  {journal} {\bibinfo
  {journal} {\apjl}\ }\textbf {\bibinfo {volume} {971}},\ \bibinfo {eid} {L29}
  (\bibinfo {year} {2024})},\ \Eprint {https://arxiv.org/abs/2407.14595}
  {arXiv:2407.14595 [astro-ph.GA]} \BibitemShut {NoStop}%
\bibitem [{\citenamefont
  {{Mingarelli}}(2026{\natexlab{a}})}]{Mingarelli_ceiling}%
  \BibitemOpen
  \bibfield  {author} {\bibinfo {author} {\bibfnamefont {C.~M.~F.}\
  \bibnamefont {{Mingarelli}}},\ }\bibfield  {title} {\bibinfo {title}
  {{Energetic Ceilings of Astrophysical Gravitational-Wave Backgrounds}},\
  }\href {https://doi.org/10.48550/arXiv.2601.18859} {\bibfield  {journal}
  {\bibinfo  {journal} {arXiv e-prints}\ ,\ \bibinfo {eid} {arXiv:2601.18859}}
  (\bibinfo {year} {2026}{\natexlab{a}})},\ \Eprint
  {https://arxiv.org/abs/2601.18859} {arXiv:2601.18859 [astro-ph.HE]}
  \BibitemShut {NoStop}%
\bibitem [{\citenamefont {{Barausse}}(2012)}]{Barausse2012}%
  \BibitemOpen
  \bibfield  {author} {\bibinfo {author} {\bibfnamefont {E.}~\bibnamefont
  {{Barausse}}},\ }\bibfield  {title} {\bibinfo {title} {{The evolution of
  massive black holes and their spins in their galactic hosts}},\ }\href
  {https://doi.org/10.1111/j.1365-2966.2012.21057.x} {\bibfield  {journal}
  {\bibinfo  {journal} {\mnras}\ }\textbf {\bibinfo {volume} {423}},\ \bibinfo
  {pages} {2533} (\bibinfo {year} {2012})},\ \Eprint
  {https://arxiv.org/abs/1201.5888} {arXiv:1201.5888 [astro-ph.CO]}
  \BibitemShut {NoStop}%
\bibitem [{\citenamefont {{Hughes}}\ and\ \citenamefont
  {{Blandford}}(2003)}]{Hughes2003}%
  \BibitemOpen
  \bibfield  {author} {\bibinfo {author} {\bibfnamefont {S.~A.}\ \bibnamefont
  {{Hughes}}}\ and\ \bibinfo {author} {\bibfnamefont {R.~D.}\ \bibnamefont
  {{Blandford}}},\ }\bibfield  {title} {\bibinfo {title} {{Black Hole Mass and
  Spin Coevolution by Mergers}},\ }\href {https://doi.org/10.1086/375495}
  {\bibfield  {journal} {\bibinfo  {journal} {\apjl}\ }\textbf {\bibinfo
  {volume} {585}},\ \bibinfo {pages} {L101} (\bibinfo {year} {2003})},\ \Eprint
  {https://arxiv.org/abs/astro-ph/0208484} {arXiv:astro-ph/0208484 [astro-ph]}
  \BibitemShut {NoStop}%
\bibitem [{\citenamefont {{Berti}}\ and\ \citenamefont
  {{Volonteri}}(2008)}]{Berti2008}%
  \BibitemOpen
  \bibfield  {author} {\bibinfo {author} {\bibfnamefont {E.}~\bibnamefont
  {{Berti}}}\ and\ \bibinfo {author} {\bibfnamefont {M.}~\bibnamefont
  {{Volonteri}}},\ }\bibfield  {title} {\bibinfo {title} {{Cosmological Black
  Hole Spin Evolution by Mergers and Accretion}},\ }\href
  {https://doi.org/10.1086/590379} {\bibfield  {journal} {\bibinfo  {journal}
  {\apj}\ }\textbf {\bibinfo {volume} {684}},\ \bibinfo {pages} {822} (\bibinfo
  {year} {2008})},\ \Eprint {https://arxiv.org/abs/0802.0025} {arXiv:0802.0025
  [astro-ph]} \BibitemShut {NoStop}%
\bibitem [{\citenamefont {{Abbott}}\ \emph {et~al.}(2025)\citenamefont
  {{Abbott}}, \citenamefont {{Abe}}, \citenamefont {{Acernese}}, \citenamefont
  {{Ackley}}, \citenamefont {{Adhikari}}, \citenamefont {{Adhikari}},
  \citenamefont {{Adkins}}, \citenamefont {{Adya}}, \citenamefont {{Affeldt}},
  \citenamefont {{Agarwal}}, \citenamefont {{Agathos}}, \citenamefont
  {{Agatsuma}}, \citenamefont {{Aggarwal}}, \citenamefont {{Aguiar}},
  \citenamefont {{Aiello}}, \citenamefont {{Ain}}, \citenamefont {{Ajith}},
  \citenamefont {{Akutsu}}, \citenamefont {{de Alarc{\'o}n}}, \citenamefont
  {{Albanesi}}, \citenamefont {{Alfaidi}}, \citenamefont {{Allocca}},
  \citenamefont {{Altin}}, \citenamefont {{Amato}}, \citenamefont {{Anand}},
  \citenamefont {{Anand}}, \citenamefont {{Ananyeva}}, \citenamefont
  {{Anderson}}, \citenamefont {{Anderson}}, \citenamefont {{Ando}},
  \citenamefont {{Andrade}}, \citenamefont {{Andres}}, \citenamefont
  {{Andr{\'e}s-Carcasona}}, \citenamefont {{Andri{\'c}}}, \citenamefont
  {{Angelova}}, \citenamefont {{Ansoldi}}, \citenamefont {{Antelis}},
  \citenamefont {{Antier}}, \citenamefont {{Apostolatos}}, \citenamefont
  {{Appavuravther}}, \citenamefont {{Appert}}, \citenamefont {{Apple}},
  \citenamefont {{Arai}}, \citenamefont {{Araya}}, \citenamefont {{Araya}},
  \citenamefont {{Areeda}}, \citenamefont {{Ar{\`e}ne}}, \citenamefont
  {{Aritomi}}, \citenamefont {{Arnaud}}, \citenamefont {{Arogeti}},
  \citenamefont {{Aronson}}, \citenamefont {{Arun}}, \citenamefont {{Asada}},
  \citenamefont {{Asali}}, \citenamefont {{Ashton}}, \citenamefont {{Aso}},
  \citenamefont {{Assiduo}}, \citenamefont {{Melo}}, \citenamefont {{Aston}},
  \citenamefont {{Astone}}, \citenamefont {{Aubin}}, \citenamefont
  {{Aultoneal}}, \citenamefont {{Austin}}, \citenamefont {{Babak}},
  \citenamefont {{Badaracco}}, \citenamefont {{Bader}}, \citenamefont
  {{Badger}}, \citenamefont {{Bae}}, \citenamefont {{Bae}}, \citenamefont
  {{Baer}}, \citenamefont {{Bagnasco}}, \citenamefont {{Bai}}, \citenamefont
  {{Baird}}, \citenamefont {{Bajpai}}, \citenamefont {{Baka}}, \citenamefont
  {{Ball}}, \citenamefont {{Ballardin}}, \citenamefont {{Ballmer}},
  \citenamefont {{Balsamo}}, \citenamefont {{Baltus}}, \citenamefont
  {{Banagiri}}, \citenamefont {{Banerjee}}, \citenamefont {{Bankar}},
  \citenamefont {{Barayoga}}, \citenamefont {{Barbieri}}, \citenamefont
  {{Barish}}, \citenamefont {{Barker}}, \citenamefont {{Barneo}}, \citenamefont
  {{Barone}}, \citenamefont {{Barr}}, \citenamefont {{Barsotti}}, \citenamefont
  {{Barsuglia}}, \citenamefont {{Barta}}, \citenamefont {{Bartlett}},
  \citenamefont {{Barton}}, \citenamefont {{Bartos}}, \citenamefont {{Basak}},
  \citenamefont {{Bassiri}}, \citenamefont {{Basti}}, \citenamefont {{Bawaj}},
  \citenamefont {{Bayley}}, \citenamefont {{Bazzan}}, \citenamefont {{Becher}},
  \citenamefont {{B{\'e}csy}}, \citenamefont {{Bedakihale}}, \citenamefont
  {{Beirnaert}}, \citenamefont {{Bejger}}, \citenamefont {{Belahcene}},
  \citenamefont {{Benedetto}}, \citenamefont {{Beniwal}}, \citenamefont
  {{Benjamin}}, \citenamefont {{Bennett}}, \citenamefont {{Bentley}},
  \citenamefont {{Benyaala}}, \citenamefont {{Bera}}, \citenamefont {{Berbel}},
  \citenamefont {{Bergamin}}, \citenamefont {{Berger}}, \citenamefont
  {{Bernuzzi}}, \citenamefont {{Berry}}, \citenamefont {{Bersanetti}},
  \citenamefont {{Bertolini}}, \citenamefont {{Betzwieser}}, \citenamefont
  {{Beveridge}}, \citenamefont {{Bhandare}}, \citenamefont {{Bhandari}},
  \citenamefont {{Bhardwaj}}, \citenamefont {{Bhatt}}, \citenamefont
  {{Bhattacharjee}}, \citenamefont {{Bhaumik}}, \citenamefont {{Bianchi}},
  \citenamefont {{Bilenko}}, \citenamefont {{Billingsley}}, \citenamefont
  {{Bini}}, \citenamefont {{Birney}}, \citenamefont {{Birnholtz}},
  \citenamefont {{Biscans}}, \citenamefont {{Bischi}}, \citenamefont
  {{Biscoveanu}}, \citenamefont {{Bisht}}, \citenamefont {{Biswas}},
  \citenamefont {{Bitossi}}, \citenamefont {{Bizouard}}, \citenamefont
  {{Blackburn}}, \citenamefont {{Blair}}, \citenamefont {{Blair}},
  \citenamefont {{Blair}}, \citenamefont {{Bobba}}, \citenamefont {{Bode}},
  \citenamefont {{Bo{\"e}r}}, \citenamefont {{Bogaert}}, \citenamefont
  {{Boldrini}}, \citenamefont {{Bolingbroke}}, \citenamefont {{Bonavena}},
  \citenamefont {{Bondu}}, \citenamefont {{Bonilla}}, \citenamefont
  {{Bonnand}}, \citenamefont {{Booker}}, \citenamefont {{Boom}}, \citenamefont
  {{Bork}}, \citenamefont {{Boschi}}, \citenamefont {{Bose}}, \citenamefont
  {{Bose}}, \citenamefont {{Bossilkov}}, \citenamefont {{Boudart}},
  \citenamefont {{Bouffanais}}, \citenamefont {{Bozzi}}, \citenamefont
  {{Bradaschia}}, \citenamefont {{Brady}}, \citenamefont {{Bramley}},
  \citenamefont {{Branch}}, \citenamefont {{Branchesi}}, \citenamefont
  {{Brau}}, \citenamefont {{Breschi}}, \citenamefont {{Briant}}, \citenamefont
  {{Briggs}}, \citenamefont {{Brillet}}, \citenamefont {{Brinkmann}},
  \citenamefont {{Brockill}}, \citenamefont {{Brooks}}, \citenamefont
  {{Brooks}}, \citenamefont {{Brown}}, \citenamefont {{Brunett}}, \citenamefont
  {{Bruno}}, \citenamefont {{Bruntz}}, \citenamefont {{Bryant}}, \citenamefont
  {{Bucci}}, \citenamefont {{Bulik}}, \citenamefont {{Bulten}}, \citenamefont
  {{Buonanno}}, \citenamefont {{Burtnyk}}, \citenamefont {{Buscicchio}},
  \citenamefont {{Buskulic}}, \citenamefont {{Buy}}, \citenamefont {{Byer}},
  \citenamefont {{Davies}}, \citenamefont {{Cabras}}, \citenamefont
  {{Cabrita}}, \citenamefont {{Cadonati}},\ and\ \citenamefont
  {{Caesar}}}]{LVK2025}%
  \BibitemOpen
  \bibfield  {author} {\bibinfo {author} {\bibfnamefont {R.}~\bibnamefont
  {{Abbott}}}, \bibinfo {author} {\bibfnamefont {H.}~\bibnamefont {{Abe}}},
  \bibinfo {author} {\bibfnamefont {F.}~\bibnamefont {{Acernese}}}, \bibinfo
  {author} {\bibfnamefont {K.}~\bibnamefont {{Ackley}}}, \bibinfo {author}
  {\bibfnamefont {N.}~\bibnamefont {{Adhikari}}}, \bibinfo {author}
  {\bibfnamefont {R.~X.}\ \bibnamefont {{Adhikari}}}, \bibinfo {author}
  {\bibfnamefont {V.~K.}\ \bibnamefont {{Adkins}}}, \bibinfo {author}
  {\bibfnamefont {V.~B.}\ \bibnamefont {{Adya}}}, \bibinfo {author}
  {\bibfnamefont {C.}~\bibnamefont {{Affeldt}}}, \bibinfo {author}
  {\bibfnamefont {D.}~\bibnamefont {{Agarwal}}}, \bibinfo {author}
  {\bibfnamefont {M.}~\bibnamefont {{Agathos}}}, \bibinfo {author}
  {\bibfnamefont {K.}~\bibnamefont {{Agatsuma}}}, \bibinfo {author}
  {\bibfnamefont {N.}~\bibnamefont {{Aggarwal}}}, \bibinfo {author}
  {\bibfnamefont {O.~D.}\ \bibnamefont {{Aguiar}}}, \bibinfo {author}
  {\bibfnamefont {L.}~\bibnamefont {{Aiello}}}, \bibinfo {author}
  {\bibfnamefont {A.}~\bibnamefont {{Ain}}}, \bibinfo {author} {\bibfnamefont
  {P.}~\bibnamefont {{Ajith}}}, \bibinfo {author} {\bibfnamefont
  {T.}~\bibnamefont {{Akutsu}}}, \bibinfo {author} {\bibfnamefont {P.~F.}\
  \bibnamefont {{de Alarc{\'o}n}}}, \bibinfo {author} {\bibfnamefont
  {S.}~\bibnamefont {{Albanesi}}}, \bibinfo {author} {\bibfnamefont {R.~A.}\
  \bibnamefont {{Alfaidi}}}, \bibinfo {author} {\bibfnamefont {A.}~\bibnamefont
  {{Allocca}}}, \bibinfo {author} {\bibfnamefont {P.~A.}\ \bibnamefont
  {{Altin}}}, \bibinfo {author} {\bibfnamefont {A.}~\bibnamefont {{Amato}}},
  \bibinfo {author} {\bibfnamefont {C.}~\bibnamefont {{Anand}}}, \bibinfo
  {author} {\bibfnamefont {S.}~\bibnamefont {{Anand}}}, \bibinfo {author}
  {\bibfnamefont {A.}~\bibnamefont {{Ananyeva}}}, \bibinfo {author}
  {\bibfnamefont {S.~B.}\ \bibnamefont {{Anderson}}}, \bibinfo {author}
  {\bibfnamefont {W.~G.}\ \bibnamefont {{Anderson}}}, \bibinfo {author}
  {\bibfnamefont {M.}~\bibnamefont {{Ando}}}, \bibinfo {author} {\bibfnamefont
  {T.}~\bibnamefont {{Andrade}}}, \bibinfo {author} {\bibfnamefont
  {N.}~\bibnamefont {{Andres}}}, \bibinfo {author} {\bibfnamefont
  {M.}~\bibnamefont {{Andr{\'e}s-Carcasona}}}, \bibinfo {author} {\bibfnamefont
  {T.}~\bibnamefont {{Andri{\'c}}}}, \bibinfo {author} {\bibfnamefont {S.~V.}\
  \bibnamefont {{Angelova}}}, \bibinfo {author} {\bibfnamefont
  {S.}~\bibnamefont {{Ansoldi}}}, \bibinfo {author} {\bibfnamefont {J.~M.}\
  \bibnamefont {{Antelis}}}, \bibinfo {author} {\bibfnamefont {S.}~\bibnamefont
  {{Antier}}}, \bibinfo {author} {\bibfnamefont {T.}~\bibnamefont
  {{Apostolatos}}}, \bibinfo {author} {\bibfnamefont {E.~Z.}\ \bibnamefont
  {{Appavuravther}}}, \bibinfo {author} {\bibfnamefont {S.}~\bibnamefont
  {{Appert}}}, \bibinfo {author} {\bibfnamefont {S.~K.}\ \bibnamefont
  {{Apple}}}, \bibinfo {author} {\bibfnamefont {K.}~\bibnamefont {{Arai}}},
  \bibinfo {author} {\bibfnamefont {A.}~\bibnamefont {{Araya}}}, \bibinfo
  {author} {\bibfnamefont {M.~C.}\ \bibnamefont {{Araya}}}, \bibinfo {author}
  {\bibfnamefont {J.~S.}\ \bibnamefont {{Areeda}}}, \bibinfo {author}
  {\bibfnamefont {M.}~\bibnamefont {{Ar{\`e}ne}}}, \bibinfo {author}
  {\bibfnamefont {N.}~\bibnamefont {{Aritomi}}}, \bibinfo {author}
  {\bibfnamefont {N.}~\bibnamefont {{Arnaud}}}, \bibinfo {author}
  {\bibfnamefont {M.}~\bibnamefont {{Arogeti}}}, \bibinfo {author}
  {\bibfnamefont {S.~M.}\ \bibnamefont {{Aronson}}}, \bibinfo {author}
  {\bibfnamefont {K.~G.}\ \bibnamefont {{Arun}}}, \bibinfo {author}
  {\bibfnamefont {H.}~\bibnamefont {{Asada}}}, \bibinfo {author} {\bibfnamefont
  {Y.}~\bibnamefont {{Asali}}}, \bibinfo {author} {\bibfnamefont
  {G.}~\bibnamefont {{Ashton}}}, \bibinfo {author} {\bibfnamefont
  {Y.}~\bibnamefont {{Aso}}}, \bibinfo {author} {\bibfnamefont
  {M.}~\bibnamefont {{Assiduo}}}, \bibinfo {author} {\bibfnamefont {S.~A.
  D.~S.}\ \bibnamefont {{Melo}}}, \bibinfo {author} {\bibfnamefont {S.~M.}\
  \bibnamefont {{Aston}}}, \bibinfo {author} {\bibfnamefont {P.}~\bibnamefont
  {{Astone}}}, \bibinfo {author} {\bibfnamefont {F.}~\bibnamefont {{Aubin}}},
  \bibinfo {author} {\bibfnamefont {K.}~\bibnamefont {{Aultoneal}}}, \bibinfo
  {author} {\bibfnamefont {C.}~\bibnamefont {{Austin}}}, \bibinfo {author}
  {\bibfnamefont {S.}~\bibnamefont {{Babak}}}, \bibinfo {author} {\bibfnamefont
  {F.}~\bibnamefont {{Badaracco}}}, \bibinfo {author} {\bibfnamefont
  {M.~K.~M.}\ \bibnamefont {{Bader}}}, \bibinfo {author} {\bibfnamefont
  {C.}~\bibnamefont {{Badger}}}, \bibinfo {author} {\bibfnamefont
  {S.}~\bibnamefont {{Bae}}}, \bibinfo {author} {\bibfnamefont
  {Y.}~\bibnamefont {{Bae}}}, \bibinfo {author} {\bibfnamefont {A.~M.}\
  \bibnamefont {{Baer}}}, \bibinfo {author} {\bibfnamefont {S.}~\bibnamefont
  {{Bagnasco}}}, \bibinfo {author} {\bibfnamefont {Y.}~\bibnamefont {{Bai}}},
  \bibinfo {author} {\bibfnamefont {J.}~\bibnamefont {{Baird}}}, \bibinfo
  {author} {\bibfnamefont {R.}~\bibnamefont {{Bajpai}}}, \bibinfo {author}
  {\bibfnamefont {T.}~\bibnamefont {{Baka}}}, \bibinfo {author} {\bibfnamefont
  {M.}~\bibnamefont {{Ball}}}, \bibinfo {author} {\bibfnamefont
  {G.}~\bibnamefont {{Ballardin}}}, \bibinfo {author} {\bibfnamefont {S.~W.}\
  \bibnamefont {{Ballmer}}}, \bibinfo {author} {\bibfnamefont {A.}~\bibnamefont
  {{Balsamo}}}, \bibinfo {author} {\bibfnamefont {G.}~\bibnamefont {{Baltus}}},
  \bibinfo {author} {\bibfnamefont {S.}~\bibnamefont {{Banagiri}}}, \bibinfo
  {author} {\bibfnamefont {B.}~\bibnamefont {{Banerjee}}}, \bibinfo {author}
  {\bibfnamefont {D.}~\bibnamefont {{Bankar}}}, \bibinfo {author}
  {\bibfnamefont {J.~C.}\ \bibnamefont {{Barayoga}}}, \bibinfo {author}
  {\bibfnamefont {C.}~\bibnamefont {{Barbieri}}}, \bibinfo {author}
  {\bibfnamefont {B.~C.}\ \bibnamefont {{Barish}}}, \bibinfo {author}
  {\bibfnamefont {D.}~\bibnamefont {{Barker}}}, \bibinfo {author}
  {\bibfnamefont {P.}~\bibnamefont {{Barneo}}}, \bibinfo {author}
  {\bibfnamefont {F.}~\bibnamefont {{Barone}}}, \bibinfo {author}
  {\bibfnamefont {B.}~\bibnamefont {{Barr}}}, \bibinfo {author} {\bibfnamefont
  {L.}~\bibnamefont {{Barsotti}}}, \bibinfo {author} {\bibfnamefont
  {M.}~\bibnamefont {{Barsuglia}}}, \bibinfo {author} {\bibfnamefont
  {D.}~\bibnamefont {{Barta}}}, \bibinfo {author} {\bibfnamefont
  {J.}~\bibnamefont {{Bartlett}}}, \bibinfo {author} {\bibfnamefont {M.~A.}\
  \bibnamefont {{Barton}}}, \bibinfo {author} {\bibfnamefont {I.}~\bibnamefont
  {{Bartos}}}, \bibinfo {author} {\bibfnamefont {S.}~\bibnamefont {{Basak}}},
  \bibinfo {author} {\bibfnamefont {R.}~\bibnamefont {{Bassiri}}}, \bibinfo
  {author} {\bibfnamefont {A.}~\bibnamefont {{Basti}}}, \bibinfo {author}
  {\bibfnamefont {M.}~\bibnamefont {{Bawaj}}}, \bibinfo {author} {\bibfnamefont
  {J.~C.}\ \bibnamefont {{Bayley}}}, \bibinfo {author} {\bibfnamefont
  {M.}~\bibnamefont {{Bazzan}}}, \bibinfo {author} {\bibfnamefont {B.~R.}\
  \bibnamefont {{Becher}}}, \bibinfo {author} {\bibfnamefont {B.}~\bibnamefont
  {{B{\'e}csy}}}, \bibinfo {author} {\bibfnamefont {V.~M.}\ \bibnamefont
  {{Bedakihale}}}, \bibinfo {author} {\bibfnamefont {F.}~\bibnamefont
  {{Beirnaert}}}, \bibinfo {author} {\bibfnamefont {M.}~\bibnamefont
  {{Bejger}}}, \bibinfo {author} {\bibfnamefont {I.}~\bibnamefont
  {{Belahcene}}}, \bibinfo {author} {\bibfnamefont {V.}~\bibnamefont
  {{Benedetto}}}, \bibinfo {author} {\bibfnamefont {D.}~\bibnamefont
  {{Beniwal}}}, \bibinfo {author} {\bibfnamefont {M.~G.}\ \bibnamefont
  {{Benjamin}}}, \bibinfo {author} {\bibfnamefont {T.~F.}\ \bibnamefont
  {{Bennett}}}, \bibinfo {author} {\bibfnamefont {J.~D.}\ \bibnamefont
  {{Bentley}}}, \bibinfo {author} {\bibfnamefont {M.}~\bibnamefont
  {{Benyaala}}}, \bibinfo {author} {\bibfnamefont {S.}~\bibnamefont {{Bera}}},
  \bibinfo {author} {\bibfnamefont {M.}~\bibnamefont {{Berbel}}}, \bibinfo
  {author} {\bibfnamefont {F.}~\bibnamefont {{Bergamin}}}, \bibinfo {author}
  {\bibfnamefont {B.~K.}\ \bibnamefont {{Berger}}}, \bibinfo {author}
  {\bibfnamefont {S.}~\bibnamefont {{Bernuzzi}}}, \bibinfo {author}
  {\bibfnamefont {C.~P.~L.}\ \bibnamefont {{Berry}}}, \bibinfo {author}
  {\bibfnamefont {D.}~\bibnamefont {{Bersanetti}}}, \bibinfo {author}
  {\bibfnamefont {A.}~\bibnamefont {{Bertolini}}}, \bibinfo {author}
  {\bibfnamefont {J.}~\bibnamefont {{Betzwieser}}}, \bibinfo {author}
  {\bibfnamefont {D.}~\bibnamefont {{Beveridge}}}, \bibinfo {author}
  {\bibfnamefont {R.}~\bibnamefont {{Bhandare}}}, \bibinfo {author}
  {\bibfnamefont {A.~V.}\ \bibnamefont {{Bhandari}}}, \bibinfo {author}
  {\bibfnamefont {U.}~\bibnamefont {{Bhardwaj}}}, \bibinfo {author}
  {\bibfnamefont {R.}~\bibnamefont {{Bhatt}}}, \bibinfo {author} {\bibfnamefont
  {D.}~\bibnamefont {{Bhattacharjee}}}, \bibinfo {author} {\bibfnamefont
  {S.}~\bibnamefont {{Bhaumik}}}, \bibinfo {author} {\bibfnamefont
  {A.}~\bibnamefont {{Bianchi}}}, \bibinfo {author} {\bibfnamefont {I.~A.}\
  \bibnamefont {{Bilenko}}}, \bibinfo {author} {\bibfnamefont {G.}~\bibnamefont
  {{Billingsley}}}, \bibinfo {author} {\bibfnamefont {S.}~\bibnamefont
  {{Bini}}}, \bibinfo {author} {\bibfnamefont {R.}~\bibnamefont {{Birney}}},
  \bibinfo {author} {\bibfnamefont {O.}~\bibnamefont {{Birnholtz}}}, \bibinfo
  {author} {\bibfnamefont {S.}~\bibnamefont {{Biscans}}}, \bibinfo {author}
  {\bibfnamefont {M.}~\bibnamefont {{Bischi}}}, \bibinfo {author}
  {\bibfnamefont {S.}~\bibnamefont {{Biscoveanu}}}, \bibinfo {author}
  {\bibfnamefont {A.}~\bibnamefont {{Bisht}}}, \bibinfo {author} {\bibfnamefont
  {B.}~\bibnamefont {{Biswas}}}, \bibinfo {author} {\bibfnamefont
  {M.}~\bibnamefont {{Bitossi}}}, \bibinfo {author} {\bibfnamefont {M.-A.}\
  \bibnamefont {{Bizouard}}}, \bibinfo {author} {\bibfnamefont {J.~K.}\
  \bibnamefont {{Blackburn}}}, \bibinfo {author} {\bibfnamefont {C.~D.}\
  \bibnamefont {{Blair}}}, \bibinfo {author} {\bibfnamefont {D.~G.}\
  \bibnamefont {{Blair}}}, \bibinfo {author} {\bibfnamefont {R.~M.}\
  \bibnamefont {{Blair}}}, \bibinfo {author} {\bibfnamefont {F.}~\bibnamefont
  {{Bobba}}}, \bibinfo {author} {\bibfnamefont {N.}~\bibnamefont {{Bode}}},
  \bibinfo {author} {\bibfnamefont {M.}~\bibnamefont {{Bo{\"e}r}}}, \bibinfo
  {author} {\bibfnamefont {G.}~\bibnamefont {{Bogaert}}}, \bibinfo {author}
  {\bibfnamefont {M.}~\bibnamefont {{Boldrini}}}, \bibinfo {author}
  {\bibfnamefont {G.~N.}\ \bibnamefont {{Bolingbroke}}}, \bibinfo {author}
  {\bibfnamefont {L.~D.}\ \bibnamefont {{Bonavena}}}, \bibinfo {author}
  {\bibfnamefont {F.}~\bibnamefont {{Bondu}}}, \bibinfo {author} {\bibfnamefont
  {E.}~\bibnamefont {{Bonilla}}}, \bibinfo {author} {\bibfnamefont
  {R.}~\bibnamefont {{Bonnand}}}, \bibinfo {author} {\bibfnamefont
  {P.}~\bibnamefont {{Booker}}}, \bibinfo {author} {\bibfnamefont {B.~A.}\
  \bibnamefont {{Boom}}}, \bibinfo {author} {\bibfnamefont {R.}~\bibnamefont
  {{Bork}}}, \bibinfo {author} {\bibfnamefont {V.}~\bibnamefont {{Boschi}}},
  \bibinfo {author} {\bibfnamefont {N.}~\bibnamefont {{Bose}}}, \bibinfo
  {author} {\bibfnamefont {S.}~\bibnamefont {{Bose}}}, \bibinfo {author}
  {\bibfnamefont {V.}~\bibnamefont {{Bossilkov}}}, \bibinfo {author}
  {\bibfnamefont {V.}~\bibnamefont {{Boudart}}}, \bibinfo {author}
  {\bibfnamefont {Y.}~\bibnamefont {{Bouffanais}}}, \bibinfo {author}
  {\bibfnamefont {A.}~\bibnamefont {{Bozzi}}}, \bibinfo {author} {\bibfnamefont
  {C.}~\bibnamefont {{Bradaschia}}}, \bibinfo {author} {\bibfnamefont {P.~R.}\
  \bibnamefont {{Brady}}}, \bibinfo {author} {\bibfnamefont {A.}~\bibnamefont
  {{Bramley}}}, \bibinfo {author} {\bibfnamefont {A.}~\bibnamefont {{Branch}}},
  \bibinfo {author} {\bibfnamefont {M.}~\bibnamefont {{Branchesi}}}, \bibinfo
  {author} {\bibfnamefont {J.~E.}\ \bibnamefont {{Brau}}}, \bibinfo {author}
  {\bibfnamefont {M.}~\bibnamefont {{Breschi}}}, \bibinfo {author}
  {\bibfnamefont {T.}~\bibnamefont {{Briant}}}, \bibinfo {author}
  {\bibfnamefont {J.~H.}\ \bibnamefont {{Briggs}}}, \bibinfo {author}
  {\bibfnamefont {A.}~\bibnamefont {{Brillet}}}, \bibinfo {author}
  {\bibfnamefont {M.}~\bibnamefont {{Brinkmann}}}, \bibinfo {author}
  {\bibfnamefont {P.}~\bibnamefont {{Brockill}}}, \bibinfo {author}
  {\bibfnamefont {A.~F.}\ \bibnamefont {{Brooks}}}, \bibinfo {author}
  {\bibfnamefont {J.}~\bibnamefont {{Brooks}}}, \bibinfo {author}
  {\bibfnamefont {D.~D.}\ \bibnamefont {{Brown}}}, \bibinfo {author}
  {\bibfnamefont {S.}~\bibnamefont {{Brunett}}}, \bibinfo {author}
  {\bibfnamefont {G.}~\bibnamefont {{Bruno}}}, \bibinfo {author} {\bibfnamefont
  {R.}~\bibnamefont {{Bruntz}}}, \bibinfo {author} {\bibfnamefont
  {J.}~\bibnamefont {{Bryant}}}, \bibinfo {author} {\bibfnamefont
  {F.}~\bibnamefont {{Bucci}}}, \bibinfo {author} {\bibfnamefont
  {T.}~\bibnamefont {{Bulik}}}, \bibinfo {author} {\bibfnamefont {H.~J.}\
  \bibnamefont {{Bulten}}}, \bibinfo {author} {\bibfnamefont {A.}~\bibnamefont
  {{Buonanno}}}, \bibinfo {author} {\bibfnamefont {K.}~\bibnamefont
  {{Burtnyk}}}, \bibinfo {author} {\bibfnamefont {R.}~\bibnamefont
  {{Buscicchio}}}, \bibinfo {author} {\bibfnamefont {D.}~\bibnamefont
  {{Buskulic}}}, \bibinfo {author} {\bibfnamefont {C.}~\bibnamefont {{Buy}}},
  \bibinfo {author} {\bibfnamefont {R.~L.}\ \bibnamefont {{Byer}}}, \bibinfo
  {author} {\bibfnamefont {G.~S.~C.}\ \bibnamefont {{Davies}}}, \bibinfo
  {author} {\bibfnamefont {G.}~\bibnamefont {{Cabras}}}, \bibinfo {author}
  {\bibfnamefont {R.}~\bibnamefont {{Cabrita}}}, \bibinfo {author}
  {\bibfnamefont {L.}~\bibnamefont {{Cadonati}}},\ and\ \bibinfo {author}
  {\bibfnamefont {M.}~\bibnamefont {{Caesar}}},\ }\bibfield  {title} {\bibinfo
  {title} {{Tests of general relativity with GWTC-3}},\ }\href
  {https://doi.org/10.1103/PhysRevD.112.084080} {\bibfield  {journal} {\bibinfo
   {journal} {\prd}\ }\textbf {\bibinfo {volume} {112}},\ \bibinfo {eid}
  {084080} (\bibinfo {year} {2025})},\ \Eprint
  {https://arxiv.org/abs/2112.06861} {arXiv:2112.06861 [gr-qc]} \BibitemShut
  {NoStop}%
\bibitem [{\citenamefont {{Amaro-Seoane}}\ \emph {et~al.}(2017)\citenamefont
  {{Amaro-Seoane}}, \citenamefont {{Audley}}, \citenamefont {{Babak}},
  \citenamefont {{Baker}}, \citenamefont {{Barausse}}, \citenamefont
  {{Bender}}, \citenamefont {{Berti}}, \citenamefont {{Binetruy}},
  \citenamefont {{Born}}, \citenamefont {{Bortoluzzi}}, \citenamefont {{Camp}},
  \citenamefont {{Caprini}}, \citenamefont {{Cardoso}}, \citenamefont
  {{Colpi}}, \citenamefont {{Conklin}}, \citenamefont {{Cornish}},
  \citenamefont {{Cutler}}, \citenamefont {{Danzmann}}, \citenamefont
  {{Dolesi}}, \citenamefont {{Ferraioli}}, \citenamefont {{Ferroni}},
  \citenamefont {{Fitzsimons}}, \citenamefont {{Gair}}, \citenamefont {{Gesa
  Bote}}, \citenamefont {{Giardini}}, \citenamefont {{Gibert}}, \citenamefont
  {{Grimani}}, \citenamefont {{Halloin}}, \citenamefont {{Heinzel}},
  \citenamefont {{Hertog}}, \citenamefont {{Hewitson}}, \citenamefont
  {{Holley-Bockelmann}}, \citenamefont {{Hollington}}, \citenamefont
  {{Hueller}}, \citenamefont {{Inchauspe}}, \citenamefont {{Jetzer}},
  \citenamefont {{Karnesis}}, \citenamefont {{Killow}}, \citenamefont
  {{Klein}}, \citenamefont {{Klipstein}}, \citenamefont {{Korsakova}},
  \citenamefont {{Larson}}, \citenamefont {{Livas}}, \citenamefont {{Lloro}},
  \citenamefont {{Man}}, \citenamefont {{Mance}}, \citenamefont {{Martino}},
  \citenamefont {{Mateos}}, \citenamefont {{McKenzie}}, \citenamefont
  {{McWilliams}}, \citenamefont {{Miller}}, \citenamefont {{Mueller}},
  \citenamefont {{Nardini}}, \citenamefont {{Nelemans}}, \citenamefont
  {{Nofrarias}}, \citenamefont {{Petiteau}}, \citenamefont {{Pivato}},
  \citenamefont {{Plagnol}}, \citenamefont {{Porter}}, \citenamefont
  {{Reiche}}, \citenamefont {{Robertson}}, \citenamefont {{Robertson}},
  \citenamefont {{Rossi}}, \citenamefont {{Russano}}, \citenamefont {{Schutz}},
  \citenamefont {{Sesana}}, \citenamefont {{Shoemaker}}, \citenamefont
  {{Slutsky}}, \citenamefont {{Sopuerta}}, \citenamefont {{Sumner}},
  \citenamefont {{Tamanini}}, \citenamefont {{Thorpe}}, \citenamefont
  {{Troebs}}, \citenamefont {{Vallisneri}}, \citenamefont {{Vecchio}},
  \citenamefont {{Vetrugno}}, \citenamefont {{Vitale}}, \citenamefont
  {{Volonteri}}, \citenamefont {{Wanner}}, \citenamefont {{Ward}},
  \citenamefont {{Wass}}, \citenamefont {{Weber}}, \citenamefont {{Ziemer}},\
  and\ \citenamefont {{Zweifel}}}]{LISA2017}%
  \BibitemOpen
  \bibfield  {author} {\bibinfo {author} {\bibfnamefont {P.}~\bibnamefont
  {{Amaro-Seoane}}}, \bibinfo {author} {\bibfnamefont {H.}~\bibnamefont
  {{Audley}}}, \bibinfo {author} {\bibfnamefont {S.}~\bibnamefont {{Babak}}},
  \bibinfo {author} {\bibfnamefont {J.}~\bibnamefont {{Baker}}}, \bibinfo
  {author} {\bibfnamefont {E.}~\bibnamefont {{Barausse}}}, \bibinfo {author}
  {\bibfnamefont {P.}~\bibnamefont {{Bender}}}, \bibinfo {author}
  {\bibfnamefont {E.}~\bibnamefont {{Berti}}}, \bibinfo {author} {\bibfnamefont
  {P.}~\bibnamefont {{Binetruy}}}, \bibinfo {author} {\bibfnamefont
  {M.}~\bibnamefont {{Born}}}, \bibinfo {author} {\bibfnamefont
  {D.}~\bibnamefont {{Bortoluzzi}}}, \bibinfo {author} {\bibfnamefont
  {J.}~\bibnamefont {{Camp}}}, \bibinfo {author} {\bibfnamefont
  {C.}~\bibnamefont {{Caprini}}}, \bibinfo {author} {\bibfnamefont
  {V.}~\bibnamefont {{Cardoso}}}, \bibinfo {author} {\bibfnamefont
  {M.}~\bibnamefont {{Colpi}}}, \bibinfo {author} {\bibfnamefont
  {J.}~\bibnamefont {{Conklin}}}, \bibinfo {author} {\bibfnamefont
  {N.}~\bibnamefont {{Cornish}}}, \bibinfo {author} {\bibfnamefont
  {C.}~\bibnamefont {{Cutler}}}, \bibinfo {author} {\bibfnamefont
  {K.}~\bibnamefont {{Danzmann}}}, \bibinfo {author} {\bibfnamefont
  {R.}~\bibnamefont {{Dolesi}}}, \bibinfo {author} {\bibfnamefont
  {L.}~\bibnamefont {{Ferraioli}}}, \bibinfo {author} {\bibfnamefont
  {V.}~\bibnamefont {{Ferroni}}}, \bibinfo {author} {\bibfnamefont
  {E.}~\bibnamefont {{Fitzsimons}}}, \bibinfo {author} {\bibfnamefont
  {J.}~\bibnamefont {{Gair}}}, \bibinfo {author} {\bibfnamefont
  {L.}~\bibnamefont {{Gesa Bote}}}, \bibinfo {author} {\bibfnamefont
  {D.}~\bibnamefont {{Giardini}}}, \bibinfo {author} {\bibfnamefont
  {F.}~\bibnamefont {{Gibert}}}, \bibinfo {author} {\bibfnamefont
  {C.}~\bibnamefont {{Grimani}}}, \bibinfo {author} {\bibfnamefont
  {H.}~\bibnamefont {{Halloin}}}, \bibinfo {author} {\bibfnamefont
  {G.}~\bibnamefont {{Heinzel}}}, \bibinfo {author} {\bibfnamefont
  {T.}~\bibnamefont {{Hertog}}}, \bibinfo {author} {\bibfnamefont
  {M.}~\bibnamefont {{Hewitson}}}, \bibinfo {author} {\bibfnamefont
  {K.}~\bibnamefont {{Holley-Bockelmann}}}, \bibinfo {author} {\bibfnamefont
  {D.}~\bibnamefont {{Hollington}}}, \bibinfo {author} {\bibfnamefont
  {M.}~\bibnamefont {{Hueller}}}, \bibinfo {author} {\bibfnamefont
  {H.}~\bibnamefont {{Inchauspe}}}, \bibinfo {author} {\bibfnamefont
  {P.}~\bibnamefont {{Jetzer}}}, \bibinfo {author} {\bibfnamefont
  {N.}~\bibnamefont {{Karnesis}}}, \bibinfo {author} {\bibfnamefont
  {C.}~\bibnamefont {{Killow}}}, \bibinfo {author} {\bibfnamefont
  {A.}~\bibnamefont {{Klein}}}, \bibinfo {author} {\bibfnamefont
  {B.}~\bibnamefont {{Klipstein}}}, \bibinfo {author} {\bibfnamefont
  {N.}~\bibnamefont {{Korsakova}}}, \bibinfo {author} {\bibfnamefont {S.~L.}\
  \bibnamefont {{Larson}}}, \bibinfo {author} {\bibfnamefont {J.}~\bibnamefont
  {{Livas}}}, \bibinfo {author} {\bibfnamefont {I.}~\bibnamefont {{Lloro}}},
  \bibinfo {author} {\bibfnamefont {N.}~\bibnamefont {{Man}}}, \bibinfo
  {author} {\bibfnamefont {D.}~\bibnamefont {{Mance}}}, \bibinfo {author}
  {\bibfnamefont {J.}~\bibnamefont {{Martino}}}, \bibinfo {author}
  {\bibfnamefont {I.}~\bibnamefont {{Mateos}}}, \bibinfo {author}
  {\bibfnamefont {K.}~\bibnamefont {{McKenzie}}}, \bibinfo {author}
  {\bibfnamefont {S.~T.}\ \bibnamefont {{McWilliams}}}, \bibinfo {author}
  {\bibfnamefont {C.}~\bibnamefont {{Miller}}}, \bibinfo {author}
  {\bibfnamefont {G.}~\bibnamefont {{Mueller}}}, \bibinfo {author}
  {\bibfnamefont {G.}~\bibnamefont {{Nardini}}}, \bibinfo {author}
  {\bibfnamefont {G.}~\bibnamefont {{Nelemans}}}, \bibinfo {author}
  {\bibfnamefont {M.}~\bibnamefont {{Nofrarias}}}, \bibinfo {author}
  {\bibfnamefont {A.}~\bibnamefont {{Petiteau}}}, \bibinfo {author}
  {\bibfnamefont {P.}~\bibnamefont {{Pivato}}}, \bibinfo {author}
  {\bibfnamefont {E.}~\bibnamefont {{Plagnol}}}, \bibinfo {author}
  {\bibfnamefont {E.}~\bibnamefont {{Porter}}}, \bibinfo {author}
  {\bibfnamefont {J.}~\bibnamefont {{Reiche}}}, \bibinfo {author}
  {\bibfnamefont {D.}~\bibnamefont {{Robertson}}}, \bibinfo {author}
  {\bibfnamefont {N.}~\bibnamefont {{Robertson}}}, \bibinfo {author}
  {\bibfnamefont {E.}~\bibnamefont {{Rossi}}}, \bibinfo {author} {\bibfnamefont
  {G.}~\bibnamefont {{Russano}}}, \bibinfo {author} {\bibfnamefont
  {B.}~\bibnamefont {{Schutz}}}, \bibinfo {author} {\bibfnamefont
  {A.}~\bibnamefont {{Sesana}}}, \bibinfo {author} {\bibfnamefont
  {D.}~\bibnamefont {{Shoemaker}}}, \bibinfo {author} {\bibfnamefont
  {J.}~\bibnamefont {{Slutsky}}}, \bibinfo {author} {\bibfnamefont {C.~F.}\
  \bibnamefont {{Sopuerta}}}, \bibinfo {author} {\bibfnamefont
  {T.}~\bibnamefont {{Sumner}}}, \bibinfo {author} {\bibfnamefont
  {N.}~\bibnamefont {{Tamanini}}}, \bibinfo {author} {\bibfnamefont
  {I.}~\bibnamefont {{Thorpe}}}, \bibinfo {author} {\bibfnamefont
  {M.}~\bibnamefont {{Troebs}}}, \bibinfo {author} {\bibfnamefont
  {M.}~\bibnamefont {{Vallisneri}}}, \bibinfo {author} {\bibfnamefont
  {A.}~\bibnamefont {{Vecchio}}}, \bibinfo {author} {\bibfnamefont
  {D.}~\bibnamefont {{Vetrugno}}}, \bibinfo {author} {\bibfnamefont
  {S.}~\bibnamefont {{Vitale}}}, \bibinfo {author} {\bibfnamefont
  {M.}~\bibnamefont {{Volonteri}}}, \bibinfo {author} {\bibfnamefont
  {G.}~\bibnamefont {{Wanner}}}, \bibinfo {author} {\bibfnamefont
  {H.}~\bibnamefont {{Ward}}}, \bibinfo {author} {\bibfnamefont
  {P.}~\bibnamefont {{Wass}}}, \bibinfo {author} {\bibfnamefont
  {W.}~\bibnamefont {{Weber}}}, \bibinfo {author} {\bibfnamefont
  {J.}~\bibnamefont {{Ziemer}}},\ and\ \bibinfo {author} {\bibfnamefont
  {P.}~\bibnamefont {{Zweifel}}},\ }\href
  {https://doi.org/10.48550/arXiv.1702.00786} {\bibinfo {title} {{Laser
  Interferometer Space Antenna}}} (\bibinfo {year} {2017}),\ \Eprint
  {https://arxiv.org/abs/1702.00786} {arXiv:1702.00786 [astro-ph.IM]}
  \BibitemShut {NoStop}%
\bibitem [{\citenamefont {{Arzoumanian}}\ \emph {et~al.}(2020)\citenamefont
  {{Arzoumanian}}, \citenamefont {{Baker}}, \citenamefont {{Brazier}},
  \citenamefont {{Brook}}, \citenamefont {{Burke-Spolaor}}, \citenamefont
  {{B{\'e}csy}}, \citenamefont {{Charisi}}, \citenamefont {{Chatterjee}},
  \citenamefont {{Cordes}}, \citenamefont {{Cornish}}, \citenamefont
  {{Crawford}}, \citenamefont {{Cromartie}}, \citenamefont {{Crowter}},
  \citenamefont {{Decesar}}, \citenamefont {{Demorest}}, \citenamefont
  {{Dolch}}, \citenamefont {{Elliott}}, \citenamefont {{Ellis}}, \citenamefont
  {{Ferdman}}, \citenamefont {{Ferrara}}, \citenamefont {{Fonseca}},
  \citenamefont {{Garver-Daniels}}, \citenamefont {{Gentile}}, \citenamefont
  {{Good}}, \citenamefont {{Hazboun}}, \citenamefont {{Islo}}, \citenamefont
  {{Jennings}}, \citenamefont {{Jones}}, \citenamefont {{Kaiser}},
  \citenamefont {{Kaplan}}, \citenamefont {{Kelley}}, \citenamefont {{Key}},
  \citenamefont {{Lam}}, \citenamefont {{Lazio}}, \citenamefont {{Levin}},
  \citenamefont {{Luo}}, \citenamefont {{Lynch}}, \citenamefont {{Madison}},
  \citenamefont {{McLaughlin}}, \citenamefont {{Mingarelli}}, \citenamefont
  {{Ng}}, \citenamefont {{Nice}}, \citenamefont {{Pennucci}}, \citenamefont
  {{Pol}}, \citenamefont {{Ransom}}, \citenamefont {{Ray}}, \citenamefont
  {{Shapiro-Albert}}, \citenamefont {{Siemens}}, \citenamefont {{Simon}},
  \citenamefont {{Spiewak}}, \citenamefont {{Stairs}}, \citenamefont
  {{Stinebring}}, \citenamefont {{Stovall}}, \citenamefont {{Swiggum}},
  \citenamefont {{Taylor}}, \citenamefont {{Vallisneri}}, \citenamefont
  {{Vigeland}}, \citenamefont {{Witt}}, \citenamefont {{Zhu}},\ and\
  \citenamefont {{NANOGrav Collaboration}}}]{Arzoumanian2020targeted}%
  \BibitemOpen
  \bibfield  {author} {\bibinfo {author} {\bibfnamefont {Z.}~\bibnamefont
  {{Arzoumanian}}}, \bibinfo {author} {\bibfnamefont {P.~T.}\ \bibnamefont
  {{Baker}}}, \bibinfo {author} {\bibfnamefont {A.}~\bibnamefont {{Brazier}}},
  \bibinfo {author} {\bibfnamefont {P.~R.}\ \bibnamefont {{Brook}}}, \bibinfo
  {author} {\bibfnamefont {S.}~\bibnamefont {{Burke-Spolaor}}}, \bibinfo
  {author} {\bibfnamefont {B.}~\bibnamefont {{B{\'e}csy}}}, \bibinfo {author}
  {\bibfnamefont {M.}~\bibnamefont {{Charisi}}}, \bibinfo {author}
  {\bibfnamefont {S.}~\bibnamefont {{Chatterjee}}}, \bibinfo {author}
  {\bibfnamefont {J.~M.}\ \bibnamefont {{Cordes}}}, \bibinfo {author}
  {\bibfnamefont {N.~J.}\ \bibnamefont {{Cornish}}}, \bibinfo {author}
  {\bibfnamefont {F.}~\bibnamefont {{Crawford}}}, \bibinfo {author}
  {\bibfnamefont {H.~T.}\ \bibnamefont {{Cromartie}}}, \bibinfo {author}
  {\bibfnamefont {K.}~\bibnamefont {{Crowter}}}, \bibinfo {author}
  {\bibfnamefont {M.~E.}\ \bibnamefont {{Decesar}}}, \bibinfo {author}
  {\bibfnamefont {P.~B.}\ \bibnamefont {{Demorest}}}, \bibinfo {author}
  {\bibfnamefont {T.}~\bibnamefont {{Dolch}}}, \bibinfo {author} {\bibfnamefont
  {R.~D.}\ \bibnamefont {{Elliott}}}, \bibinfo {author} {\bibfnamefont {J.~A.}\
  \bibnamefont {{Ellis}}}, \bibinfo {author} {\bibfnamefont {R.~D.}\
  \bibnamefont {{Ferdman}}}, \bibinfo {author} {\bibfnamefont {E.~C.}\
  \bibnamefont {{Ferrara}}}, \bibinfo {author} {\bibfnamefont {E.}~\bibnamefont
  {{Fonseca}}}, \bibinfo {author} {\bibfnamefont {N.}~\bibnamefont
  {{Garver-Daniels}}}, \bibinfo {author} {\bibfnamefont {P.~A.}\ \bibnamefont
  {{Gentile}}}, \bibinfo {author} {\bibfnamefont {D.~C.}\ \bibnamefont
  {{Good}}}, \bibinfo {author} {\bibfnamefont {J.~S.}\ \bibnamefont
  {{Hazboun}}}, \bibinfo {author} {\bibfnamefont {K.}~\bibnamefont {{Islo}}},
  \bibinfo {author} {\bibfnamefont {R.~J.}\ \bibnamefont {{Jennings}}},
  \bibinfo {author} {\bibfnamefont {M.~L.}\ \bibnamefont {{Jones}}}, \bibinfo
  {author} {\bibfnamefont {A.~R.}\ \bibnamefont {{Kaiser}}}, \bibinfo {author}
  {\bibfnamefont {D.~L.}\ \bibnamefont {{Kaplan}}}, \bibinfo {author}
  {\bibfnamefont {L.~Z.}\ \bibnamefont {{Kelley}}}, \bibinfo {author}
  {\bibfnamefont {J.~S.}\ \bibnamefont {{Key}}}, \bibinfo {author}
  {\bibfnamefont {M.~T.}\ \bibnamefont {{Lam}}}, \bibinfo {author}
  {\bibfnamefont {T.~J.~W.}\ \bibnamefont {{Lazio}}}, \bibinfo {author}
  {\bibfnamefont {L.}~\bibnamefont {{Levin}}}, \bibinfo {author} {\bibfnamefont
  {J.}~\bibnamefont {{Luo}}}, \bibinfo {author} {\bibfnamefont {R.~S.}\
  \bibnamefont {{Lynch}}}, \bibinfo {author} {\bibfnamefont {D.~R.}\
  \bibnamefont {{Madison}}}, \bibinfo {author} {\bibfnamefont {M.~A.}\
  \bibnamefont {{McLaughlin}}}, \bibinfo {author} {\bibfnamefont {C.~M.~F.}\
  \bibnamefont {{Mingarelli}}}, \bibinfo {author} {\bibfnamefont
  {C.}~\bibnamefont {{Ng}}}, \bibinfo {author} {\bibfnamefont {D.~J.}\
  \bibnamefont {{Nice}}}, \bibinfo {author} {\bibfnamefont {T.~T.}\
  \bibnamefont {{Pennucci}}}, \bibinfo {author} {\bibfnamefont {N.~S.}\
  \bibnamefont {{Pol}}}, \bibinfo {author} {\bibfnamefont {S.~M.}\ \bibnamefont
  {{Ransom}}}, \bibinfo {author} {\bibfnamefont {P.~S.}\ \bibnamefont {{Ray}}},
  \bibinfo {author} {\bibfnamefont {B.~J.}\ \bibnamefont {{Shapiro-Albert}}},
  \bibinfo {author} {\bibfnamefont {X.}~\bibnamefont {{Siemens}}}, \bibinfo
  {author} {\bibfnamefont {J.}~\bibnamefont {{Simon}}}, \bibinfo {author}
  {\bibfnamefont {R.}~\bibnamefont {{Spiewak}}}, \bibinfo {author}
  {\bibfnamefont {I.~H.}\ \bibnamefont {{Stairs}}}, \bibinfo {author}
  {\bibfnamefont {D.~R.}\ \bibnamefont {{Stinebring}}}, \bibinfo {author}
  {\bibfnamefont {K.}~\bibnamefont {{Stovall}}}, \bibinfo {author}
  {\bibfnamefont {J.~K.}\ \bibnamefont {{Swiggum}}}, \bibinfo {author}
  {\bibfnamefont {S.~R.}\ \bibnamefont {{Taylor}}}, \bibinfo {author}
  {\bibfnamefont {M.}~\bibnamefont {{Vallisneri}}}, \bibinfo {author}
  {\bibfnamefont {S.~J.}\ \bibnamefont {{Vigeland}}}, \bibinfo {author}
  {\bibfnamefont {C.~A.}\ \bibnamefont {{Witt}}}, \bibinfo {author}
  {\bibfnamefont {W.}~\bibnamefont {{Zhu}}},\ and\ \bibinfo {author}
  {\bibnamefont {{NANOGrav Collaboration}}},\ }\bibfield  {title} {\bibinfo
  {title} {{Multimessenger Gravitational-wave Searches with Pulsar Timing
  Arrays: Application to 3C 66B Using the NANOGrav 11-year Data Set}},\ }\href
  {https://doi.org/10.3847/1538-4357/ababa1} {\bibfield  {journal} {\bibinfo
  {journal} {\apj}\ }\textbf {\bibinfo {volume} {900}},\ \bibinfo {eid} {102}
  (\bibinfo {year} {2020})},\ \Eprint {https://arxiv.org/abs/2005.07123}
  {arXiv:2005.07123 [astro-ph.GA]} \BibitemShut {NoStop}%
\bibitem [{\citenamefont {{Agarwal}}\ \emph {et~al.}(2026)\citenamefont
  {{Agarwal}}, \citenamefont {{Agazie}}, \citenamefont {{Anumarlapudi}},
  \citenamefont {{Archibald}}, \citenamefont {{Arzoumanian}}, \citenamefont
  {{Baier}}, \citenamefont {{Baker}}, \citenamefont {{B{\'e}csy}},
  \citenamefont {{Blecha}}, \citenamefont {{Brazier}}, \citenamefont {{Brook}},
  \citenamefont {{Burke-Spolaor}}, \citenamefont {{Burnette}}, \citenamefont
  {{Case}}, \citenamefont {{Casey-Clyde}}, \citenamefont {{Chang}},
  \citenamefont {{Charisi}}, \citenamefont {{Chatterjee}}, \citenamefont
  {{Cohen}}, \citenamefont {{Coppi}}, \citenamefont {{Cordes}}, \citenamefont
  {{Cornish}}, \citenamefont {{Crawford}}, \citenamefont {{Cromartie}},
  \citenamefont {{Crowter}}, \citenamefont {{Decesar}}, \citenamefont
  {{Demorest}}, \citenamefont {{Deng}}, \citenamefont {{Dey}}, \citenamefont
  {{Dolch}}, \citenamefont {{D'Orazio}}, \citenamefont {{Eisenberg}},
  \citenamefont {{Ferrara}}, \citenamefont {{Doskoch}}, \citenamefont
  {{Fiore}}, \citenamefont {{Fonseca}}, \citenamefont {{Freedman}},
  \citenamefont {{Gardiner}}, \citenamefont {{Garver-Daniels}}, \citenamefont
  {{Gentile}}, \citenamefont {{Gersbach}}, \citenamefont {{Glaser}},
  \citenamefont {{Graham}}, \citenamefont {{Good}}, \citenamefont
  {{G{\"u}ltekin}}, \citenamefont {{Harris}}, \citenamefont {{Hazboun}},
  \citenamefont {{Hutchison}}, \citenamefont {{Jennings}}, \citenamefont
  {{Johnson}}, \citenamefont {{Jones}}, \citenamefont {{Kaplan}}, \citenamefont
  {{Kelley}}, \citenamefont {{Kerr}}, \citenamefont {{Key}}, \citenamefont
  {{Laal}}, \citenamefont {{Lam}}, \citenamefont {{Lamb}}, \citenamefont
  {{Larsen}}, \citenamefont {{Lazio}}, \citenamefont {{Lewandowska}},
  \citenamefont {{Liu}}, \citenamefont {{Lorimer}}, \citenamefont {{Luo}},
  \citenamefont {{Lynch}}, \citenamefont {{Ma}}, \citenamefont {{Madison}},
  \citenamefont {{Matt}}, \citenamefont {{McEwen}}, \citenamefont {{McKee}},
  \citenamefont {{McLaughlin}}, \citenamefont {{McMann}}, \citenamefont
  {{Meyers}}, \citenamefont {{Meyers}}, \citenamefont {{Mingarelli}},
  \citenamefont {{Mitridate}}, \citenamefont {{Natarajan}}, \citenamefont
  {{Ng}}, \citenamefont {{Nice}}, \citenamefont {{Nichols}}, \citenamefont
  {{Ocker}}, \citenamefont {{Olum}}, \citenamefont {{Pennucci}}, \citenamefont
  {{Perera}}, \citenamefont {{Petrov}}, \citenamefont {{Pol}}, \citenamefont
  {{Radovan}}, \citenamefont {{Ransom}}, \citenamefont {{Ray}}, \citenamefont
  {{Romano}}, \citenamefont {{Runnoe}}, \citenamefont {{Saffer}}, \citenamefont
  {{Sardesai}}, \citenamefont {{Schmiedekamp}}, \citenamefont {{Schmiedekamp}},
  \citenamefont {{Schmitz}}, \citenamefont {{Semenzato}}, \citenamefont
  {{Shapiro-Albert}}, \citenamefont {{Shivakumar}}, \citenamefont {{Siemens}},
  \citenamefont {{Simon}}, \citenamefont {{Sosa Fiscella}}, \citenamefont
  {{Stairs}}, \citenamefont {{Stinebring}}, \citenamefont {{Stovall}},
  \citenamefont {{Susobhanan}}, \citenamefont {{Swiggum}}, \citenamefont
  {{Taylor}}, \citenamefont {{Taylor}}, \citenamefont {{Thompson}},
  \citenamefont {{Turner}}, \citenamefont {{Vallisneri}}, \citenamefont {{van
  Haasteren}}, \citenamefont {{Vigeland}}, \citenamefont {{Wahl}},
  \citenamefont {{Willson}}, \citenamefont {{Wilson}}, \citenamefont {{Witt}},
  \citenamefont {{Wright}}, \citenamefont {{Young}}, \citenamefont {{Zheng}},\
  and\ \citenamefont {{Nanograv Collaboration}}}]{NG15targeted}%
  \BibitemOpen
  \bibfield  {author} {\bibinfo {author} {\bibfnamefont {N.}~\bibnamefont
  {{Agarwal}}}, \bibinfo {author} {\bibfnamefont {G.}~\bibnamefont {{Agazie}}},
  \bibinfo {author} {\bibfnamefont {A.}~\bibnamefont {{Anumarlapudi}}},
  \bibinfo {author} {\bibfnamefont {A.~M.}\ \bibnamefont {{Archibald}}},
  \bibinfo {author} {\bibfnamefont {Z.}~\bibnamefont {{Arzoumanian}}}, \bibinfo
  {author} {\bibfnamefont {J.~G.}\ \bibnamefont {{Baier}}}, \bibinfo {author}
  {\bibfnamefont {P.~T.}\ \bibnamefont {{Baker}}}, \bibinfo {author}
  {\bibfnamefont {B.}~\bibnamefont {{B{\'e}csy}}}, \bibinfo {author}
  {\bibfnamefont {L.}~\bibnamefont {{Blecha}}}, \bibinfo {author}
  {\bibfnamefont {A.}~\bibnamefont {{Brazier}}}, \bibinfo {author}
  {\bibfnamefont {P.~R.}\ \bibnamefont {{Brook}}}, \bibinfo {author}
  {\bibfnamefont {S.}~\bibnamefont {{Burke-Spolaor}}}, \bibinfo {author}
  {\bibfnamefont {R.}~\bibnamefont {{Burnette}}}, \bibinfo {author}
  {\bibfnamefont {R.}~\bibnamefont {{Case}}}, \bibinfo {author} {\bibfnamefont
  {J.~A.}\ \bibnamefont {{Casey-Clyde}}}, \bibinfo {author} {\bibfnamefont
  {Y.-T.}\ \bibnamefont {{Chang}}}, \bibinfo {author} {\bibfnamefont
  {M.}~\bibnamefont {{Charisi}}}, \bibinfo {author} {\bibfnamefont
  {S.}~\bibnamefont {{Chatterjee}}}, \bibinfo {author} {\bibfnamefont
  {T.}~\bibnamefont {{Cohen}}}, \bibinfo {author} {\bibfnamefont
  {P.}~\bibnamefont {{Coppi}}}, \bibinfo {author} {\bibfnamefont {J.~M.}\
  \bibnamefont {{Cordes}}}, \bibinfo {author} {\bibfnamefont {N.~J.}\
  \bibnamefont {{Cornish}}}, \bibinfo {author} {\bibfnamefont {F.}~\bibnamefont
  {{Crawford}}}, \bibinfo {author} {\bibfnamefont {H.~T.}\ \bibnamefont
  {{Cromartie}}}, \bibinfo {author} {\bibfnamefont {K.}~\bibnamefont
  {{Crowter}}}, \bibinfo {author} {\bibfnamefont {M.~E.}\ \bibnamefont
  {{Decesar}}}, \bibinfo {author} {\bibfnamefont {P.~B.}\ \bibnamefont
  {{Demorest}}}, \bibinfo {author} {\bibfnamefont {H.}~\bibnamefont {{Deng}}},
  \bibinfo {author} {\bibfnamefont {L.}~\bibnamefont {{Dey}}}, \bibinfo
  {author} {\bibfnamefont {T.}~\bibnamefont {{Dolch}}}, \bibinfo {author}
  {\bibfnamefont {D.~J.}\ \bibnamefont {{D'Orazio}}}, \bibinfo {author}
  {\bibfnamefont {E.}~\bibnamefont {{Eisenberg}}}, \bibinfo {author}
  {\bibfnamefont {E.~C.}\ \bibnamefont {{Ferrara}}}, \bibinfo {author}
  {\bibfnamefont {G.}~\bibnamefont {{Doskoch}}}, \bibinfo {author}
  {\bibfnamefont {W.}~\bibnamefont {{Fiore}}}, \bibinfo {author} {\bibfnamefont
  {E.}~\bibnamefont {{Fonseca}}}, \bibinfo {author} {\bibfnamefont {G.~E.}\
  \bibnamefont {{Freedman}}}, \bibinfo {author} {\bibfnamefont {E.~C.}\
  \bibnamefont {{Gardiner}}}, \bibinfo {author} {\bibfnamefont
  {N.}~\bibnamefont {{Garver-Daniels}}}, \bibinfo {author} {\bibfnamefont
  {P.~A.}\ \bibnamefont {{Gentile}}}, \bibinfo {author} {\bibfnamefont {K.~A.}\
  \bibnamefont {{Gersbach}}}, \bibinfo {author} {\bibfnamefont
  {J.}~\bibnamefont {{Glaser}}}, \bibinfo {author} {\bibfnamefont {M.~J.}\
  \bibnamefont {{Graham}}}, \bibinfo {author} {\bibfnamefont {D.~C.}\
  \bibnamefont {{Good}}}, \bibinfo {author} {\bibfnamefont {K.}~\bibnamefont
  {{G{\"u}ltekin}}}, \bibinfo {author} {\bibfnamefont {C.~J.}\ \bibnamefont
  {{Harris}}}, \bibinfo {author} {\bibfnamefont {J.~S.}\ \bibnamefont
  {{Hazboun}}}, \bibinfo {author} {\bibfnamefont {F.}~\bibnamefont
  {{Hutchison}}}, \bibinfo {author} {\bibfnamefont {R.~J.}\ \bibnamefont
  {{Jennings}}}, \bibinfo {author} {\bibfnamefont {A.~D.}\ \bibnamefont
  {{Johnson}}}, \bibinfo {author} {\bibfnamefont {M.~L.}\ \bibnamefont
  {{Jones}}}, \bibinfo {author} {\bibfnamefont {D.~L.}\ \bibnamefont
  {{Kaplan}}}, \bibinfo {author} {\bibfnamefont {L.~Z.}\ \bibnamefont
  {{Kelley}}}, \bibinfo {author} {\bibfnamefont {M.}~\bibnamefont {{Kerr}}},
  \bibinfo {author} {\bibfnamefont {J.~S.}\ \bibnamefont {{Key}}}, \bibinfo
  {author} {\bibfnamefont {N.}~\bibnamefont {{Laal}}}, \bibinfo {author}
  {\bibfnamefont {M.~T.}\ \bibnamefont {{Lam}}}, \bibinfo {author}
  {\bibfnamefont {W.~G.}\ \bibnamefont {{Lamb}}}, \bibinfo {author}
  {\bibfnamefont {B.}~\bibnamefont {{Larsen}}}, \bibinfo {author}
  {\bibfnamefont {T.~J.~W.}\ \bibnamefont {{Lazio}}}, \bibinfo {author}
  {\bibfnamefont {N.}~\bibnamefont {{Lewandowska}}}, \bibinfo {author}
  {\bibfnamefont {T.}~\bibnamefont {{Liu}}}, \bibinfo {author} {\bibfnamefont
  {D.~R.}\ \bibnamefont {{Lorimer}}}, \bibinfo {author} {\bibfnamefont
  {J.}~\bibnamefont {{Luo}}}, \bibinfo {author} {\bibfnamefont {R.~S.}\
  \bibnamefont {{Lynch}}}, \bibinfo {author} {\bibfnamefont {C.-P.}\
  \bibnamefont {{Ma}}}, \bibinfo {author} {\bibfnamefont {D.~R.}\ \bibnamefont
  {{Madison}}}, \bibinfo {author} {\bibfnamefont {C.}~\bibnamefont {{Matt}}},
  \bibinfo {author} {\bibfnamefont {A.}~\bibnamefont {{McEwen}}}, \bibinfo
  {author} {\bibfnamefont {J.~W.}\ \bibnamefont {{McKee}}}, \bibinfo {author}
  {\bibfnamefont {M.~A.}\ \bibnamefont {{McLaughlin}}}, \bibinfo {author}
  {\bibfnamefont {N.}~\bibnamefont {{McMann}}}, \bibinfo {author}
  {\bibfnamefont {B.~W.}\ \bibnamefont {{Meyers}}}, \bibinfo {author}
  {\bibfnamefont {P.~M.}\ \bibnamefont {{Meyers}}}, \bibinfo {author}
  {\bibfnamefont {C.~M.~F.}\ \bibnamefont {{Mingarelli}}}, \bibinfo {author}
  {\bibfnamefont {A.}~\bibnamefont {{Mitridate}}}, \bibinfo {author}
  {\bibfnamefont {P.}~\bibnamefont {{Natarajan}}}, \bibinfo {author}
  {\bibfnamefont {C.}~\bibnamefont {{Ng}}}, \bibinfo {author} {\bibfnamefont
  {D.~J.}\ \bibnamefont {{Nice}}}, \bibinfo {author} {\bibfnamefont
  {S.}~\bibnamefont {{Nichols}}}, \bibinfo {author} {\bibfnamefont {S.~K.}\
  \bibnamefont {{Ocker}}}, \bibinfo {author} {\bibfnamefont {K.~D.}\
  \bibnamefont {{Olum}}}, \bibinfo {author} {\bibfnamefont {T.~T.}\
  \bibnamefont {{Pennucci}}}, \bibinfo {author} {\bibfnamefont {B.~B.~P.}\
  \bibnamefont {{Perera}}}, \bibinfo {author} {\bibfnamefont {P.}~\bibnamefont
  {{Petrov}}}, \bibinfo {author} {\bibfnamefont {N.~S.}\ \bibnamefont {{Pol}}},
  \bibinfo {author} {\bibfnamefont {H.~A.}\ \bibnamefont {{Radovan}}}, \bibinfo
  {author} {\bibfnamefont {S.~M.}\ \bibnamefont {{Ransom}}}, \bibinfo {author}
  {\bibfnamefont {P.~S.}\ \bibnamefont {{Ray}}}, \bibinfo {author}
  {\bibfnamefont {J.~D.}\ \bibnamefont {{Romano}}}, \bibinfo {author}
  {\bibfnamefont {J.~C.}\ \bibnamefont {{Runnoe}}}, \bibinfo {author}
  {\bibfnamefont {A.}~\bibnamefont {{Saffer}}}, \bibinfo {author}
  {\bibfnamefont {S.~C.}\ \bibnamefont {{Sardesai}}}, \bibinfo {author}
  {\bibfnamefont {A.}~\bibnamefont {{Schmiedekamp}}}, \bibinfo {author}
  {\bibfnamefont {C.}~\bibnamefont {{Schmiedekamp}}}, \bibinfo {author}
  {\bibfnamefont {K.}~\bibnamefont {{Schmitz}}}, \bibinfo {author}
  {\bibfnamefont {F.}~\bibnamefont {{Semenzato}}}, \bibinfo {author}
  {\bibfnamefont {B.~J.}\ \bibnamefont {{Shapiro-Albert}}}, \bibinfo {author}
  {\bibfnamefont {R.}~\bibnamefont {{Shivakumar}}}, \bibinfo {author}
  {\bibfnamefont {X.}~\bibnamefont {{Siemens}}}, \bibinfo {author}
  {\bibfnamefont {J.}~\bibnamefont {{Simon}}}, \bibinfo {author} {\bibfnamefont
  {S.~V.}\ \bibnamefont {{Sosa Fiscella}}}, \bibinfo {author} {\bibfnamefont
  {I.~H.}\ \bibnamefont {{Stairs}}}, \bibinfo {author} {\bibfnamefont {D.~R.}\
  \bibnamefont {{Stinebring}}}, \bibinfo {author} {\bibfnamefont
  {K.}~\bibnamefont {{Stovall}}}, \bibinfo {author} {\bibfnamefont
  {A.}~\bibnamefont {{Susobhanan}}}, \bibinfo {author} {\bibfnamefont {J.~K.}\
  \bibnamefont {{Swiggum}}}, \bibinfo {author} {\bibfnamefont {J.~A.}\
  \bibnamefont {{Taylor}}}, \bibinfo {author} {\bibfnamefont {S.~R.}\
  \bibnamefont {{Taylor}}}, \bibinfo {author} {\bibfnamefont {M.~S.}\
  \bibnamefont {{Thompson}}}, \bibinfo {author} {\bibfnamefont {J.~E.}\
  \bibnamefont {{Turner}}}, \bibinfo {author} {\bibfnamefont {M.}~\bibnamefont
  {{Vallisneri}}}, \bibinfo {author} {\bibfnamefont {R.}~\bibnamefont {{van
  Haasteren}}}, \bibinfo {author} {\bibfnamefont {S.~J.}\ \bibnamefont
  {{Vigeland}}}, \bibinfo {author} {\bibfnamefont {H.~M.}\ \bibnamefont
  {{Wahl}}}, \bibinfo {author} {\bibfnamefont {L.}~\bibnamefont {{Willson}}},
  \bibinfo {author} {\bibfnamefont {K.~P.}\ \bibnamefont {{Wilson}}}, \bibinfo
  {author} {\bibfnamefont {C.~A.}\ \bibnamefont {{Witt}}}, \bibinfo {author}
  {\bibfnamefont {D.}~\bibnamefont {{Wright}}}, \bibinfo {author}
  {\bibfnamefont {O.}~\bibnamefont {{Young}}}, \bibinfo {author} {\bibfnamefont
  {Q.}~\bibnamefont {{Zheng}}},\ and\ \bibinfo {author} {\bibnamefont
  {{Nanograv Collaboration}}},\ }\bibfield  {title} {\bibinfo {title} {{The
  NANOGrav 15 yr Dataset: Targeted Searches for Supermassive Black Hole
  Binaries}},\ }\href {https://doi.org/10.3847/2041-8213/ae3719} {\bibfield
  {journal} {\bibinfo  {journal} {\apjl}\ }\textbf {\bibinfo {volume} {998}},\
  \bibinfo {eid} {L11} (\bibinfo {year} {2026})},\ \Eprint
  {https://arxiv.org/abs/2508.16534} {arXiv:2508.16534 [astro-ph.HE]}
  \BibitemShut {NoStop}%
\bibitem [{\citenamefont {{Reardon}}\ \emph {et~al.}(2024)\citenamefont
  {{Reardon}}, \citenamefont {{Bailes}}, \citenamefont {{Shannon}},
  \citenamefont {{Flynn}}, \citenamefont {{Askew}}, \citenamefont {{Bhat}},
  \citenamefont {{Chen}}, \citenamefont {{Cury{\l}o}}, \citenamefont {{Feng}},
  \citenamefont {{Hobbs}}, \citenamefont {{Kapur}}, \citenamefont {{Kerr}},
  \citenamefont {{Liu}}, \citenamefont {{Manchester}}, \citenamefont
  {{Mandow}}, \citenamefont {{Mishra}}, \citenamefont {{Russell}},
  \citenamefont {{Shamohammadi}}, \citenamefont {{Zhang}},\ and\ \citenamefont
  {{Zic}}}]{Reardon2024}%
  \BibitemOpen
  \bibfield  {author} {\bibinfo {author} {\bibfnamefont {D.~J.}\ \bibnamefont
  {{Reardon}}}, \bibinfo {author} {\bibfnamefont {M.}~\bibnamefont {{Bailes}}},
  \bibinfo {author} {\bibfnamefont {R.~M.}\ \bibnamefont {{Shannon}}}, \bibinfo
  {author} {\bibfnamefont {C.}~\bibnamefont {{Flynn}}}, \bibinfo {author}
  {\bibfnamefont {J.}~\bibnamefont {{Askew}}}, \bibinfo {author} {\bibfnamefont
  {N.~D.~R.}\ \bibnamefont {{Bhat}}}, \bibinfo {author} {\bibfnamefont {Z.-C.}\
  \bibnamefont {{Chen}}}, \bibinfo {author} {\bibfnamefont {M.}~\bibnamefont
  {{Cury{\l}o}}}, \bibinfo {author} {\bibfnamefont {Y.}~\bibnamefont {{Feng}}},
  \bibinfo {author} {\bibfnamefont {G.~B.}\ \bibnamefont {{Hobbs}}}, \bibinfo
  {author} {\bibfnamefont {A.}~\bibnamefont {{Kapur}}}, \bibinfo {author}
  {\bibfnamefont {M.}~\bibnamefont {{Kerr}}}, \bibinfo {author} {\bibfnamefont
  {X.}~\bibnamefont {{Liu}}}, \bibinfo {author} {\bibfnamefont {R.~N.}\
  \bibnamefont {{Manchester}}}, \bibinfo {author} {\bibfnamefont
  {R.}~\bibnamefont {{Mandow}}}, \bibinfo {author} {\bibfnamefont
  {S.}~\bibnamefont {{Mishra}}}, \bibinfo {author} {\bibfnamefont {C.~J.}\
  \bibnamefont {{Russell}}}, \bibinfo {author} {\bibfnamefont {M.}~\bibnamefont
  {{Shamohammadi}}}, \bibinfo {author} {\bibfnamefont {L.}~\bibnamefont
  {{Zhang}}},\ and\ \bibinfo {author} {\bibfnamefont {A.}~\bibnamefont
  {{Zic}}},\ }\bibfield  {title} {\bibinfo {title} {{The Neutron Star Mass,
  Distance, and Inclination from Precision Timing of the Brilliant Millisecond
  Pulsar J0437-4715}},\ }\href {https://doi.org/10.3847/2041-8213/ad614a}
  {\bibfield  {journal} {\bibinfo  {journal} {\apjl}\ }\textbf {\bibinfo
  {volume} {971}},\ \bibinfo {eid} {L18} (\bibinfo {year} {2024})},\ \Eprint
  {https://arxiv.org/abs/2407.07132} {arXiv:2407.07132 [astro-ph.HE]}
  \BibitemShut {NoStop}%
\bibitem [{\citenamefont {{Deller}}\ \emph {et~al.}(2019)\citenamefont
  {{Deller}}, \citenamefont {{Goss}}, \citenamefont {{Brisken}}, \citenamefont
  {{Chatterjee}}, \citenamefont {{Cordes}}, \citenamefont {{Janssen}},
  \citenamefont {{Kovalev}}, \citenamefont {{Lazio}}, \citenamefont {{Petrov}},
  \citenamefont {{Stappers}},\ and\ \citenamefont {{Lyne}}}]{Deller2019}%
  \BibitemOpen
  \bibfield  {author} {\bibinfo {author} {\bibfnamefont {A.~T.}\ \bibnamefont
  {{Deller}}}, \bibinfo {author} {\bibfnamefont {W.~M.}\ \bibnamefont
  {{Goss}}}, \bibinfo {author} {\bibfnamefont {W.~F.}\ \bibnamefont
  {{Brisken}}}, \bibinfo {author} {\bibfnamefont {S.}~\bibnamefont
  {{Chatterjee}}}, \bibinfo {author} {\bibfnamefont {J.~M.}\ \bibnamefont
  {{Cordes}}}, \bibinfo {author} {\bibfnamefont {G.~H.}\ \bibnamefont
  {{Janssen}}}, \bibinfo {author} {\bibfnamefont {Y.~Y.}\ \bibnamefont
  {{Kovalev}}}, \bibinfo {author} {\bibfnamefont {T.~J.~W.}\ \bibnamefont
  {{Lazio}}}, \bibinfo {author} {\bibfnamefont {L.}~\bibnamefont {{Petrov}}},
  \bibinfo {author} {\bibfnamefont {B.~W.}\ \bibnamefont {{Stappers}}},\ and\
  \bibinfo {author} {\bibfnamefont {A.}~\bibnamefont {{Lyne}}},\ }\bibfield
  {title} {\bibinfo {title} {{Microarcsecond VLBI Pulsar Astrometry with
  PSR{\ensuremath{\pi}} II. Parallax Distances for 57 Pulsars}},\ }\href
  {https://doi.org/10.3847/1538-4357/ab11c7} {\bibfield  {journal} {\bibinfo
  {journal} {\apj}\ }\textbf {\bibinfo {volume} {875}},\ \bibinfo {eid} {100}
  (\bibinfo {year} {2019})},\ \Eprint {https://arxiv.org/abs/1808.09046}
  {arXiv:1808.09046 [astro-ph.IM]} \BibitemShut {NoStop}%
\bibitem [{\citenamefont {{Ding}}\ \emph {et~al.}(2023)\citenamefont {{Ding}},
  \citenamefont {{Deller}}, \citenamefont {{Stappers}}, \citenamefont
  {{Lazio}}, \citenamefont {{Kaplan}}, \citenamefont {{Chatterjee}},
  \citenamefont {{Brisken}}, \citenamefont {{Cordes}}, \citenamefont
  {{Freire}}, \citenamefont {{Fonseca}}, \citenamefont {{Stairs}},
  \citenamefont {{Guillemot}}, \citenamefont {{Lyne}}, \citenamefont
  {{Cognard}}, \citenamefont {{Reardon}},\ and\ \citenamefont
  {{Theureau}}}]{Ding2023}%
  \BibitemOpen
  \bibfield  {author} {\bibinfo {author} {\bibfnamefont {H.}~\bibnamefont
  {{Ding}}}, \bibinfo {author} {\bibfnamefont {A.~T.}\ \bibnamefont
  {{Deller}}}, \bibinfo {author} {\bibfnamefont {B.~W.}\ \bibnamefont
  {{Stappers}}}, \bibinfo {author} {\bibfnamefont {T.~J.~W.}\ \bibnamefont
  {{Lazio}}}, \bibinfo {author} {\bibfnamefont {D.}~\bibnamefont {{Kaplan}}},
  \bibinfo {author} {\bibfnamefont {S.}~\bibnamefont {{Chatterjee}}}, \bibinfo
  {author} {\bibfnamefont {W.}~\bibnamefont {{Brisken}}}, \bibinfo {author}
  {\bibfnamefont {J.}~\bibnamefont {{Cordes}}}, \bibinfo {author}
  {\bibfnamefont {P.~C.~C.}\ \bibnamefont {{Freire}}}, \bibinfo {author}
  {\bibfnamefont {E.}~\bibnamefont {{Fonseca}}}, \bibinfo {author}
  {\bibfnamefont {I.}~\bibnamefont {{Stairs}}}, \bibinfo {author}
  {\bibfnamefont {L.}~\bibnamefont {{Guillemot}}}, \bibinfo {author}
  {\bibfnamefont {A.}~\bibnamefont {{Lyne}}}, \bibinfo {author} {\bibfnamefont
  {I.}~\bibnamefont {{Cognard}}}, \bibinfo {author} {\bibfnamefont {D.~J.}\
  \bibnamefont {{Reardon}}},\ and\ \bibinfo {author} {\bibfnamefont
  {G.}~\bibnamefont {{Theureau}}},\ }\bibfield  {title} {\bibinfo {title} {{The
  MSPSR{\ensuremath{\pi}} catalogue: VLBA astrometry of 18 millisecond
  pulsars}},\ }\href {https://doi.org/10.1093/mnras/stac3725} {\bibfield
  {journal} {\bibinfo  {journal} {\mnras}\ }\textbf {\bibinfo {volume} {519}},\
  \bibinfo {pages} {4982} (\bibinfo {year} {2023})},\ \Eprint
  {https://arxiv.org/abs/2212.06351} {arXiv:2212.06351 [astro-ph.HE]}
  \BibitemShut {NoStop}%
\bibitem [{\citenamefont {{Agazie}}\ \emph
  {et~al.}(2023{\natexlab{b}})\citenamefont {{Agazie}}, \citenamefont
  {{Anumarlapudi}}, \citenamefont {{Archibald}}, \citenamefont {{Arzoumanian}},
  \citenamefont {{Baker}}, \citenamefont {{B{\'e}csy}}, \citenamefont
  {{Blecha}}, \citenamefont {{Brazier}}, \citenamefont {{Brook}}, \citenamefont
  {{Burke-Spolaor}}, \citenamefont {{Case}}, \citenamefont {{Casey-Clyde}},
  \citenamefont {{Charisi}}, \citenamefont {{Chatterjee}}, \citenamefont
  {{Cohen}}, \citenamefont {{Cordes}}, \citenamefont {{Cornish}}, \citenamefont
  {{Crawford}}, \citenamefont {{Cromartie}}, \citenamefont {{Crowter}},
  \citenamefont {{Decesar}}, \citenamefont {{Demorest}}, \citenamefont
  {{Digman}}, \citenamefont {{Dolch}}, \citenamefont {{Drachler}},
  \citenamefont {{Ferrara}}, \citenamefont {{Fiore}}, \citenamefont
  {{Fonseca}}, \citenamefont {{Freedman}}, \citenamefont {{Garver-Daniels}},
  \citenamefont {{Gentile}}, \citenamefont {{Glaser}}, \citenamefont {{Good}},
  \citenamefont {{G{\"u}ltekin}}, \citenamefont {{Hazboun}}, \citenamefont
  {{Hourihane}}, \citenamefont {{Jennings}}, \citenamefont {{Johnson}},
  \citenamefont {{Jones}}, \citenamefont {{Kaiser}}, \citenamefont {{Kaplan}},
  \citenamefont {{Kelley}}, \citenamefont {{Kerr}}, \citenamefont {{Key}},
  \citenamefont {{Laal}}, \citenamefont {{Lam}}, \citenamefont {{Lamb}},
  \citenamefont {{Lazio}}, \citenamefont {{Lewandowska}}, \citenamefont
  {{Liu}}, \citenamefont {{Lorimer}}, \citenamefont {{Luo}}, \citenamefont
  {{Lynch}}, \citenamefont {{Ma}}, \citenamefont {{Madison}}, \citenamefont
  {{McEwen}}, \citenamefont {{McKee}}, \citenamefont {{McLaughlin}},
  \citenamefont {{McMann}}, \citenamefont {{Meyers}}, \citenamefont {{Meyers}},
  \citenamefont {{Mingarelli}}, \citenamefont {{Mitridate}}, \citenamefont
  {{Ng}}, \citenamefont {{Nice}}, \citenamefont {{Ocker}}, \citenamefont
  {{Olum}}, \citenamefont {{Pennucci}}, \citenamefont {{Perera}}, \citenamefont
  {{Petrov}}, \citenamefont {{Pol}}, \citenamefont {{Radovan}}, \citenamefont
  {{Ransom}}, \citenamefont {{Ray}}, \citenamefont {{Romano}}, \citenamefont
  {{Sardesai}}, \citenamefont {{Schmiedekamp}}, \citenamefont {{Schmiedekamp}},
  \citenamefont {{Schmitz}}, \citenamefont {{Shapiro-Albert}}, \citenamefont
  {{Siemens}}, \citenamefont {{Simon}}, \citenamefont {{Siwek}}, \citenamefont
  {{Stairs}}, \citenamefont {{Stinebring}}, \citenamefont {{Stovall}},
  \citenamefont {{Susobhanan}}, \citenamefont {{Swiggum}}, \citenamefont
  {{Taylor}}, \citenamefont {{Taylor}}, \citenamefont {{Turner}}, \citenamefont
  {{Unal}}, \citenamefont {{Vallisneri}}, \citenamefont {{van Haasteren}},
  \citenamefont {{Vigeland}}, \citenamefont {{Wahl}}, \citenamefont {{Witt}},
  \citenamefont {{Young}},\ and\ \citenamefont {{Nanograv
  Collaboration}}}]{NG15-continuous}%
  \BibitemOpen
  \bibfield  {author} {\bibinfo {author} {\bibfnamefont {G.}~\bibnamefont
  {{Agazie}}}, \bibinfo {author} {\bibfnamefont {A.}~\bibnamefont
  {{Anumarlapudi}}}, \bibinfo {author} {\bibfnamefont {A.~M.}\ \bibnamefont
  {{Archibald}}}, \bibinfo {author} {\bibfnamefont {Z.}~\bibnamefont
  {{Arzoumanian}}}, \bibinfo {author} {\bibfnamefont {P.~T.}\ \bibnamefont
  {{Baker}}}, \bibinfo {author} {\bibfnamefont {B.}~\bibnamefont
  {{B{\'e}csy}}}, \bibinfo {author} {\bibfnamefont {L.}~\bibnamefont
  {{Blecha}}}, \bibinfo {author} {\bibfnamefont {A.}~\bibnamefont {{Brazier}}},
  \bibinfo {author} {\bibfnamefont {P.~R.}\ \bibnamefont {{Brook}}}, \bibinfo
  {author} {\bibfnamefont {S.}~\bibnamefont {{Burke-Spolaor}}}, \bibinfo
  {author} {\bibfnamefont {R.}~\bibnamefont {{Case}}}, \bibinfo {author}
  {\bibfnamefont {J.~A.}\ \bibnamefont {{Casey-Clyde}}}, \bibinfo {author}
  {\bibfnamefont {M.}~\bibnamefont {{Charisi}}}, \bibinfo {author}
  {\bibfnamefont {S.}~\bibnamefont {{Chatterjee}}}, \bibinfo {author}
  {\bibfnamefont {T.}~\bibnamefont {{Cohen}}}, \bibinfo {author} {\bibfnamefont
  {J.~M.}\ \bibnamefont {{Cordes}}}, \bibinfo {author} {\bibfnamefont {N.~J.}\
  \bibnamefont {{Cornish}}}, \bibinfo {author} {\bibfnamefont {F.}~\bibnamefont
  {{Crawford}}}, \bibinfo {author} {\bibfnamefont {H.~T.}\ \bibnamefont
  {{Cromartie}}}, \bibinfo {author} {\bibfnamefont {K.}~\bibnamefont
  {{Crowter}}}, \bibinfo {author} {\bibfnamefont {M.~E.}\ \bibnamefont
  {{Decesar}}}, \bibinfo {author} {\bibfnamefont {P.~B.}\ \bibnamefont
  {{Demorest}}}, \bibinfo {author} {\bibfnamefont {M.~C.}\ \bibnamefont
  {{Digman}}}, \bibinfo {author} {\bibfnamefont {T.}~\bibnamefont {{Dolch}}},
  \bibinfo {author} {\bibfnamefont {B.}~\bibnamefont {{Drachler}}}, \bibinfo
  {author} {\bibfnamefont {E.~C.}\ \bibnamefont {{Ferrara}}}, \bibinfo {author}
  {\bibfnamefont {W.}~\bibnamefont {{Fiore}}}, \bibinfo {author} {\bibfnamefont
  {E.}~\bibnamefont {{Fonseca}}}, \bibinfo {author} {\bibfnamefont {G.~E.}\
  \bibnamefont {{Freedman}}}, \bibinfo {author} {\bibfnamefont
  {N.}~\bibnamefont {{Garver-Daniels}}}, \bibinfo {author} {\bibfnamefont
  {P.~A.}\ \bibnamefont {{Gentile}}}, \bibinfo {author} {\bibfnamefont
  {J.}~\bibnamefont {{Glaser}}}, \bibinfo {author} {\bibfnamefont {D.~C.}\
  \bibnamefont {{Good}}}, \bibinfo {author} {\bibfnamefont {K.}~\bibnamefont
  {{G{\"u}ltekin}}}, \bibinfo {author} {\bibfnamefont {J.~S.}\ \bibnamefont
  {{Hazboun}}}, \bibinfo {author} {\bibfnamefont {S.}~\bibnamefont
  {{Hourihane}}}, \bibinfo {author} {\bibfnamefont {R.~J.}\ \bibnamefont
  {{Jennings}}}, \bibinfo {author} {\bibfnamefont {A.~D.}\ \bibnamefont
  {{Johnson}}}, \bibinfo {author} {\bibfnamefont {M.~L.}\ \bibnamefont
  {{Jones}}}, \bibinfo {author} {\bibfnamefont {A.~R.}\ \bibnamefont
  {{Kaiser}}}, \bibinfo {author} {\bibfnamefont {D.~L.}\ \bibnamefont
  {{Kaplan}}}, \bibinfo {author} {\bibfnamefont {L.~Z.}\ \bibnamefont
  {{Kelley}}}, \bibinfo {author} {\bibfnamefont {M.}~\bibnamefont {{Kerr}}},
  \bibinfo {author} {\bibfnamefont {J.~S.}\ \bibnamefont {{Key}}}, \bibinfo
  {author} {\bibfnamefont {N.}~\bibnamefont {{Laal}}}, \bibinfo {author}
  {\bibfnamefont {M.~T.}\ \bibnamefont {{Lam}}}, \bibinfo {author}
  {\bibfnamefont {W.~G.}\ \bibnamefont {{Lamb}}}, \bibinfo {author}
  {\bibfnamefont {T.~J.~W.}\ \bibnamefont {{Lazio}}}, \bibinfo {author}
  {\bibfnamefont {N.}~\bibnamefont {{Lewandowska}}}, \bibinfo {author}
  {\bibfnamefont {T.}~\bibnamefont {{Liu}}}, \bibinfo {author} {\bibfnamefont
  {D.~R.}\ \bibnamefont {{Lorimer}}}, \bibinfo {author} {\bibfnamefont
  {J.}~\bibnamefont {{Luo}}}, \bibinfo {author} {\bibfnamefont {R.~S.}\
  \bibnamefont {{Lynch}}}, \bibinfo {author} {\bibfnamefont {C.-P.}\
  \bibnamefont {{Ma}}}, \bibinfo {author} {\bibfnamefont {D.~R.}\ \bibnamefont
  {{Madison}}}, \bibinfo {author} {\bibfnamefont {A.}~\bibnamefont {{McEwen}}},
  \bibinfo {author} {\bibfnamefont {J.~W.}\ \bibnamefont {{McKee}}}, \bibinfo
  {author} {\bibfnamefont {M.~A.}\ \bibnamefont {{McLaughlin}}}, \bibinfo
  {author} {\bibfnamefont {N.}~\bibnamefont {{McMann}}}, \bibinfo {author}
  {\bibfnamefont {B.~W.}\ \bibnamefont {{Meyers}}}, \bibinfo {author}
  {\bibfnamefont {P.~M.}\ \bibnamefont {{Meyers}}}, \bibinfo {author}
  {\bibfnamefont {C.~M.~F.}\ \bibnamefont {{Mingarelli}}}, \bibinfo {author}
  {\bibfnamefont {A.}~\bibnamefont {{Mitridate}}}, \bibinfo {author}
  {\bibfnamefont {C.}~\bibnamefont {{Ng}}}, \bibinfo {author} {\bibfnamefont
  {D.~J.}\ \bibnamefont {{Nice}}}, \bibinfo {author} {\bibfnamefont {S.~K.}\
  \bibnamefont {{Ocker}}}, \bibinfo {author} {\bibfnamefont {K.~D.}\
  \bibnamefont {{Olum}}}, \bibinfo {author} {\bibfnamefont {T.~T.}\
  \bibnamefont {{Pennucci}}}, \bibinfo {author} {\bibfnamefont {B.~B.~P.}\
  \bibnamefont {{Perera}}}, \bibinfo {author} {\bibfnamefont {P.}~\bibnamefont
  {{Petrov}}}, \bibinfo {author} {\bibfnamefont {N.~S.}\ \bibnamefont {{Pol}}},
  \bibinfo {author} {\bibfnamefont {H.~A.}\ \bibnamefont {{Radovan}}}, \bibinfo
  {author} {\bibfnamefont {S.~M.}\ \bibnamefont {{Ransom}}}, \bibinfo {author}
  {\bibfnamefont {P.~S.}\ \bibnamefont {{Ray}}}, \bibinfo {author}
  {\bibfnamefont {J.~D.}\ \bibnamefont {{Romano}}}, \bibinfo {author}
  {\bibfnamefont {S.~C.}\ \bibnamefont {{Sardesai}}}, \bibinfo {author}
  {\bibfnamefont {A.}~\bibnamefont {{Schmiedekamp}}}, \bibinfo {author}
  {\bibfnamefont {C.}~\bibnamefont {{Schmiedekamp}}}, \bibinfo {author}
  {\bibfnamefont {K.}~\bibnamefont {{Schmitz}}}, \bibinfo {author}
  {\bibfnamefont {B.~J.}\ \bibnamefont {{Shapiro-Albert}}}, \bibinfo {author}
  {\bibfnamefont {X.}~\bibnamefont {{Siemens}}}, \bibinfo {author}
  {\bibfnamefont {J.}~\bibnamefont {{Simon}}}, \bibinfo {author} {\bibfnamefont
  {M.~S.}\ \bibnamefont {{Siwek}}}, \bibinfo {author} {\bibfnamefont {I.~H.}\
  \bibnamefont {{Stairs}}}, \bibinfo {author} {\bibfnamefont {D.~R.}\
  \bibnamefont {{Stinebring}}}, \bibinfo {author} {\bibfnamefont
  {K.}~\bibnamefont {{Stovall}}}, \bibinfo {author} {\bibfnamefont
  {A.}~\bibnamefont {{Susobhanan}}}, \bibinfo {author} {\bibfnamefont {J.~K.}\
  \bibnamefont {{Swiggum}}}, \bibinfo {author} {\bibfnamefont {J.}~\bibnamefont
  {{Taylor}}}, \bibinfo {author} {\bibfnamefont {S.~R.}\ \bibnamefont
  {{Taylor}}}, \bibinfo {author} {\bibfnamefont {J.~E.}\ \bibnamefont
  {{Turner}}}, \bibinfo {author} {\bibfnamefont {C.}~\bibnamefont {{Unal}}},
  \bibinfo {author} {\bibfnamefont {M.}~\bibnamefont {{Vallisneri}}}, \bibinfo
  {author} {\bibfnamefont {R.}~\bibnamefont {{van Haasteren}}}, \bibinfo
  {author} {\bibfnamefont {S.~J.}\ \bibnamefont {{Vigeland}}}, \bibinfo
  {author} {\bibfnamefont {H.~M.}\ \bibnamefont {{Wahl}}}, \bibinfo {author}
  {\bibfnamefont {C.~A.}\ \bibnamefont {{Witt}}}, \bibinfo {author}
  {\bibfnamefont {O.}~\bibnamefont {{Young}}},\ and\ \bibinfo {author}
  {\bibnamefont {{Nanograv Collaboration}}},\ }\bibfield  {title} {\bibinfo
  {title} {{The NANOGrav 15 yr Data Set: Bayesian Limits on Gravitational Waves
  from Individual Supermassive Black Hole Binaries}},\ }\href
  {https://doi.org/10.3847/2041-8213/ace18a} {\bibfield  {journal} {\bibinfo
  {journal} {\apjl}\ }\textbf {\bibinfo {volume} {951}},\ \bibinfo {eid} {L50}
  (\bibinfo {year} {2023}{\natexlab{b}})},\ \Eprint
  {https://arxiv.org/abs/2306.16222} {arXiv:2306.16222 [astro-ph.HE]}
  \BibitemShut {NoStop}%
\bibitem [{\citenamefont
  {{Mingarelli}}(2026{\natexlab{b}})}]{Mingarelli_AGN_IMRI}%
  \BibitemOpen
  \bibfield  {author} {\bibinfo {author} {\bibfnamefont {C.~M.~F.}\
  \bibnamefont {{Mingarelli}}},\ }\bibfield  {title} {\bibinfo {title} {{A
  Gravitational Wave Background from Intermediate Mass Black Holes in AGN
  Disks}},\ }\href {https://doi.org/10.48550/arXiv.2602.09217} {\bibfield
  {journal} {\bibinfo  {journal} {arXiv e-prints}\ ,\ \bibinfo {eid}
  {arXiv:2602.09217}} (\bibinfo {year} {2026}{\natexlab{b}})},\ \Eprint
  {https://arxiv.org/abs/2602.09217} {arXiv:2602.09217 [astro-ph.GA]}
  \BibitemShut {NoStop}%
\bibitem [{\citenamefont {{Bonetti}}\ and\ \citenamefont
  {{Sesana}}(2020)}]{BonettiSesana2020}%
  \BibitemOpen
  \bibfield  {author} {\bibinfo {author} {\bibfnamefont {M.}~\bibnamefont
  {{Bonetti}}}\ and\ \bibinfo {author} {\bibfnamefont {A.}~\bibnamefont
  {{Sesana}}},\ }\bibfield  {title} {\bibinfo {title} {{Gravitational wave
  background from extreme mass ratio inspirals}},\ }\href
  {https://doi.org/10.1103/PhysRevD.102.103023} {\bibfield  {journal} {\bibinfo
   {journal} {\prd}\ }\textbf {\bibinfo {volume} {102}},\ \bibinfo {eid}
  {103023} (\bibinfo {year} {2020})},\ \Eprint
  {https://arxiv.org/abs/2007.14403} {arXiv:2007.14403 [astro-ph.GA]}
  \BibitemShut {NoStop}%
\bibitem [{\citenamefont {{Misner}}\ \emph {et~al.}(1973)\citenamefont
  {{Misner}}, \citenamefont {{Thorne}},\ and\ \citenamefont {{Wheeler}}}]{MTW}%
  \BibitemOpen
  \bibfield  {author} {\bibinfo {author} {\bibfnamefont {C.~W.}\ \bibnamefont
  {{Misner}}}, \bibinfo {author} {\bibfnamefont {K.~S.}\ \bibnamefont
  {{Thorne}}},\ and\ \bibinfo {author} {\bibfnamefont {J.~A.}\ \bibnamefont
  {{Wheeler}}},\ }\href@noop {} {\emph {\bibinfo {title} {{Gravitation}}}}\
  (\bibinfo {year} {1973})\BibitemShut {NoStop}%
\bibitem [{\citenamefont {{Jaranowski}}\ and\ \citenamefont
  {{Kr{\'o}lak}}(2012)}]{JK2012}%
  \BibitemOpen
  \bibfield  {author} {\bibinfo {author} {\bibfnamefont {P.}~\bibnamefont
  {{Jaranowski}}}\ and\ \bibinfo {author} {\bibfnamefont {A.}~\bibnamefont
  {{Kr{\'o}lak}}},\ }\bibfield  {title} {\bibinfo {title} {{Gravitational-Wave
  Data Analysis. Formalism and Sample Applications: The Gaussian Case}},\
  }\href {https://doi.org/10.12942/lrr-2012-4} {\bibfield  {journal} {\bibinfo
  {journal} {Living Reviews in Relativity}\ }\textbf {\bibinfo {volume} {15}},\
  \bibinfo {eid} {4} (\bibinfo {year} {2012})},\ \Eprint
  {https://arxiv.org/abs/0711.1115} {arXiv:0711.1115 [gr-qc]} \BibitemShut
  {NoStop}%
\bibitem [{\citenamefont {{Moore}}\ \emph {et~al.}(2015)\citenamefont
  {{Moore}}, \citenamefont {{Cole}},\ and\ \citenamefont
  {{Berry}}}]{MooreCB2015}%
  \BibitemOpen
  \bibfield  {author} {\bibinfo {author} {\bibfnamefont {C.~J.}\ \bibnamefont
  {{Moore}}}, \bibinfo {author} {\bibfnamefont {R.~H.}\ \bibnamefont
  {{Cole}}},\ and\ \bibinfo {author} {\bibfnamefont {C.~P.~L.}\ \bibnamefont
  {{Berry}}},\ }\bibfield  {title} {\bibinfo {title} {{Gravitational-wave
  sensitivity curves}},\ }\href {https://doi.org/10.1088/0264-9381/32/1/015014}
  {\bibfield  {journal} {\bibinfo  {journal} {Classical and Quantum Gravity}\
  }\textbf {\bibinfo {volume} {32}},\ \bibinfo {eid} {015014} (\bibinfo {year}
  {2015})},\ \Eprint {https://arxiv.org/abs/1408.0740} {arXiv:1408.0740
  [gr-qc]} \BibitemShut {NoStop}%
\bibitem [{\citenamefont {{Hazboun}}\ \emph
  {et~al.}(2019{\natexlab{a}})\citenamefont {{Hazboun}}, \citenamefont
  {{Romano}},\ and\ \citenamefont {{Smith}}}]{HazbounRS2019}%
  \BibitemOpen
  \bibfield  {author} {\bibinfo {author} {\bibfnamefont {J.~S.}\ \bibnamefont
  {{Hazboun}}}, \bibinfo {author} {\bibfnamefont {J.~D.}\ \bibnamefont
  {{Romano}}},\ and\ \bibinfo {author} {\bibfnamefont {T.~L.}\ \bibnamefont
  {{Smith}}},\ }\bibfield  {title} {\bibinfo {title} {{Realistic sensitivity
  curves for pulsar timing arrays}},\ }\href
  {https://doi.org/10.1103/PhysRevD.100.104028} {\bibfield  {journal} {\bibinfo
   {journal} {\prd}\ }\textbf {\bibinfo {volume} {100}},\ \bibinfo {eid}
  {104028} (\bibinfo {year} {2019}{\natexlab{a}})},\ \Eprint
  {https://arxiv.org/abs/1907.04341} {arXiv:1907.04341 [gr-qc]} \BibitemShut
  {NoStop}%
\bibitem [{\citenamefont {{Larsen}}\ \emph {et~al.}(2026)\citenamefont
  {{Larsen}}, \citenamefont {{Baier}}, \citenamefont {{Hazboun}}, \citenamefont
  {{Simon}},\ and\ \citenamefont {{NANOGrav Collaboration}}}]{NG15noise}%
  \BibitemOpen
  \bibfield  {author} {\bibinfo {author} {\bibfnamefont {B.}~\bibnamefont
  {{Larsen}}}, \bibinfo {author} {\bibfnamefont {J.}~\bibnamefont {{Baier}}},
  \bibinfo {author} {\bibfnamefont {J.~S.}\ \bibnamefont {{Hazboun}}}, \bibinfo
  {author} {\bibfnamefont {J.}~\bibnamefont {{Simon}}},\ and\ \bibinfo {author}
  {\bibnamefont {{NANOGrav Collaboration}}},\ }\bibfield  {title} {\bibinfo
  {title} {{The NANOGrav 15-year Data Set: Custom Chromatic Noise Models for
  All 67 Pulsars}},\ }\href@noop {} {\bibfield  {journal} {\bibinfo  {journal}
  {in preparation}\ } (\bibinfo {year} {2026})}\BibitemShut {NoStop}%
\bibitem [{\citenamefont {{Hazboun}}\ \emph
  {et~al.}(2019{\natexlab{b}})\citenamefont {{Hazboun}}, \citenamefont
  {{Romano}},\ and\ \citenamefont {{Smith}}}]{Hazboun2019hasasia}%
  \BibitemOpen
  \bibfield  {author} {\bibinfo {author} {\bibfnamefont {J.}~\bibnamefont
  {{Hazboun}}}, \bibinfo {author} {\bibfnamefont {J.}~\bibnamefont
  {{Romano}}},\ and\ \bibinfo {author} {\bibfnamefont {T.}~\bibnamefont
  {{Smith}}},\ }\bibfield  {title} {\bibinfo {title} {{Hasasia: A Python
  package for Pulsar Timing Array Sensitivity Curves}},\ }\href
  {https://doi.org/10.21105/joss.01775} {\bibfield  {journal} {\bibinfo
  {journal} {The Journal of Open Source Software}\ }\textbf {\bibinfo {volume}
  {4}},\ \bibinfo {eid} {1775} (\bibinfo {year}
  {2019}{\natexlab{b}})}\BibitemShut {NoStop}%
\bibitem [{\citenamefont {{Sesana}}\ \emph {et~al.}(2008)\citenamefont
  {{Sesana}}, \citenamefont {{Vecchio}},\ and\ \citenamefont
  {{Colacino}}}]{Sesana2008}%
  \BibitemOpen
  \bibfield  {author} {\bibinfo {author} {\bibfnamefont {A.}~\bibnamefont
  {{Sesana}}}, \bibinfo {author} {\bibfnamefont {A.}~\bibnamefont
  {{Vecchio}}},\ and\ \bibinfo {author} {\bibfnamefont {C.~N.}\ \bibnamefont
  {{Colacino}}},\ }\bibfield  {title} {\bibinfo {title} {{The stochastic
  gravitational-wave background from massive black hole binary systems:
  implications for observations with Pulsar Timing Arrays}},\ }\href
  {https://doi.org/10.1111/j.1365-2966.2008.13682.x} {\bibfield  {journal}
  {\bibinfo  {journal} {\mnras}\ }\textbf {\bibinfo {volume} {390}},\ \bibinfo
  {pages} {192} (\bibinfo {year} {2008})},\ \Eprint
  {https://arxiv.org/abs/0804.4476} {arXiv:0804.4476 [astro-ph]} \BibitemShut
  {NoStop}%
\bibitem [{\citenamefont {{Agazie}}\ \emph {et~al.}(2025)\citenamefont
  {{Agazie}}, \citenamefont {{Anumarlapudi}}, \citenamefont {{Archibald}},
  \citenamefont {{Arzoumanian}}, \citenamefont {{Baier}}, \citenamefont
  {{Baker}}, \citenamefont {{B{\'e}csy}}, \citenamefont {{Blecha}},
  \citenamefont {{Brazier}}, \citenamefont {{Brook}}, \citenamefont {{Brown}},
  \citenamefont {{Burke-Spolaor}}, \citenamefont {{Casey-Clyde}}, \citenamefont
  {{Charisi}}, \citenamefont {{Chatterjee}}, \citenamefont {{Cohen}},
  \citenamefont {{Cordes}}, \citenamefont {{Cornish}}, \citenamefont
  {{Crawford}}, \citenamefont {{Cromartie}}, \citenamefont {{Crowter}},
  \citenamefont {{DeCesar}}, \citenamefont {{Demorest}}, \citenamefont
  {{Deng}}, \citenamefont {{Dolch}}, \citenamefont {{Ferrara}}, \citenamefont
  {{Fiore}}, \citenamefont {{Fonseca}}, \citenamefont {{Freedman}},
  \citenamefont {{Garver-Daniels}}, \citenamefont {{Gentile}}, \citenamefont
  {{Glaser}}, \citenamefont {{Good}}, \citenamefont {{G{\"u}ltekin}},
  \citenamefont {{Hazboun}}, \citenamefont {{Jennings}}, \citenamefont
  {{Johnson}}, \citenamefont {{Jones}}, \citenamefont {{Kaiser}}, \citenamefont
  {{Kaplan}}, \citenamefont {{Kelley}}, \citenamefont {{Kerr}}, \citenamefont
  {{Key}}, \citenamefont {{Laal}}, \citenamefont {{Lam}}, \citenamefont
  {{Lamb}}, \citenamefont {{Larsen}}, \citenamefont {{Lazio}}, \citenamefont
  {{Lewandowska}}, \citenamefont {{Liu}}, \citenamefont {{Lorimer}},
  \citenamefont {{Luo}}, \citenamefont {{Lynch}}, \citenamefont {{Ma}},
  \citenamefont {{Madison}}, \citenamefont {{McEwen}}, \citenamefont {{McKee}},
  \citenamefont {{McLaughlin}}, \citenamefont {{McMann}}, \citenamefont
  {{Meyers}}, \citenamefont {{Meyers}}, \citenamefont {{Mingarelli}},
  \citenamefont {{Mitridate}}, \citenamefont {{Natarajan}}, \citenamefont
  {{Ng}}, \citenamefont {{Nice}}, \citenamefont {{Ocker}}, \citenamefont
  {{Olum}}, \citenamefont {{Pennucci}}, \citenamefont {{Perera}}, \citenamefont
  {{Pol}}, \citenamefont {{Radovan}}, \citenamefont {{Ransom}}, \citenamefont
  {{Ray}}, \citenamefont {{Romano}}, \citenamefont {{Runnoe}}, \citenamefont
  {{Sardesai}}, \citenamefont {{Schmiedekamp}}, \citenamefont {{Schmiedekamp}},
  \citenamefont {{Schmitz}}, \citenamefont {{Shapiro-Albert}}, \citenamefont
  {{Siemens}}, \citenamefont {{Simon}}, \citenamefont {{Siwek}}, \citenamefont
  {{Sosa Fiscella}}, \citenamefont {{Stairs}}, \citenamefont {{Stinebring}},
  \citenamefont {{Stovall}}, \citenamefont {{Susobhanan}}, \citenamefont
  {{Swiggum}}, \citenamefont {{Taylor}}, \citenamefont {{Turner}},
  \citenamefont {{Unal}}, \citenamefont {{Vallisneri}}, \citenamefont
  {{Vigeland}}, \citenamefont {{Wahl}}, \citenamefont {{Willson}},
  \citenamefont {{Witt}}, \citenamefont {{Wright}},\ and\ \citenamefont
  {{Young}}}]{NG15discreteness}%
  \BibitemOpen
  \bibfield  {author} {\bibinfo {author} {\bibfnamefont {G.}~\bibnamefont
  {{Agazie}}}, \bibinfo {author} {\bibfnamefont {A.}~\bibnamefont
  {{Anumarlapudi}}}, \bibinfo {author} {\bibfnamefont {A.~M.}\ \bibnamefont
  {{Archibald}}}, \bibinfo {author} {\bibfnamefont {Z.}~\bibnamefont
  {{Arzoumanian}}}, \bibinfo {author} {\bibfnamefont {J.~G.}\ \bibnamefont
  {{Baier}}}, \bibinfo {author} {\bibfnamefont {P.~T.}\ \bibnamefont
  {{Baker}}}, \bibinfo {author} {\bibfnamefont {B.}~\bibnamefont
  {{B{\'e}csy}}}, \bibinfo {author} {\bibfnamefont {L.}~\bibnamefont
  {{Blecha}}}, \bibinfo {author} {\bibfnamefont {A.}~\bibnamefont {{Brazier}}},
  \bibinfo {author} {\bibfnamefont {P.~R.}\ \bibnamefont {{Brook}}}, \bibinfo
  {author} {\bibfnamefont {L.}~\bibnamefont {{Brown}}}, \bibinfo {author}
  {\bibfnamefont {S.}~\bibnamefont {{Burke-Spolaor}}}, \bibinfo {author}
  {\bibfnamefont {J.~A.}\ \bibnamefont {{Casey-Clyde}}}, \bibinfo {author}
  {\bibfnamefont {M.}~\bibnamefont {{Charisi}}}, \bibinfo {author}
  {\bibfnamefont {S.}~\bibnamefont {{Chatterjee}}}, \bibinfo {author}
  {\bibfnamefont {T.}~\bibnamefont {{Cohen}}}, \bibinfo {author} {\bibfnamefont
  {J.~M.}\ \bibnamefont {{Cordes}}}, \bibinfo {author} {\bibfnamefont {N.~J.}\
  \bibnamefont {{Cornish}}}, \bibinfo {author} {\bibfnamefont {F.}~\bibnamefont
  {{Crawford}}}, \bibinfo {author} {\bibfnamefont {H.~T.}\ \bibnamefont
  {{Cromartie}}}, \bibinfo {author} {\bibfnamefont {K.}~\bibnamefont
  {{Crowter}}}, \bibinfo {author} {\bibfnamefont {M.~E.}\ \bibnamefont
  {{DeCesar}}}, \bibinfo {author} {\bibfnamefont {P.~B.}\ \bibnamefont
  {{Demorest}}}, \bibinfo {author} {\bibfnamefont {H.}~\bibnamefont {{Deng}}},
  \bibinfo {author} {\bibfnamefont {T.}~\bibnamefont {{Dolch}}}, \bibinfo
  {author} {\bibfnamefont {E.~C.}\ \bibnamefont {{Ferrara}}}, \bibinfo {author}
  {\bibfnamefont {W.}~\bibnamefont {{Fiore}}}, \bibinfo {author} {\bibfnamefont
  {E.}~\bibnamefont {{Fonseca}}}, \bibinfo {author} {\bibfnamefont {G.~E.}\
  \bibnamefont {{Freedman}}}, \bibinfo {author} {\bibfnamefont
  {N.}~\bibnamefont {{Garver-Daniels}}}, \bibinfo {author} {\bibfnamefont
  {P.~A.}\ \bibnamefont {{Gentile}}}, \bibinfo {author} {\bibfnamefont
  {J.}~\bibnamefont {{Glaser}}}, \bibinfo {author} {\bibfnamefont {D.~C.}\
  \bibnamefont {{Good}}}, \bibinfo {author} {\bibfnamefont {K.}~\bibnamefont
  {{G{\"u}ltekin}}}, \bibinfo {author} {\bibfnamefont {J.~S.}\ \bibnamefont
  {{Hazboun}}}, \bibinfo {author} {\bibfnamefont {R.~J.}\ \bibnamefont
  {{Jennings}}}, \bibinfo {author} {\bibfnamefont {A.~D.}\ \bibnamefont
  {{Johnson}}}, \bibinfo {author} {\bibfnamefont {M.~L.}\ \bibnamefont
  {{Jones}}}, \bibinfo {author} {\bibfnamefont {A.~R.}\ \bibnamefont
  {{Kaiser}}}, \bibinfo {author} {\bibfnamefont {D.~L.}\ \bibnamefont
  {{Kaplan}}}, \bibinfo {author} {\bibfnamefont {L.~Z.}\ \bibnamefont
  {{Kelley}}}, \bibinfo {author} {\bibfnamefont {M.}~\bibnamefont {{Kerr}}},
  \bibinfo {author} {\bibfnamefont {J.~S.}\ \bibnamefont {{Key}}}, \bibinfo
  {author} {\bibfnamefont {N.}~\bibnamefont {{Laal}}}, \bibinfo {author}
  {\bibfnamefont {M.~T.}\ \bibnamefont {{Lam}}}, \bibinfo {author}
  {\bibfnamefont {W.~G.}\ \bibnamefont {{Lamb}}}, \bibinfo {author}
  {\bibfnamefont {B.}~\bibnamefont {{Larsen}}}, \bibinfo {author}
  {\bibfnamefont {T.~J.~W.}\ \bibnamefont {{Lazio}}}, \bibinfo {author}
  {\bibfnamefont {N.}~\bibnamefont {{Lewandowska}}}, \bibinfo {author}
  {\bibfnamefont {T.}~\bibnamefont {{Liu}}}, \bibinfo {author} {\bibfnamefont
  {D.~R.}\ \bibnamefont {{Lorimer}}}, \bibinfo {author} {\bibfnamefont
  {J.}~\bibnamefont {{Luo}}}, \bibinfo {author} {\bibfnamefont {R.~S.}\
  \bibnamefont {{Lynch}}}, \bibinfo {author} {\bibfnamefont {C.-P.}\
  \bibnamefont {{Ma}}}, \bibinfo {author} {\bibfnamefont {D.~R.}\ \bibnamefont
  {{Madison}}}, \bibinfo {author} {\bibfnamefont {A.}~\bibnamefont {{McEwen}}},
  \bibinfo {author} {\bibfnamefont {J.~W.}\ \bibnamefont {{McKee}}}, \bibinfo
  {author} {\bibfnamefont {M.~A.}\ \bibnamefont {{McLaughlin}}}, \bibinfo
  {author} {\bibfnamefont {N.}~\bibnamefont {{McMann}}}, \bibinfo {author}
  {\bibfnamefont {B.~W.}\ \bibnamefont {{Meyers}}}, \bibinfo {author}
  {\bibfnamefont {P.~M.}\ \bibnamefont {{Meyers}}}, \bibinfo {author}
  {\bibfnamefont {C.~M.~F.}\ \bibnamefont {{Mingarelli}}}, \bibinfo {author}
  {\bibfnamefont {A.}~\bibnamefont {{Mitridate}}}, \bibinfo {author}
  {\bibfnamefont {P.}~\bibnamefont {{Natarajan}}}, \bibinfo {author}
  {\bibfnamefont {C.}~\bibnamefont {{Ng}}}, \bibinfo {author} {\bibfnamefont
  {D.~J.}\ \bibnamefont {{Nice}}}, \bibinfo {author} {\bibfnamefont {S.~K.}\
  \bibnamefont {{Ocker}}}, \bibinfo {author} {\bibfnamefont {K.~D.}\
  \bibnamefont {{Olum}}}, \bibinfo {author} {\bibfnamefont {T.~T.}\
  \bibnamefont {{Pennucci}}}, \bibinfo {author} {\bibfnamefont {B.~B.~P.}\
  \bibnamefont {{Perera}}}, \bibinfo {author} {\bibfnamefont {N.~S.}\
  \bibnamefont {{Pol}}}, \bibinfo {author} {\bibfnamefont {H.~A.}\ \bibnamefont
  {{Radovan}}}, \bibinfo {author} {\bibfnamefont {S.~M.}\ \bibnamefont
  {{Ransom}}}, \bibinfo {author} {\bibfnamefont {P.~S.}\ \bibnamefont {{Ray}}},
  \bibinfo {author} {\bibfnamefont {J.~D.}\ \bibnamefont {{Romano}}}, \bibinfo
  {author} {\bibfnamefont {J.~C.}\ \bibnamefont {{Runnoe}}}, \bibinfo {author}
  {\bibfnamefont {S.~C.}\ \bibnamefont {{Sardesai}}}, \bibinfo {author}
  {\bibfnamefont {A.}~\bibnamefont {{Schmiedekamp}}}, \bibinfo {author}
  {\bibfnamefont {C.}~\bibnamefont {{Schmiedekamp}}}, \bibinfo {author}
  {\bibfnamefont {K.}~\bibnamefont {{Schmitz}}}, \bibinfo {author}
  {\bibfnamefont {B.~J.}\ \bibnamefont {{Shapiro-Albert}}}, \bibinfo {author}
  {\bibfnamefont {X.}~\bibnamefont {{Siemens}}}, \bibinfo {author}
  {\bibfnamefont {J.}~\bibnamefont {{Simon}}}, \bibinfo {author} {\bibfnamefont
  {M.~S.}\ \bibnamefont {{Siwek}}}, \bibinfo {author} {\bibfnamefont {S.~V.}\
  \bibnamefont {{Sosa Fiscella}}}, \bibinfo {author} {\bibfnamefont {I.~H.}\
  \bibnamefont {{Stairs}}}, \bibinfo {author} {\bibfnamefont {D.~R.}\
  \bibnamefont {{Stinebring}}}, \bibinfo {author} {\bibfnamefont
  {K.}~\bibnamefont {{Stovall}}}, \bibinfo {author} {\bibfnamefont
  {A.}~\bibnamefont {{Susobhanan}}}, \bibinfo {author} {\bibfnamefont {J.~K.}\
  \bibnamefont {{Swiggum}}}, \bibinfo {author} {\bibfnamefont {S.~R.}\
  \bibnamefont {{Taylor}}}, \bibinfo {author} {\bibfnamefont {J.~E.}\
  \bibnamefont {{Turner}}}, \bibinfo {author} {\bibfnamefont {C.}~\bibnamefont
  {{Unal}}}, \bibinfo {author} {\bibfnamefont {M.}~\bibnamefont
  {{Vallisneri}}}, \bibinfo {author} {\bibfnamefont {S.~J.}\ \bibnamefont
  {{Vigeland}}}, \bibinfo {author} {\bibfnamefont {H.~M.}\ \bibnamefont
  {{Wahl}}}, \bibinfo {author} {\bibfnamefont {L.}~\bibnamefont {{Willson}}},
  \bibinfo {author} {\bibfnamefont {C.~A.}\ \bibnamefont {{Witt}}}, \bibinfo
  {author} {\bibfnamefont {D.}~\bibnamefont {{Wright}}},\ and\ \bibinfo
  {author} {\bibfnamefont {O.}~\bibnamefont {{Young}}},\ }\bibfield  {title}
  {\bibinfo {title} {{The NANOGrav 15 yr Data Set: Looking for Signs of
  Discreteness in the Gravitational-wave Background}},\ }\href
  {https://doi.org/10.3847/1538-4357/ad93d5} {\bibfield  {journal} {\bibinfo
  {journal} {\apj}\ }\textbf {\bibinfo {volume} {978}},\ \bibinfo {eid} {31}
  (\bibinfo {year} {2025})},\ \Eprint {https://arxiv.org/abs/2404.07020}
  {arXiv:2404.07020 [astro-ph.HE]} \BibitemShut {NoStop}%
\bibitem [{\citenamefont {{Mingarelli}}\ \emph {et~al.}(2013)\citenamefont
  {{Mingarelli}}, \citenamefont {{Sidery}}, \citenamefont {{Mandel}},\ and\
  \citenamefont {{Vecchio}}}]{Mingarelli2013}%
  \BibitemOpen
  \bibfield  {author} {\bibinfo {author} {\bibfnamefont {C.~M.~F.}\
  \bibnamefont {{Mingarelli}}}, \bibinfo {author} {\bibfnamefont
  {T.}~\bibnamefont {{Sidery}}}, \bibinfo {author} {\bibfnamefont
  {I.}~\bibnamefont {{Mandel}}},\ and\ \bibinfo {author} {\bibfnamefont
  {A.}~\bibnamefont {{Vecchio}}},\ }\bibfield  {title} {\bibinfo {title}
  {{Characterizing gravitational wave stochastic background anisotropy with
  pulsar timing arrays}},\ }\href {https://doi.org/10.1103/PhysRevD.88.062005}
  {\bibfield  {journal} {\bibinfo  {journal} {Physical Review D}\ }\textbf
  {\bibinfo {volume} {88}},\ \bibinfo {eid} {062005} (\bibinfo {year}
  {2013})},\ \Eprint {https://arxiv.org/abs/1306.5394} {arXiv:1306.5394
  [astro-ph.HE]} \BibitemShut {NoStop}%
\bibitem [{\citenamefont {{Kaiser}}\ and\ \citenamefont
  {{McWilliams}}(2021)}]{Kaiser2021}%
  \BibitemOpen
  \bibfield  {author} {\bibinfo {author} {\bibfnamefont {A.~R.}\ \bibnamefont
  {{Kaiser}}}\ and\ \bibinfo {author} {\bibfnamefont {S.~T.}\ \bibnamefont
  {{McWilliams}}},\ }\bibfield  {title} {\bibinfo {title} {{Sensitivity of
  present and future detectors across the black-hole binary gravitational wave
  spectrum}},\ }\href {https://doi.org/10.1088/1361-6382/abd4f6} {\bibfield
  {journal} {\bibinfo  {journal} {\cqg}\ }\textbf {\bibinfo {volume} {38}},\
  \bibinfo {eid} {055009} (\bibinfo {year} {2021})},\ \Eprint
  {https://arxiv.org/abs/2010.02135} {arXiv:2010.02135 [gr-qc]} \BibitemShut
  {NoStop}%
\bibitem [{\citenamefont {{Mingarelli}}\ \emph {et~al.}(2026)\citenamefont
  {{Mingarelli}}, \citenamefont {{Casey-Clyde}}, \citenamefont {{Chang}},
  \citenamefont {{Eisenberg}}, \citenamefont {{Hutchison}}, \citenamefont
  {{Khusid}}, \citenamefont {{Larsen}}, \citenamefont {{Moran}}, \citenamefont
  {{Semenzato}}, \citenamefont {{Willson}},\ and\ \citenamefont
  {{Zheng}}}]{Mingarelli2026review}%
  \BibitemOpen
  \bibfield  {author} {\bibinfo {author} {\bibfnamefont {C.~M.~F.}\
  \bibnamefont {{Mingarelli}}}, \bibinfo {author} {\bibfnamefont {J.~A.}\
  \bibnamefont {{Casey-Clyde}}}, \bibinfo {author} {\bibfnamefont {Y.~T.}\
  \bibnamefont {{Chang}}}, \bibinfo {author} {\bibfnamefont {E.}~\bibnamefont
  {{Eisenberg}}}, \bibinfo {author} {\bibfnamefont {F.}~\bibnamefont
  {{Hutchison}}}, \bibinfo {author} {\bibfnamefont {N.}~\bibnamefont
  {{Khusid}}}, \bibinfo {author} {\bibfnamefont {B.}~\bibnamefont {{Larsen}}},
  \bibinfo {author} {\bibfnamefont {A.}~\bibnamefont {{Moran}}}, \bibinfo
  {author} {\bibfnamefont {F.}~\bibnamefont {{Semenzato}}}, \bibinfo {author}
  {\bibfnamefont {L.}~\bibnamefont {{Willson}}},\ and\ \bibinfo {author}
  {\bibfnamefont {Q.}~\bibnamefont {{Zheng}}},\ }\bibfield  {title} {\bibinfo
  {title} {{Pulsar timing arrays: the emerging gravitational-wave landscape}},\
  }\href {https://doi.org/10.48550/arXiv.2603.13643} {\bibfield  {journal}
  {\bibinfo  {journal} {arXiv e-prints}\ ,\ \bibinfo {eid} {arXiv:2603.13643}}
  (\bibinfo {year} {2026})},\ \Eprint {https://arxiv.org/abs/2603.13643}
  {arXiv:2603.13643 [astro-ph.HE]} \BibitemShut {NoStop}%
\bibitem [{\citenamefont {{Desvignes}}\ \emph {et~al.}(2016)\citenamefont
  {{Desvignes}}, \citenamefont {{Caballero}}, \citenamefont {{Lentati}},
  \citenamefont {{Verbiest}}, \citenamefont {{Champion}}, \citenamefont
  {{Stappers}}, \citenamefont {{Janssen}}, \citenamefont {{Lazarus}},
  \citenamefont {{Os{\l}owski}}, \citenamefont {{Babak}}, \citenamefont
  {{Bassa}}, \citenamefont {{Brem}}, \citenamefont {{Burgay}}, \citenamefont
  {{Cognard}}, \citenamefont {{Gair}}, \citenamefont {{Graikou}}, \citenamefont
  {{Guillemot}}, \citenamefont {{Hessels}}, \citenamefont {{Jessner}},
  \citenamefont {{Jordan}}, \citenamefont {{Karuppusamy}}, \citenamefont
  {{Kramer}}, \citenamefont {{Lassus}}, \citenamefont {{Lazaridis}},
  \citenamefont {{Lee}}, \citenamefont {{Liu}}, \citenamefont {{Lyne}},
  \citenamefont {{McKee}}, \citenamefont {{Mingarelli}}, \citenamefont
  {{Perrodin}}, \citenamefont {{Petiteau}}, \citenamefont {{Possenti}},
  \citenamefont {{Purver}}, \citenamefont {{Rosado}}, \citenamefont
  {{Sanidas}}, \citenamefont {{Sesana}}, \citenamefont {{Shaifullah}},
  \citenamefont {{Smits}}, \citenamefont {{Taylor}}, \citenamefont
  {{Theureau}}, \citenamefont {{Tiburzi}}, \citenamefont {{van Haasteren}},\
  and\ \citenamefont {{Vecchio}}}]{Desvignes2016}%
  \BibitemOpen
  \bibfield  {author} {\bibinfo {author} {\bibfnamefont {G.}~\bibnamefont
  {{Desvignes}}}, \bibinfo {author} {\bibfnamefont {R.~N.}\ \bibnamefont
  {{Caballero}}}, \bibinfo {author} {\bibfnamefont {L.}~\bibnamefont
  {{Lentati}}}, \bibinfo {author} {\bibfnamefont {J.~P.~W.}\ \bibnamefont
  {{Verbiest}}}, \bibinfo {author} {\bibfnamefont {D.~J.}\ \bibnamefont
  {{Champion}}}, \bibinfo {author} {\bibfnamefont {B.~W.}\ \bibnamefont
  {{Stappers}}}, \bibinfo {author} {\bibfnamefont {G.~H.}\ \bibnamefont
  {{Janssen}}}, \bibinfo {author} {\bibfnamefont {P.}~\bibnamefont
  {{Lazarus}}}, \bibinfo {author} {\bibfnamefont {S.}~\bibnamefont
  {{Os{\l}owski}}}, \bibinfo {author} {\bibfnamefont {S.}~\bibnamefont
  {{Babak}}}, \bibinfo {author} {\bibfnamefont {C.~G.}\ \bibnamefont
  {{Bassa}}}, \bibinfo {author} {\bibfnamefont {P.}~\bibnamefont {{Brem}}},
  \bibinfo {author} {\bibfnamefont {M.}~\bibnamefont {{Burgay}}}, \bibinfo
  {author} {\bibfnamefont {I.}~\bibnamefont {{Cognard}}}, \bibinfo {author}
  {\bibfnamefont {J.~R.}\ \bibnamefont {{Gair}}}, \bibinfo {author}
  {\bibfnamefont {E.}~\bibnamefont {{Graikou}}}, \bibinfo {author}
  {\bibfnamefont {L.}~\bibnamefont {{Guillemot}}}, \bibinfo {author}
  {\bibfnamefont {J.~W.~T.}\ \bibnamefont {{Hessels}}}, \bibinfo {author}
  {\bibfnamefont {A.}~\bibnamefont {{Jessner}}}, \bibinfo {author}
  {\bibfnamefont {C.}~\bibnamefont {{Jordan}}}, \bibinfo {author}
  {\bibfnamefont {R.}~\bibnamefont {{Karuppusamy}}}, \bibinfo {author}
  {\bibfnamefont {M.}~\bibnamefont {{Kramer}}}, \bibinfo {author}
  {\bibfnamefont {A.}~\bibnamefont {{Lassus}}}, \bibinfo {author}
  {\bibfnamefont {K.}~\bibnamefont {{Lazaridis}}}, \bibinfo {author}
  {\bibfnamefont {K.~J.}\ \bibnamefont {{Lee}}}, \bibinfo {author}
  {\bibfnamefont {K.}~\bibnamefont {{Liu}}}, \bibinfo {author} {\bibfnamefont
  {A.~G.}\ \bibnamefont {{Lyne}}}, \bibinfo {author} {\bibfnamefont
  {J.}~\bibnamefont {{McKee}}}, \bibinfo {author} {\bibfnamefont {C.~M.~F.}\
  \bibnamefont {{Mingarelli}}}, \bibinfo {author} {\bibfnamefont
  {D.}~\bibnamefont {{Perrodin}}}, \bibinfo {author} {\bibfnamefont
  {A.}~\bibnamefont {{Petiteau}}}, \bibinfo {author} {\bibfnamefont
  {A.}~\bibnamefont {{Possenti}}}, \bibinfo {author} {\bibfnamefont {M.~B.}\
  \bibnamefont {{Purver}}}, \bibinfo {author} {\bibfnamefont {P.~A.}\
  \bibnamefont {{Rosado}}}, \bibinfo {author} {\bibfnamefont {S.}~\bibnamefont
  {{Sanidas}}}, \bibinfo {author} {\bibfnamefont {A.}~\bibnamefont {{Sesana}}},
  \bibinfo {author} {\bibfnamefont {G.}~\bibnamefont {{Shaifullah}}}, \bibinfo
  {author} {\bibfnamefont {R.}~\bibnamefont {{Smits}}}, \bibinfo {author}
  {\bibfnamefont {S.~R.}\ \bibnamefont {{Taylor}}}, \bibinfo {author}
  {\bibfnamefont {G.}~\bibnamefont {{Theureau}}}, \bibinfo {author}
  {\bibfnamefont {C.}~\bibnamefont {{Tiburzi}}}, \bibinfo {author}
  {\bibfnamefont {R.}~\bibnamefont {{van Haasteren}}},\ and\ \bibinfo {author}
  {\bibfnamefont {A.}~\bibnamefont {{Vecchio}}},\ }\bibfield  {title} {\bibinfo
  {title} {{High-precision timing of 42 millisecond pulsars with the European
  Pulsar Timing Array}},\ }\href {https://doi.org/10.1093/mnras/stw483}
  {\bibfield  {journal} {\bibinfo  {journal} {\mnras}\ }\textbf {\bibinfo
  {volume} {458}},\ \bibinfo {pages} {3341} (\bibinfo {year} {2016})},\ \Eprint
  {https://arxiv.org/abs/1602.08511} {arXiv:1602.08511 [astro-ph.HE]}
  \BibitemShut {NoStop}%
\bibitem [{\citenamefont {{Shannon}}\ \emph {et~al.}(2025)\citenamefont
  {{Shannon}}, \citenamefont {{Bhat}}, \citenamefont {{Chalueau}},
  \citenamefont {{Chen}}, \citenamefont {{Cromartie}}, \citenamefont
  {{Gopukumar}}, \citenamefont {{Grunthal}}, \citenamefont {{Hazboun}},
  \citenamefont {{Iraci}}, \citenamefont {{Joshi}}, \citenamefont {{Kato}},
  \citenamefont {{Keith}}, \citenamefont {{Lee}}, \citenamefont {{Liu}},
  \citenamefont {{Middleton}}, \citenamefont {{Miles}}, \citenamefont
  {{Mingarelli}}, \citenamefont {{Parthasarathy}}, \citenamefont {{Reardon}},
  \citenamefont {{Shaifullah}}, \citenamefont {{Takahashi}}, \citenamefont
  {{Tiburzi}}, \citenamefont {{Truant}}, \citenamefont {{Xue}},\ and\
  \citenamefont {{Zic}}}]{SKAPTA2025}%
  \BibitemOpen
  \bibfield  {author} {\bibinfo {author} {\bibfnamefont {R.~M.}\ \bibnamefont
  {{Shannon}}}, \bibinfo {author} {\bibfnamefont {N.~D.~R.}\ \bibnamefont
  {{Bhat}}}, \bibinfo {author} {\bibfnamefont {A.}~\bibnamefont {{Chalueau}}},
  \bibinfo {author} {\bibfnamefont {S.}~\bibnamefont {{Chen}}}, \bibinfo
  {author} {\bibfnamefont {H.~T.}\ \bibnamefont {{Cromartie}}}, \bibinfo
  {author} {\bibfnamefont {A.}~\bibnamefont {{Gopukumar}}}, \bibinfo {author}
  {\bibfnamefont {K.}~\bibnamefont {{Grunthal}}}, \bibinfo {author}
  {\bibfnamefont {J.~S.}\ \bibnamefont {{Hazboun}}}, \bibinfo {author}
  {\bibfnamefont {F.}~\bibnamefont {{Iraci}}}, \bibinfo {author} {\bibfnamefont
  {B.~C.}\ \bibnamefont {{Joshi}}}, \bibinfo {author} {\bibfnamefont
  {R.}~\bibnamefont {{Kato}}}, \bibinfo {author} {\bibfnamefont {M.~J.}\
  \bibnamefont {{Keith}}}, \bibinfo {author} {\bibfnamefont {K.}~\bibnamefont
  {{Lee}}}, \bibinfo {author} {\bibfnamefont {K.}~\bibnamefont {{Liu}}},
  \bibinfo {author} {\bibfnamefont {H.}~\bibnamefont {{Middleton}}}, \bibinfo
  {author} {\bibfnamefont {M.~T.}\ \bibnamefont {{Miles}}}, \bibinfo {author}
  {\bibfnamefont {C.~M.~F.}\ \bibnamefont {{Mingarelli}}}, \bibinfo {author}
  {\bibfnamefont {A.}~\bibnamefont {{Parthasarathy}}}, \bibinfo {author}
  {\bibfnamefont {D.~J.}\ \bibnamefont {{Reardon}}}, \bibinfo {author}
  {\bibfnamefont {G.~M.}\ \bibnamefont {{Shaifullah}}}, \bibinfo {author}
  {\bibfnamefont {K.}~\bibnamefont {{Takahashi}}}, \bibinfo {author}
  {\bibfnamefont {C.}~\bibnamefont {{Tiburzi}}}, \bibinfo {author}
  {\bibfnamefont {R.}~\bibnamefont {{Truant}}}, \bibinfo {author}
  {\bibfnamefont {X.}~\bibnamefont {{Xue}}},\ and\ \bibinfo {author}
  {\bibfnamefont {A.}~\bibnamefont {{Zic}}},\ }\bibfield  {title} {\bibinfo
  {title} {{The SKAO Pulsar Timing Array}},\ }\href
  {https://doi.org/10.33232/001c.154243} {\bibfield  {journal} {\bibinfo
  {journal} {The Open Journal of Astrophysics}\ }\textbf {\bibinfo {volume}
  {8}},\ \bibinfo {pages} {54243} (\bibinfo {year} {2025})},\ \Eprint
  {https://arxiv.org/abs/2512.16163} {arXiv:2512.16163 [astro-ph.HE]}
  \BibitemShut {NoStop}%
\bibitem [{\citenamefont {{Keane}}\ \emph {et~al.}(2025)\citenamefont
  {{Keane}}, \citenamefont {{Graber}}, \citenamefont {{Levin}}, \citenamefont
  {{Tan}}, \citenamefont {{Johnson}}, \citenamefont {{Ng}}, \citenamefont
  {{Pardo-Araujo}}, \citenamefont {{Ronchi}}, \citenamefont {{Vohl}},\ and\
  \citenamefont {{Xue}}}]{Keane2025}%
  \BibitemOpen
  \bibfield  {author} {\bibinfo {author} {\bibfnamefont {E.~F.}\ \bibnamefont
  {{Keane}}}, \bibinfo {author} {\bibfnamefont {V.}~\bibnamefont {{Graber}}},
  \bibinfo {author} {\bibfnamefont {L.}~\bibnamefont {{Levin}}}, \bibinfo
  {author} {\bibfnamefont {C.~M.}\ \bibnamefont {{Tan}}}, \bibinfo {author}
  {\bibfnamefont {O.~A.}\ \bibnamefont {{Johnson}}}, \bibinfo {author}
  {\bibfnamefont {C.}~\bibnamefont {{Ng}}}, \bibinfo {author} {\bibfnamefont
  {C.}~\bibnamefont {{Pardo-Araujo}}}, \bibinfo {author} {\bibfnamefont
  {M.}~\bibnamefont {{Ronchi}}}, \bibinfo {author} {\bibfnamefont
  {D.}~\bibnamefont {{Vohl}}},\ and\ \bibinfo {author} {\bibfnamefont
  {M.}~\bibnamefont {{Xue}}},\ }\bibfield  {title} {\bibinfo {title} {{A Square
  Kilometre Array Pulsar Census}},\ }\href
  {https://doi.org/10.33232/001c.154256} {\bibfield  {journal} {\bibinfo
  {journal} {The Open Journal of Astrophysics}\ }\textbf {\bibinfo {volume}
  {8}},\ \bibinfo {pages} {54256} (\bibinfo {year} {2025})},\ \Eprint
  {https://arxiv.org/abs/2512.16153} {arXiv:2512.16153 [astro-ph.HE]}
  \BibitemShut {NoStop}%
\bibitem [{\citenamefont {McLaughlin}\ \emph {et~al.}(2025)\citenamefont
  {McLaughlin}, \citenamefont {Walter}, \citenamefont {Hallinan},\ and\
  \citenamefont {Ravi}}]{DSA2000}%
  \BibitemOpen
  \bibfield  {author} {\bibinfo {author} {\bibfnamefont {M.}~\bibnamefont
  {McLaughlin}}, \bibinfo {author} {\bibfnamefont {F.}~\bibnamefont {Walter}},
  \bibinfo {author} {\bibfnamefont {G.}~\bibnamefont {Hallinan}},\ and\
  \bibinfo {author} {\bibfnamefont {V.}~\bibnamefont {Ravi}},\ }\href
  {https://www.deepsynoptic.org/overview} {\bibinfo {title} {{DSA-2000}
  community science document}} (\bibinfo {year} {2025})\BibitemShut {NoStop}%
\bibitem [{\citenamefont {{Agazie}}\ \emph
  {et~al.}(2023{\natexlab{c}})\citenamefont {{Agazie}}, \citenamefont {{Alam}},
  \citenamefont {{Anumarlapudi}}, \citenamefont {{Archibald}}, \citenamefont
  {{Arzoumanian}}, \citenamefont {{Baker}}, \citenamefont {{Blecha}},
  \citenamefont {{Bonidie}}, \citenamefont {{Brazier}}, \citenamefont
  {{Brook}}, \citenamefont {{Burke-Spolaor}}, \citenamefont {{B{\'e}csy}},
  \citenamefont {{Chapman}}, \citenamefont {{Charisi}}, \citenamefont
  {{Chatterjee}}, \citenamefont {{Cohen}}, \citenamefont {{Cordes}},
  \citenamefont {{Cornish}}, \citenamefont {{Crawford}}, \citenamefont
  {{Cromartie}}, \citenamefont {{Crowter}}, \citenamefont {{Decesar}},
  \citenamefont {{Demorest}}, \citenamefont {{Dolch}}, \citenamefont
  {{Drachler}}, \citenamefont {{Ferrara}}, \citenamefont {{Fiore}},
  \citenamefont {{Fonseca}}, \citenamefont {{Freedman}}, \citenamefont
  {{Garver-Daniels}}, \citenamefont {{Gentile}}, \citenamefont {{Glaser}},
  \citenamefont {{Good}}, \citenamefont {{G{\"u}ltekin}}, \citenamefont
  {{Hazboun}}, \citenamefont {{Jennings}}, \citenamefont {{Jessup}},
  \citenamefont {{Johnson}}, \citenamefont {{Jones}}, \citenamefont {{Kaiser}},
  \citenamefont {{Kaplan}}, \citenamefont {{Kelley}}, \citenamefont {{Kerr}},
  \citenamefont {{Key}}, \citenamefont {{Kuske}}, \citenamefont {{Laal}},
  \citenamefont {{Lam}}, \citenamefont {{Lamb}}, \citenamefont {{Lazio}},
  \citenamefont {{Lewandowska}}, \citenamefont {{Lin}}, \citenamefont {{Liu}},
  \citenamefont {{Lorimer}}, \citenamefont {{Luo}}, \citenamefont {{Lynch}},
  \citenamefont {{Ma}}, \citenamefont {{Madison}}, \citenamefont {{Maraccini}},
  \citenamefont {{McEwen}}, \citenamefont {{McKee}}, \citenamefont
  {{McLaughlin}}, \citenamefont {{McMann}}, \citenamefont {{Meyers}},
  \citenamefont {{Mingarelli}}, \citenamefont {{Mitridate}}, \citenamefont
  {{Ng}}, \citenamefont {{Nice}}, \citenamefont {{Ocker}}, \citenamefont
  {{Olum}}, \citenamefont {{Panciu}}, \citenamefont {{Pennucci}}, \citenamefont
  {{Perera}}, \citenamefont {{Pol}}, \citenamefont {{Radovan}}, \citenamefont
  {{Ransom}}, \citenamefont {{Ray}}, \citenamefont {{Romano}}, \citenamefont
  {{Salo}}, \citenamefont {{Sardesai}}, \citenamefont {{Schmiedekamp}},
  \citenamefont {{Schmiedekamp}}, \citenamefont {{Schmitz}}, \citenamefont
  {{Shapiro-Albert}}, \citenamefont {{Siemens}}, \citenamefont {{Simon}},
  \citenamefont {{Siwek}}, \citenamefont {{Stairs}}, \citenamefont
  {{Stinebring}}, \citenamefont {{Stovall}}, \citenamefont {{Susobhanan}},
  \citenamefont {{Swiggum}}, \citenamefont {{Taylor}}, \citenamefont
  {{Turner}}, \citenamefont {{Unal}}, \citenamefont {{Vallisneri}},
  \citenamefont {{Vigeland}}, \citenamefont {{Wahl}}, \citenamefont {{Wang}},
  \citenamefont {{Witt}}, \citenamefont {{Young}},\ and\ \citenamefont
  {{Nanograv Collaboration}}}]{NG15-timing}%
  \BibitemOpen
  \bibfield  {author} {\bibinfo {author} {\bibfnamefont {G.}~\bibnamefont
  {{Agazie}}}, \bibinfo {author} {\bibfnamefont {M.~F.}\ \bibnamefont
  {{Alam}}}, \bibinfo {author} {\bibfnamefont {A.}~\bibnamefont
  {{Anumarlapudi}}}, \bibinfo {author} {\bibfnamefont {A.~M.}\ \bibnamefont
  {{Archibald}}}, \bibinfo {author} {\bibfnamefont {Z.}~\bibnamefont
  {{Arzoumanian}}}, \bibinfo {author} {\bibfnamefont {P.~T.}\ \bibnamefont
  {{Baker}}}, \bibinfo {author} {\bibfnamefont {L.}~\bibnamefont {{Blecha}}},
  \bibinfo {author} {\bibfnamefont {V.}~\bibnamefont {{Bonidie}}}, \bibinfo
  {author} {\bibfnamefont {A.}~\bibnamefont {{Brazier}}}, \bibinfo {author}
  {\bibfnamefont {P.~R.}\ \bibnamefont {{Brook}}}, \bibinfo {author}
  {\bibfnamefont {S.}~\bibnamefont {{Burke-Spolaor}}}, \bibinfo {author}
  {\bibfnamefont {B.}~\bibnamefont {{B{\'e}csy}}}, \bibinfo {author}
  {\bibfnamefont {C.}~\bibnamefont {{Chapman}}}, \bibinfo {author}
  {\bibfnamefont {M.}~\bibnamefont {{Charisi}}}, \bibinfo {author}
  {\bibfnamefont {S.}~\bibnamefont {{Chatterjee}}}, \bibinfo {author}
  {\bibfnamefont {T.}~\bibnamefont {{Cohen}}}, \bibinfo {author} {\bibfnamefont
  {J.~M.}\ \bibnamefont {{Cordes}}}, \bibinfo {author} {\bibfnamefont {N.~J.}\
  \bibnamefont {{Cornish}}}, \bibinfo {author} {\bibfnamefont {F.}~\bibnamefont
  {{Crawford}}}, \bibinfo {author} {\bibfnamefont {H.~T.}\ \bibnamefont
  {{Cromartie}}}, \bibinfo {author} {\bibfnamefont {K.}~\bibnamefont
  {{Crowter}}}, \bibinfo {author} {\bibfnamefont {M.~E.}\ \bibnamefont
  {{Decesar}}}, \bibinfo {author} {\bibfnamefont {P.~B.}\ \bibnamefont
  {{Demorest}}}, \bibinfo {author} {\bibfnamefont {T.}~\bibnamefont {{Dolch}}},
  \bibinfo {author} {\bibfnamefont {B.}~\bibnamefont {{Drachler}}}, \bibinfo
  {author} {\bibfnamefont {E.~C.}\ \bibnamefont {{Ferrara}}}, \bibinfo {author}
  {\bibfnamefont {W.}~\bibnamefont {{Fiore}}}, \bibinfo {author} {\bibfnamefont
  {E.}~\bibnamefont {{Fonseca}}}, \bibinfo {author} {\bibfnamefont {G.~E.}\
  \bibnamefont {{Freedman}}}, \bibinfo {author} {\bibfnamefont
  {N.}~\bibnamefont {{Garver-Daniels}}}, \bibinfo {author} {\bibfnamefont
  {P.~A.}\ \bibnamefont {{Gentile}}}, \bibinfo {author} {\bibfnamefont
  {J.}~\bibnamefont {{Glaser}}}, \bibinfo {author} {\bibfnamefont {D.~C.}\
  \bibnamefont {{Good}}}, \bibinfo {author} {\bibfnamefont {K.}~\bibnamefont
  {{G{\"u}ltekin}}}, \bibinfo {author} {\bibfnamefont {J.~S.}\ \bibnamefont
  {{Hazboun}}}, \bibinfo {author} {\bibfnamefont {R.~J.}\ \bibnamefont
  {{Jennings}}}, \bibinfo {author} {\bibfnamefont {C.}~\bibnamefont
  {{Jessup}}}, \bibinfo {author} {\bibfnamefont {A.~D.}\ \bibnamefont
  {{Johnson}}}, \bibinfo {author} {\bibfnamefont {M.~L.}\ \bibnamefont
  {{Jones}}}, \bibinfo {author} {\bibfnamefont {A.~R.}\ \bibnamefont
  {{Kaiser}}}, \bibinfo {author} {\bibfnamefont {D.~L.}\ \bibnamefont
  {{Kaplan}}}, \bibinfo {author} {\bibfnamefont {L.~Z.}\ \bibnamefont
  {{Kelley}}}, \bibinfo {author} {\bibfnamefont {M.}~\bibnamefont {{Kerr}}},
  \bibinfo {author} {\bibfnamefont {J.~S.}\ \bibnamefont {{Key}}}, \bibinfo
  {author} {\bibfnamefont {A.}~\bibnamefont {{Kuske}}}, \bibinfo {author}
  {\bibfnamefont {N.}~\bibnamefont {{Laal}}}, \bibinfo {author} {\bibfnamefont
  {M.~T.}\ \bibnamefont {{Lam}}}, \bibinfo {author} {\bibfnamefont {W.~G.}\
  \bibnamefont {{Lamb}}}, \bibinfo {author} {\bibfnamefont {T.~J.~W.}\
  \bibnamefont {{Lazio}}}, \bibinfo {author} {\bibfnamefont {N.}~\bibnamefont
  {{Lewandowska}}}, \bibinfo {author} {\bibfnamefont {Y.}~\bibnamefont
  {{Lin}}}, \bibinfo {author} {\bibfnamefont {T.}~\bibnamefont {{Liu}}},
  \bibinfo {author} {\bibfnamefont {D.~R.}\ \bibnamefont {{Lorimer}}}, \bibinfo
  {author} {\bibfnamefont {J.}~\bibnamefont {{Luo}}}, \bibinfo {author}
  {\bibfnamefont {R.~S.}\ \bibnamefont {{Lynch}}}, \bibinfo {author}
  {\bibfnamefont {C.-P.}\ \bibnamefont {{Ma}}}, \bibinfo {author}
  {\bibfnamefont {D.~R.}\ \bibnamefont {{Madison}}}, \bibinfo {author}
  {\bibfnamefont {K.}~\bibnamefont {{Maraccini}}}, \bibinfo {author}
  {\bibfnamefont {A.}~\bibnamefont {{McEwen}}}, \bibinfo {author}
  {\bibfnamefont {J.~W.}\ \bibnamefont {{McKee}}}, \bibinfo {author}
  {\bibfnamefont {M.~A.}\ \bibnamefont {{McLaughlin}}}, \bibinfo {author}
  {\bibfnamefont {N.}~\bibnamefont {{McMann}}}, \bibinfo {author}
  {\bibfnamefont {B.~W.}\ \bibnamefont {{Meyers}}}, \bibinfo {author}
  {\bibfnamefont {C.~M.~F.}\ \bibnamefont {{Mingarelli}}}, \bibinfo {author}
  {\bibfnamefont {A.}~\bibnamefont {{Mitridate}}}, \bibinfo {author}
  {\bibfnamefont {C.}~\bibnamefont {{Ng}}}, \bibinfo {author} {\bibfnamefont
  {D.~J.}\ \bibnamefont {{Nice}}}, \bibinfo {author} {\bibfnamefont {S.~K.}\
  \bibnamefont {{Ocker}}}, \bibinfo {author} {\bibfnamefont {K.~D.}\
  \bibnamefont {{Olum}}}, \bibinfo {author} {\bibfnamefont {E.}~\bibnamefont
  {{Panciu}}}, \bibinfo {author} {\bibfnamefont {T.~T.}\ \bibnamefont
  {{Pennucci}}}, \bibinfo {author} {\bibfnamefont {B.~B.~P.}\ \bibnamefont
  {{Perera}}}, \bibinfo {author} {\bibfnamefont {N.~S.}\ \bibnamefont {{Pol}}},
  \bibinfo {author} {\bibfnamefont {H.~A.}\ \bibnamefont {{Radovan}}}, \bibinfo
  {author} {\bibfnamefont {S.~M.}\ \bibnamefont {{Ransom}}}, \bibinfo {author}
  {\bibfnamefont {P.~S.}\ \bibnamefont {{Ray}}}, \bibinfo {author}
  {\bibfnamefont {J.~D.}\ \bibnamefont {{Romano}}}, \bibinfo {author}
  {\bibfnamefont {L.}~\bibnamefont {{Salo}}}, \bibinfo {author} {\bibfnamefont
  {S.~C.}\ \bibnamefont {{Sardesai}}}, \bibinfo {author} {\bibfnamefont
  {C.}~\bibnamefont {{Schmiedekamp}}}, \bibinfo {author} {\bibfnamefont
  {A.}~\bibnamefont {{Schmiedekamp}}}, \bibinfo {author} {\bibfnamefont
  {K.}~\bibnamefont {{Schmitz}}}, \bibinfo {author} {\bibfnamefont {B.~J.}\
  \bibnamefont {{Shapiro-Albert}}}, \bibinfo {author} {\bibfnamefont
  {X.}~\bibnamefont {{Siemens}}}, \bibinfo {author} {\bibfnamefont
  {J.}~\bibnamefont {{Simon}}}, \bibinfo {author} {\bibfnamefont {M.~S.}\
  \bibnamefont {{Siwek}}}, \bibinfo {author} {\bibfnamefont {I.~H.}\
  \bibnamefont {{Stairs}}}, \bibinfo {author} {\bibfnamefont {D.~R.}\
  \bibnamefont {{Stinebring}}}, \bibinfo {author} {\bibfnamefont
  {K.}~\bibnamefont {{Stovall}}}, \bibinfo {author} {\bibfnamefont
  {A.}~\bibnamefont {{Susobhanan}}}, \bibinfo {author} {\bibfnamefont {J.~K.}\
  \bibnamefont {{Swiggum}}}, \bibinfo {author} {\bibfnamefont {S.~R.}\
  \bibnamefont {{Taylor}}}, \bibinfo {author} {\bibfnamefont {J.~E.}\
  \bibnamefont {{Turner}}}, \bibinfo {author} {\bibfnamefont {C.}~\bibnamefont
  {{Unal}}}, \bibinfo {author} {\bibfnamefont {M.}~\bibnamefont
  {{Vallisneri}}}, \bibinfo {author} {\bibfnamefont {S.~J.}\ \bibnamefont
  {{Vigeland}}}, \bibinfo {author} {\bibfnamefont {H.~M.}\ \bibnamefont
  {{Wahl}}}, \bibinfo {author} {\bibfnamefont {Q.}~\bibnamefont {{Wang}}},
  \bibinfo {author} {\bibfnamefont {C.~A.}\ \bibnamefont {{Witt}}}, \bibinfo
  {author} {\bibfnamefont {O.}~\bibnamefont {{Young}}},\ and\ \bibinfo {author}
  {\bibnamefont {{Nanograv Collaboration}}},\ }\bibfield  {title} {\bibinfo
  {title} {{The NANOGrav 15 yr Data Set: Observations and Timing of 68
  Millisecond Pulsars}},\ }\href {https://doi.org/10.3847/2041-8213/acda9a}
  {\bibfield  {journal} {\bibinfo  {journal} {\apjl}\ }\textbf {\bibinfo
  {volume} {951}},\ \bibinfo {eid} {L9} (\bibinfo {year}
  {2023}{\natexlab{c}})},\ \Eprint {https://arxiv.org/abs/2306.16217}
  {arXiv:2306.16217 [astro-ph.HE]} \BibitemShut {NoStop}%
\bibitem [{\citenamefont {{Ma}}\ \emph {et~al.}(2014)\citenamefont {{Ma}},
  \citenamefont {{Greene}}, \citenamefont {{McConnell}}, \citenamefont
  {{Janish}}, \citenamefont {{Blakeslee}}, \citenamefont {{Thomas}},\ and\
  \citenamefont {{Murphy}}}]{Ma2014}%
  \BibitemOpen
  \bibfield  {author} {\bibinfo {author} {\bibfnamefont {C.-P.}\ \bibnamefont
  {{Ma}}}, \bibinfo {author} {\bibfnamefont {J.~E.}\ \bibnamefont {{Greene}}},
  \bibinfo {author} {\bibfnamefont {N.}~\bibnamefont {{McConnell}}}, \bibinfo
  {author} {\bibfnamefont {R.}~\bibnamefont {{Janish}}}, \bibinfo {author}
  {\bibfnamefont {J.~P.}\ \bibnamefont {{Blakeslee}}}, \bibinfo {author}
  {\bibfnamefont {J.}~\bibnamefont {{Thomas}}},\ and\ \bibinfo {author}
  {\bibfnamefont {J.~D.}\ \bibnamefont {{Murphy}}},\ }\bibfield  {title}
  {\bibinfo {title} {{The MASSIVE Survey. I. A Volume-limited Integral-field
  Spectroscopic Study of the Most Massive Early-type Galaxies within 108
  Mpc}},\ }\href {https://doi.org/10.1088/0004-637X/795/2/158} {\bibfield
  {journal} {\bibinfo  {journal} {\apj}\ }\textbf {\bibinfo {volume} {795}},\
  \bibinfo {eid} {158} (\bibinfo {year} {2014})},\ \Eprint
  {https://arxiv.org/abs/1407.1054} {arXiv:1407.1054 [astro-ph.GA]}
  \BibitemShut {NoStop}%
\bibitem [{\citenamefont {{Casey-Clyde}}\ \emph {et~al.}(2025)\citenamefont
  {{Casey-Clyde}}, \citenamefont {{Mingarelli}}, \citenamefont {{Greene}},
  \citenamefont {{Goulding}}, \citenamefont {{Chen}},\ and\ \citenamefont
  {{Trump}}}]{CaseyClyde2025}%
  \BibitemOpen
  \bibfield  {author} {\bibinfo {author} {\bibfnamefont {J.~A.}\ \bibnamefont
  {{Casey-Clyde}}}, \bibinfo {author} {\bibfnamefont {C.~M.~F.}\ \bibnamefont
  {{Mingarelli}}}, \bibinfo {author} {\bibfnamefont {J.~E.}\ \bibnamefont
  {{Greene}}}, \bibinfo {author} {\bibfnamefont {A.~D.}\ \bibnamefont
  {{Goulding}}}, \bibinfo {author} {\bibfnamefont {S.}~\bibnamefont {{Chen}}},\
  and\ \bibinfo {author} {\bibfnamefont {J.~R.}\ \bibnamefont {{Trump}}},\
  }\bibfield  {title} {\bibinfo {title} {{Quasars Can Signpost Supermassive
  Black Hole Binaries}},\ }\href {https://doi.org/10.3847/1538-4357/adce05}
  {\bibfield  {journal} {\bibinfo  {journal} {\apj}\ }\textbf {\bibinfo
  {volume} {987}},\ \bibinfo {eid} {106} (\bibinfo {year} {2025})},\ \Eprint
  {https://arxiv.org/abs/2405.19406} {arXiv:2405.19406 [astro-ph.HE]}
  \BibitemShut {NoStop}%
\bibitem [{\citenamefont {Phinney}(2001)}]{Phinney2001}%
  \BibitemOpen
  \bibfield  {author} {\bibinfo {author} {\bibfnamefont {E.~S.}\ \bibnamefont
  {Phinney}},\ }\href@noop {} {\bibinfo {title} {A practical theorem on
  gravitational wave backgrounds}} (\bibinfo {year} {2001}),\ \Eprint
  {https://arxiv.org/abs/astro-ph/0108028} {arXiv:astro-ph/0108028}
  \BibitemShut {NoStop}%
\bibitem [{\citenamefont {{Reynolds}}(2021)}]{Reynolds2021}%
  \BibitemOpen
  \bibfield  {author} {\bibinfo {author} {\bibfnamefont {C.~S.}\ \bibnamefont
  {{Reynolds}}},\ }\bibfield  {title} {\bibinfo {title} {{Observational
  Constraints on Black Hole Spin}},\ }\href
  {https://doi.org/10.1146/annurev-astro-112420-035022} {\bibfield  {journal}
  {\bibinfo  {journal} {\araa}\ }\textbf {\bibinfo {volume} {59}},\ \bibinfo
  {pages} {117} (\bibinfo {year} {2021})},\ \Eprint
  {https://arxiv.org/abs/2011.08948} {arXiv:2011.08948 [astro-ph.HE]}
  \BibitemShut {NoStop}%
\bibitem [{\citenamefont {{Blanchet}}(2014)}]{Blanchet2006}%
  \BibitemOpen
  \bibfield  {author} {\bibinfo {author} {\bibfnamefont {L.}~\bibnamefont
  {{Blanchet}}},\ }\bibfield  {title} {\bibinfo {title} {{Gravitational
  Radiation from Post-Newtonian Sources and Inspiralling Compact Binaries}},\
  }\href {https://doi.org/10.12942/lrr-2014-2} {\bibfield  {journal} {\bibinfo
  {journal} {Living Reviews in Relativity}\ }\textbf {\bibinfo {volume} {17}},\
  \bibinfo {eid} {2} (\bibinfo {year} {2014})},\ \Eprint
  {https://arxiv.org/abs/1310.1528} {arXiv:1310.1528 [gr-qc]} \BibitemShut
  {NoStop}%
\bibitem [{\citenamefont {{Buonanno}}\ \emph {et~al.}(2009)\citenamefont
  {{Buonanno}}, \citenamefont {{Iyer}}, \citenamefont {{Ochsner}},
  \citenamefont {{Pan}},\ and\ \citenamefont {{Sathyaprakash}}}]{BIOPS2009}%
  \BibitemOpen
  \bibfield  {author} {\bibinfo {author} {\bibfnamefont {A.}~\bibnamefont
  {{Buonanno}}}, \bibinfo {author} {\bibfnamefont {B.~R.}\ \bibnamefont
  {{Iyer}}}, \bibinfo {author} {\bibfnamefont {E.}~\bibnamefont {{Ochsner}}},
  \bibinfo {author} {\bibfnamefont {Y.}~\bibnamefont {{Pan}}},\ and\ \bibinfo
  {author} {\bibfnamefont {B.~S.}\ \bibnamefont {{Sathyaprakash}}},\ }\bibfield
   {title} {\bibinfo {title} {{Comparison of post-Newtonian templates for
  compact binary inspiral signals in gravitational-wave detectors}},\ }\href
  {https://doi.org/10.1103/PhysRevD.80.084043} {\bibfield  {journal} {\bibinfo
  {journal} {\prd}\ }\textbf {\bibinfo {volume} {80}},\ \bibinfo {eid} {084043}
  (\bibinfo {year} {2009})},\ \Eprint {https://arxiv.org/abs/0907.0700}
  {arXiv:0907.0700 [gr-qc]} \BibitemShut {NoStop}%
\bibitem [{\citenamefont {{Mik{\'o}czi}}\ \emph {et~al.}(2005)\citenamefont
  {{Mik{\'o}czi}}, \citenamefont {{Vas{\'u}th}},\ and\ \citenamefont
  {{Gergely}}}]{MBMG2005}%
  \BibitemOpen
  \bibfield  {author} {\bibinfo {author} {\bibfnamefont {B.}~\bibnamefont
  {{Mik{\'o}czi}}}, \bibinfo {author} {\bibfnamefont {M.}~\bibnamefont
  {{Vas{\'u}th}}},\ and\ \bibinfo {author} {\bibfnamefont {L.~{\'A}.}\
  \bibnamefont {{Gergely}}},\ }\bibfield  {title} {\bibinfo {title}
  {{Self-interaction spin effects in inspiralling compact binaries}},\ }\href
  {https://doi.org/10.1103/PhysRevD.71.124043} {\bibfield  {journal} {\bibinfo
  {journal} {\prd}\ }\textbf {\bibinfo {volume} {71}},\ \bibinfo {eid} {124043}
  (\bibinfo {year} {2005})},\ \Eprint {https://arxiv.org/abs/astro-ph/0504538}
  {arXiv:astro-ph/0504538 [astro-ph]} \BibitemShut {NoStop}%
\bibitem [{\citenamefont {{Apostolatos}}\ \emph {et~al.}(1994)\citenamefont
  {{Apostolatos}}, \citenamefont {{Cutler}}, \citenamefont {{Sussman}},\ and\
  \citenamefont {{Thorne}}}]{Apostolatos1994}%
  \BibitemOpen
  \bibfield  {author} {\bibinfo {author} {\bibfnamefont {T.~A.}\ \bibnamefont
  {{Apostolatos}}}, \bibinfo {author} {\bibfnamefont {C.}~\bibnamefont
  {{Cutler}}}, \bibinfo {author} {\bibfnamefont {G.~J.}\ \bibnamefont
  {{Sussman}}},\ and\ \bibinfo {author} {\bibfnamefont {K.~S.}\ \bibnamefont
  {{Thorne}}},\ }\bibfield  {title} {\bibinfo {title} {{Spin-induced orbital
  precession and its modulation of the gravitational waveforms from merging
  binaries}},\ }\href {https://doi.org/10.1103/PhysRevD.49.6274} {\bibfield
  {journal} {\bibinfo  {journal} {\prd}\ }\textbf {\bibinfo {volume} {49}},\
  \bibinfo {pages} {6274} (\bibinfo {year} {1994})}\BibitemShut {NoStop}%
\bibitem [{\citenamefont {{Kidder}}(1995)}]{Kidder1995}%
  \BibitemOpen
  \bibfield  {author} {\bibinfo {author} {\bibfnamefont {L.~E.}\ \bibnamefont
  {{Kidder}}},\ }\bibfield  {title} {\bibinfo {title} {{Coalescing binary
  systems of compact objects to (post)$^{5/2}$-Newtonian order. V. Spin
  effects}},\ }\href {https://doi.org/10.1103/PhysRevD.52.821} {\bibfield
  {journal} {\bibinfo  {journal} {\prd}\ }\textbf {\bibinfo {volume} {52}},\
  \bibinfo {pages} {821} (\bibinfo {year} {1995})},\ \Eprint
  {https://arxiv.org/abs/gr-qc/9506022} {arXiv:gr-qc/9506022 [gr-qc]}
  \BibitemShut {NoStop}%
\bibitem [{\citenamefont {{Moran}}\ \emph {et~al.}(2023)\citenamefont
  {{Moran}}, \citenamefont {{Mingarelli}}, \citenamefont {{Bedell}},
  \citenamefont {{Good}},\ and\ \citenamefont {{Spergel}}}]{Moran2023}%
  \BibitemOpen
  \bibfield  {author} {\bibinfo {author} {\bibfnamefont {A.}~\bibnamefont
  {{Moran}}}, \bibinfo {author} {\bibfnamefont {C.~M.~F.}\ \bibnamefont
  {{Mingarelli}}}, \bibinfo {author} {\bibfnamefont {M.}~\bibnamefont
  {{Bedell}}}, \bibinfo {author} {\bibfnamefont {D.}~\bibnamefont {{Good}}},\
  and\ \bibinfo {author} {\bibfnamefont {D.~N.}\ \bibnamefont {{Spergel}}},\
  }\bibfield  {title} {\bibinfo {title} {{Improving Distances to Binary
  Millisecond Pulsars with Gaia}},\ }\href
  {https://doi.org/10.3847/1538-4357/acec75} {\bibfield  {journal} {\bibinfo
  {journal} {\apj}\ }\textbf {\bibinfo {volume} {954}},\ \bibinfo {eid} {89}
  (\bibinfo {year} {2023})},\ \Eprint {https://arxiv.org/abs/2210.10816}
  {arXiv:2210.10816 [astro-ph.IM]} \BibitemShut {NoStop}%
\bibitem [{\citenamefont {{McKinnon}}\ and\ \citenamefont {{van der
  Marel}}(2026)}]{McKinnon2026}%
  \BibitemOpen
  \bibfield  {author} {\bibinfo {author} {\bibfnamefont {K.~A.}\ \bibnamefont
  {{McKinnon}}}\ and\ \bibinfo {author} {\bibfnamefont {R.~P.}\ \bibnamefont
  {{van der Marel}}},\ }\href {https://doi.org/10.48550/arXiv.2602.00310}
  {\bibinfo {title} {{Simulating Roman+Gaia Combined Astrometry, Parallaxes,
  and Proper Motions}}} (\bibinfo {year} {2026}),\ \Eprint
  {https://arxiv.org/abs/2602.00310} {arXiv:2602.00310 [astro-ph.IM]}
  \BibitemShut {NoStop}%
\bibitem [{\citenamefont {{Haiman}}\ \emph {et~al.}(2009)\citenamefont
  {{Haiman}}, \citenamefont {{Kocsis}},\ and\ \citenamefont
  {{Menou}}}]{HaimanKocsisMenou2009}%
  \BibitemOpen
  \bibfield  {author} {\bibinfo {author} {\bibfnamefont {Z.}~\bibnamefont
  {{Haiman}}}, \bibinfo {author} {\bibfnamefont {B.}~\bibnamefont {{Kocsis}}},\
  and\ \bibinfo {author} {\bibfnamefont {K.}~\bibnamefont {{Menou}}},\
  }\bibfield  {title} {\bibinfo {title} {{The Population of Viscosity- and
  Gravitational Wave-driven Supermassive Black Hole Binaries Among Luminous
  Active Galactic Nuclei}},\ }\href
  {https://doi.org/10.1088/0004-637X/700/2/1952} {\bibfield  {journal}
  {\bibinfo  {journal} {\apj}\ }\textbf {\bibinfo {volume} {700}},\ \bibinfo
  {pages} {1952} (\bibinfo {year} {2009})},\ \Eprint
  {https://arxiv.org/abs/0904.1383} {arXiv:0904.1383 [astro-ph.CO]}
  \BibitemShut {NoStop}%
\bibitem [{\citenamefont {{Duffell}}\ \emph {et~al.}(2020)\citenamefont
  {{Duffell}}, \citenamefont {{D'Orazio}}, \citenamefont {{Derdzinski}},
  \citenamefont {{Haiman}}, \citenamefont {{MacFadyen}}, \citenamefont
  {{Rosen}},\ and\ \citenamefont {{Zrake}}}]{Duffell2020}%
  \BibitemOpen
  \bibfield  {author} {\bibinfo {author} {\bibfnamefont {P.~C.}\ \bibnamefont
  {{Duffell}}}, \bibinfo {author} {\bibfnamefont {D.}~\bibnamefont
  {{D'Orazio}}}, \bibinfo {author} {\bibfnamefont {A.}~\bibnamefont
  {{Derdzinski}}}, \bibinfo {author} {\bibfnamefont {Z.}~\bibnamefont
  {{Haiman}}}, \bibinfo {author} {\bibfnamefont {A.}~\bibnamefont
  {{MacFadyen}}}, \bibinfo {author} {\bibfnamefont {A.~L.}\ \bibnamefont
  {{Rosen}}},\ and\ \bibinfo {author} {\bibfnamefont {J.}~\bibnamefont
  {{Zrake}}},\ }\bibfield  {title} {\bibinfo {title} {{Circumbinary Disks:
  Accretion and Torque as a Function of Mass Ratio and Disk Viscosity}},\
  }\href {https://doi.org/10.3847/1538-4357/abab95} {\bibfield  {journal}
  {\bibinfo  {journal} {\apj}\ }\textbf {\bibinfo {volume} {901}},\ \bibinfo
  {eid} {25} (\bibinfo {year} {2020})},\ \Eprint
  {https://arxiv.org/abs/1911.05506} {arXiv:1911.05506 [astro-ph.HE]}
  \BibitemShut {NoStop}%
\bibitem [{\citenamefont {{Siwek}}\ \emph {et~al.}(2024)\citenamefont
  {{Siwek}}, \citenamefont {{Kelley}},\ and\ \citenamefont
  {{Hernquist}}}]{Siwek2024}%
  \BibitemOpen
  \bibfield  {author} {\bibinfo {author} {\bibfnamefont {M.}~\bibnamefont
  {{Siwek}}}, \bibinfo {author} {\bibfnamefont {L.~Z.}\ \bibnamefont
  {{Kelley}}},\ and\ \bibinfo {author} {\bibfnamefont {L.}~\bibnamefont
  {{Hernquist}}},\ }\bibfield  {title} {\bibinfo {title} {{Signatures of
  circumbinary disc dynamics in multimessenger population studies of massive
  black hole binaries}},\ }\href {https://doi.org/10.1093/mnras/stae2251}
  {\bibfield  {journal} {\bibinfo  {journal} {\mnras}\ }\textbf {\bibinfo
  {volume} {534}},\ \bibinfo {pages} {2609} (\bibinfo {year} {2024})},\ \Eprint
  {https://arxiv.org/abs/2403.08871} {arXiv:2403.08871 [astro-ph.HE]}
  \BibitemShut {NoStop}%
\bibitem [{\citenamefont {{Siwek}}\ \emph {et~al.}(2023)\citenamefont
  {{Siwek}}, \citenamefont {{Weinberger}},\ and\ \citenamefont
  {{Hernquist}}}]{SiwekWH2023}%
  \BibitemOpen
  \bibfield  {author} {\bibinfo {author} {\bibfnamefont {M.}~\bibnamefont
  {{Siwek}}}, \bibinfo {author} {\bibfnamefont {R.}~\bibnamefont
  {{Weinberger}}},\ and\ \bibinfo {author} {\bibfnamefont {L.}~\bibnamefont
  {{Hernquist}}},\ }\bibfield  {title} {\bibinfo {title} {{Orbital evolution of
  binaries in circumbinary discs}},\ }\href
  {https://doi.org/10.1093/mnras/stad1131} {\bibfield  {journal} {\bibinfo
  {journal} {\mnras}\ }\textbf {\bibinfo {volume} {522}},\ \bibinfo {pages}
  {2707} (\bibinfo {year} {2023})},\ \Eprint {https://arxiv.org/abs/2302.01785}
  {arXiv:2302.01785 [astro-ph.HE]} \BibitemShut {NoStop}%
\bibitem [{\citenamefont {{Tiede}}\ \emph {et~al.}(2025)\citenamefont
  {{Tiede}}, \citenamefont {{Zrake}}, \citenamefont {{MacFadyen}},\ and\
  \citenamefont {{Haiman}}}]{Tiede2025}%
  \BibitemOpen
  \bibfield  {author} {\bibinfo {author} {\bibfnamefont {C.}~\bibnamefont
  {{Tiede}}}, \bibinfo {author} {\bibfnamefont {J.}~\bibnamefont {{Zrake}}},
  \bibinfo {author} {\bibfnamefont {A.}~\bibnamefont {{MacFadyen}}},\ and\
  \bibinfo {author} {\bibfnamefont {Z.}~\bibnamefont {{Haiman}}},\ }\bibfield
  {title} {\bibinfo {title} {{Suppressed Accretion onto Massive Black Hole
  Binaries Surrounded by Thin Disks}},\ }\href
  {https://doi.org/10.3847/1538-4357/adc727} {\bibfield  {journal} {\bibinfo
  {journal} {\apj}\ }\textbf {\bibinfo {volume} {984}},\ \bibinfo {eid} {144}
  (\bibinfo {year} {2025})},\ \Eprint {https://arxiv.org/abs/2410.03830}
  {arXiv:2410.03830 [astro-ph.GA]} \BibitemShut {NoStop}%
\bibitem [{\citenamefont {{Arvanitaki}}\ \emph {et~al.}(2010)\citenamefont
  {{Arvanitaki}}, \citenamefont {{Dimopoulos}}, \citenamefont {{Dubovsky}},
  \citenamefont {{Kaloper}},\ and\ \citenamefont
  {{March-Russell}}}]{Arvanitaki2010}%
  \BibitemOpen
  \bibfield  {author} {\bibinfo {author} {\bibfnamefont {A.}~\bibnamefont
  {{Arvanitaki}}}, \bibinfo {author} {\bibfnamefont {S.}~\bibnamefont
  {{Dimopoulos}}}, \bibinfo {author} {\bibfnamefont {S.}~\bibnamefont
  {{Dubovsky}}}, \bibinfo {author} {\bibfnamefont {N.}~\bibnamefont
  {{Kaloper}}},\ and\ \bibinfo {author} {\bibfnamefont {J.}~\bibnamefont
  {{March-Russell}}},\ }\bibfield  {title} {\bibinfo {title} {{String
  axiverse}},\ }\href {https://doi.org/10.1103/PhysRevD.81.123530} {\bibfield
  {journal} {\bibinfo  {journal} {\prd}\ }\textbf {\bibinfo {volume} {81}},\
  \bibinfo {eid} {123530} (\bibinfo {year} {2010})},\ \Eprint
  {https://arxiv.org/abs/0905.4720} {arXiv:0905.4720 [hep-th]} \BibitemShut
  {NoStop}%
\bibitem [{\citenamefont {{Brito}}\ \emph {et~al.}(2015)\citenamefont
  {{Brito}}, \citenamefont {{Cardoso}},\ and\ \citenamefont
  {{Pani}}}]{Brito2015}%
  \BibitemOpen
  \bibfield  {author} {\bibinfo {author} {\bibfnamefont {R.}~\bibnamefont
  {{Brito}}}, \bibinfo {author} {\bibfnamefont {V.}~\bibnamefont {{Cardoso}}},\
  and\ \bibinfo {author} {\bibfnamefont {P.}~\bibnamefont {{Pani}}},\ }\href
  {https://doi.org/10.1007/978-3-030-46622-0} {\emph {\bibinfo {title}
  {{Superradiance}}}},\ \bibinfo {series} {Lecture Notes in Physics}, Vol.\
  \bibinfo {volume} {906}\ (\bibinfo  {publisher} {Springer},\ \bibinfo {year}
  {2015})\ \Eprint {https://arxiv.org/abs/1501.06570} {arXiv:1501.06570
  [gr-qc]} \BibitemShut {NoStop}%
\bibitem [{\citenamefont {{Baumann}}\ \emph {et~al.}(2019)\citenamefont
  {{Baumann}}, \citenamefont {{Chia}},\ and\ \citenamefont
  {{Porto}}}]{Baumann2019}%
  \BibitemOpen
  \bibfield  {author} {\bibinfo {author} {\bibfnamefont {D.}~\bibnamefont
  {{Baumann}}}, \bibinfo {author} {\bibfnamefont {H.~S.}\ \bibnamefont
  {{Chia}}},\ and\ \bibinfo {author} {\bibfnamefont {R.~A.}\ \bibnamefont
  {{Porto}}},\ }\bibfield  {title} {\bibinfo {title} {{Probing ultralight
  bosons with binary black holes}},\ }\href
  {https://doi.org/10.1103/PhysRevD.99.044001} {\bibfield  {journal} {\bibinfo
  {journal} {\prd}\ }\textbf {\bibinfo {volume} {99}},\ \bibinfo {eid} {044001}
  (\bibinfo {year} {2019})},\ \Eprint {https://arxiv.org/abs/1804.03208}
  {arXiv:1804.03208 [gr-qc]} \BibitemShut {NoStop}%
\bibitem [{\citenamefont {{Gondolo}}\ and\ \citenamefont
  {{Silk}}(1999)}]{Gondolo1999}%
  \BibitemOpen
  \bibfield  {author} {\bibinfo {author} {\bibfnamefont {P.}~\bibnamefont
  {{Gondolo}}}\ and\ \bibinfo {author} {\bibfnamefont {J.}~\bibnamefont
  {{Silk}}},\ }\bibfield  {title} {\bibinfo {title} {{Dark Matter Annihilation
  at the Galactic Center}},\ }\href
  {https://doi.org/10.1103/PhysRevLett.83.1719} {\bibfield  {journal} {\bibinfo
   {journal} {\prl}\ }\textbf {\bibinfo {volume} {83}},\ \bibinfo {pages}
  {1719} (\bibinfo {year} {1999})},\ \Eprint
  {https://arxiv.org/abs/astro-ph/9906391} {arXiv:astro-ph/9906391 [astro-ph]}
  \BibitemShut {NoStop}%
\bibitem [{\citenamefont {{Alonso-{\'A}lvarez}}\ \emph
  {et~al.}(2024)\citenamefont {{Alonso-{\'A}lvarez}}, \citenamefont {{Cline}},\
  and\ \citenamefont {{Dewar}}}]{AlonsoAlvarez2024}%
  \BibitemOpen
  \bibfield  {author} {\bibinfo {author} {\bibfnamefont {G.}~\bibnamefont
  {{Alonso-{\'A}lvarez}}}, \bibinfo {author} {\bibfnamefont {J.~M.}\
  \bibnamefont {{Cline}}},\ and\ \bibinfo {author} {\bibfnamefont
  {C.}~\bibnamefont {{Dewar}}},\ }\bibfield  {title} {\bibinfo {title}
  {{Self-Interacting Dark Matter Solves the Final Parsec Problem of
  Supermassive Black Hole Mergers}},\ }\href
  {https://doi.org/10.1103/PhysRevLett.133.021401} {\bibfield  {journal}
  {\bibinfo  {journal} {\prl}\ }\textbf {\bibinfo {volume} {133}},\ \bibinfo
  {eid} {021401} (\bibinfo {year} {2024})},\ \Eprint
  {https://arxiv.org/abs/2401.14450} {arXiv:2401.14450 [astro-ph.CO]}
  \BibitemShut {NoStop}%
\bibitem [{\citenamefont {{Peters}}(1964)}]{Peters1964}%
  \BibitemOpen
  \bibfield  {author} {\bibinfo {author} {\bibfnamefont {P.~C.}\ \bibnamefont
  {{Peters}}},\ }\bibfield  {title} {\bibinfo {title} {{Gravitational Radiation
  and the Motion of Two Point Masses}},\ }\href
  {https://doi.org/10.1103/PhysRev.136.B1224} {\bibfield  {journal} {\bibinfo
  {journal} {Physical Review}\ }\textbf {\bibinfo {volume} {136}},\ \bibinfo
  {pages} {B1224} (\bibinfo {year} {1964})}\BibitemShut {NoStop}%
\bibitem [{\citenamefont {{D'Orazio}}\ and\ \citenamefont
  {{Duffell}}(2021)}]{DOrazio2021}%
  \BibitemOpen
  \bibfield  {author} {\bibinfo {author} {\bibfnamefont {D.~J.}\ \bibnamefont
  {{D'Orazio}}}\ and\ \bibinfo {author} {\bibfnamefont {P.~C.}\ \bibnamefont
  {{Duffell}}},\ }\bibfield  {title} {\bibinfo {title} {{Orbital Evolution of
  Equal-mass Eccentric Binaries due to a Gas Disk: Eccentric Inspirals and
  Circular Outspirals}},\ }\href {https://doi.org/10.3847/2041-8213/ac0621}
  {\bibfield  {journal} {\bibinfo  {journal} {\apjl}\ }\textbf {\bibinfo
  {volume} {914}},\ \bibinfo {eid} {L21} (\bibinfo {year} {2021})},\ \Eprint
  {https://arxiv.org/abs/2103.09251} {arXiv:2103.09251 [astro-ph.HE]}
  \BibitemShut {NoStop}%
\bibitem [{\citenamefont {{Taylor}}\ \emph {et~al.}(2016)\citenamefont
  {{Taylor}}, \citenamefont {{Huerta}}, \citenamefont {{Gair}},\ and\
  \citenamefont {{McWilliams}}}]{Taylor2016}%
  \BibitemOpen
  \bibfield  {author} {\bibinfo {author} {\bibfnamefont {S.~R.}\ \bibnamefont
  {{Taylor}}}, \bibinfo {author} {\bibfnamefont {E.~A.}\ \bibnamefont
  {{Huerta}}}, \bibinfo {author} {\bibfnamefont {J.~R.}\ \bibnamefont
  {{Gair}}},\ and\ \bibinfo {author} {\bibfnamefont {S.~T.}\ \bibnamefont
  {{McWilliams}}},\ }\bibfield  {title} {\bibinfo {title} {{Detecting Eccentric
  Supermassive Black Hole Binaries with Pulsar Timing Arrays: Resolvable Source
  Strategies}},\ }\href {https://doi.org/10.3847/0004-637X/817/1/70} {\bibfield
   {journal} {\bibinfo  {journal} {\apj}\ }\textbf {\bibinfo {volume} {817}},\
  \bibinfo {eid} {70} (\bibinfo {year} {2016})},\ \Eprint
  {https://arxiv.org/abs/1505.06208} {arXiv:1505.06208 [gr-qc]} \BibitemShut
  {NoStop}%
\bibitem [{\citenamefont {{Yunes}}\ and\ \citenamefont
  {{Pretorius}}(2009)}]{Yunes2009}%
  \BibitemOpen
  \bibfield  {author} {\bibinfo {author} {\bibfnamefont {N.}~\bibnamefont
  {{Yunes}}}\ and\ \bibinfo {author} {\bibfnamefont {F.}~\bibnamefont
  {{Pretorius}}},\ }\bibfield  {title} {\bibinfo {title} {{Fundamental
  theoretical bias in gravitational wave astrophysics and the parametrized
  post-Einsteinian framework}},\ }\href
  {https://doi.org/10.1103/PhysRevD.80.122003} {\bibfield  {journal} {\bibinfo
  {journal} {\prd}\ }\textbf {\bibinfo {volume} {80}},\ \bibinfo {eid} {122003}
  (\bibinfo {year} {2009})},\ \Eprint {https://arxiv.org/abs/0909.3328}
  {arXiv:0909.3328 [gr-qc]} \BibitemShut {NoStop}%
\bibitem [{\citenamefont {{Barausse}}\ \emph {et~al.}(2016)\citenamefont
  {{Barausse}}, \citenamefont {{Yunes}},\ and\ \citenamefont
  {{Chamberlain}}}]{Barausse2016}%
  \BibitemOpen
  \bibfield  {author} {\bibinfo {author} {\bibfnamefont {E.}~\bibnamefont
  {{Barausse}}}, \bibinfo {author} {\bibfnamefont {N.}~\bibnamefont
  {{Yunes}}},\ and\ \bibinfo {author} {\bibfnamefont {K.}~\bibnamefont
  {{Chamberlain}}},\ }\bibfield  {title} {\bibinfo {title} {{Theory-Agnostic
  Constraints on Black-Hole Dipole Radiation with Multiband Gravitational-Wave
  Astrophysics}},\ }\href {https://doi.org/10.1103/PhysRevLett.116.241104}
  {\bibfield  {journal} {\bibinfo  {journal} {\prl}\ }\textbf {\bibinfo
  {volume} {116}},\ \bibinfo {eid} {241104} (\bibinfo {year} {2016})},\ \Eprint
  {https://arxiv.org/abs/1603.04075} {arXiv:1603.04075 [gr-qc]} \BibitemShut
  {NoStop}%
\bibitem [{\citenamefont {{Will}}(1998)}]{Will1998}%
  \BibitemOpen
  \bibfield  {author} {\bibinfo {author} {\bibfnamefont {C.~M.}\ \bibnamefont
  {{Will}}},\ }\bibfield  {title} {\bibinfo {title} {{Bounding the mass of the
  graviton using gravitational-wave observations of inspiralling compact
  binaries}},\ }\href {https://doi.org/10.1103/PhysRevD.57.2061} {\bibfield
  {journal} {\bibinfo  {journal} {\prd}\ }\textbf {\bibinfo {volume} {57}},\
  \bibinfo {pages} {2061} (\bibinfo {year} {1998})},\ \Eprint
  {https://arxiv.org/abs/gr-qc/9709011} {arXiv:gr-qc/9709011 [gr-qc]}
  \BibitemShut {NoStop}%
\bibitem [{\citenamefont {{Zheng}}\ \emph {et~al.}(2026)\citenamefont
  {{Zheng}}, \citenamefont {{Larsen}}, \citenamefont {{Eisenberg}},\ and\
  \citenamefont {{Mingarelli}}}]{ZhengLarsenEisenbergMingarelli2026}%
  \BibitemOpen
  \bibfield  {author} {\bibinfo {author} {\bibfnamefont {Q.}~\bibnamefont
  {{Zheng}}}, \bibinfo {author} {\bibfnamefont {B.}~\bibnamefont {{Larsen}}},
  \bibinfo {author} {\bibfnamefont {E.}~\bibnamefont {{Eisenberg}}},\ and\
  \bibinfo {author} {\bibfnamefont {C.~M.~F.}\ \bibnamefont {{Mingarelli}}},\
  }\bibfield  {title} {\bibinfo {title} {{Testing General Relativity with
  Individual Supermassive Black Hole Binaries}}} (\bibinfo {year} {2026}),\
  \bibinfo {note} {in preparation}\BibitemShut {NoStop}%
\bibitem [{\citenamefont {{Agazie}}\ \emph
  {et~al.}(2023{\natexlab{d}})\citenamefont {{Agazie}}, \citenamefont
  {{Anumarlapudi}}, \citenamefont {{Archibald}}, \citenamefont {{Arzoumanian}},
  \citenamefont {{Baker}}, \citenamefont {{B{\'e}csy}}, \citenamefont
  {{Blecha}}, \citenamefont {{Brazier}}, \citenamefont {{Brook}}, \citenamefont
  {{Burke-Spolaor}}, \citenamefont {{Case}}, \citenamefont {{Casey-Clyde}},
  \citenamefont {{Charisi}}, \citenamefont {{Chatterjee}}, \citenamefont
  {{Cohen}}, \citenamefont {{Cordes}}, \citenamefont {{Cornish}}, \citenamefont
  {{Crawford}}, \citenamefont {{Cromartie}}, \citenamefont {{Crowter}},
  \citenamefont {{Decesar}}, \citenamefont {{Demorest}}, \citenamefont
  {{Digman}}, \citenamefont {{Dolch}}, \citenamefont {{Drachler}},
  \citenamefont {{Ferrara}}, \citenamefont {{Fiore}}, \citenamefont
  {{Fonseca}}, \citenamefont {{Freedman}}, \citenamefont {{Garver-Daniels}},
  \citenamefont {{Gentile}}, \citenamefont {{Glaser}}, \citenamefont {{Good}},
  \citenamefont {{G{\"u}ltekin}}, \citenamefont {{Hazboun}}, \citenamefont
  {{Hourihane}}, \citenamefont {{Jennings}}, \citenamefont {{Johnson}},
  \citenamefont {{Jones}}, \citenamefont {{Kaiser}}, \citenamefont {{Kaplan}},
  \citenamefont {{Kelley}}, \citenamefont {{Kerr}}, \citenamefont {{Key}},
  \citenamefont {{Laal}}, \citenamefont {{Lam}}, \citenamefont {{Lamb}},
  \citenamefont {{Lazio}}, \citenamefont {{Lewandowska}}, \citenamefont
  {{Liu}}, \citenamefont {{Lorimer}}, \citenamefont {{Luo}}, \citenamefont
  {{Lynch}}, \citenamefont {{Ma}}, \citenamefont {{Madison}}, \citenamefont
  {{McEwen}}, \citenamefont {{McKee}}, \citenamefont {{McLaughlin}},
  \citenamefont {{McMann}}, \citenamefont {{Meyers}}, \citenamefont {{Meyers}},
  \citenamefont {{Mingarelli}}, \citenamefont {{Mitridate}}, \citenamefont
  {{Ng}}, \citenamefont {{Nice}}, \citenamefont {{Ocker}}, \citenamefont
  {{Olum}}, \citenamefont {{Pennucci}}, \citenamefont {{Perera}}, \citenamefont
  {{Petrov}}, \citenamefont {{Pol}}, \citenamefont {{Radovan}}, \citenamefont
  {{Ransom}}, \citenamefont {{Ray}}, \citenamefont {{Romano}}, \citenamefont
  {{Sardesai}}, \citenamefont {{Schmiedekamp}}, \citenamefont {{Schmiedekamp}},
  \citenamefont {{Schmitz}}, \citenamefont {{Shapiro-Albert}}, \citenamefont
  {{Siemens}}, \citenamefont {{Simon}}, \citenamefont {{Siwek}}, \citenamefont
  {{Stairs}}, \citenamefont {{Stinebring}}, \citenamefont {{Stovall}},
  \citenamefont {{Susobhanan}}, \citenamefont {{Swiggum}}, \citenamefont
  {{Taylor}}, \citenamefont {{Taylor}}, \citenamefont {{Turner}}, \citenamefont
  {{Unal}}, \citenamefont {{Vallisneri}}, \citenamefont {{van Haasteren}},
  \citenamefont {{Vigeland}}, \citenamefont {{Wahl}}, \citenamefont {{Witt}},
  \citenamefont {{Young}},\ and\ \citenamefont {{Nanograv
  Collaboration}}}]{NG15cw}%
  \BibitemOpen
  \bibfield  {author} {\bibinfo {author} {\bibfnamefont {G.}~\bibnamefont
  {{Agazie}}}, \bibinfo {author} {\bibfnamefont {A.}~\bibnamefont
  {{Anumarlapudi}}}, \bibinfo {author} {\bibfnamefont {A.~M.}\ \bibnamefont
  {{Archibald}}}, \bibinfo {author} {\bibfnamefont {Z.}~\bibnamefont
  {{Arzoumanian}}}, \bibinfo {author} {\bibfnamefont {P.~T.}\ \bibnamefont
  {{Baker}}}, \bibinfo {author} {\bibfnamefont {B.}~\bibnamefont
  {{B{\'e}csy}}}, \bibinfo {author} {\bibfnamefont {L.}~\bibnamefont
  {{Blecha}}}, \bibinfo {author} {\bibfnamefont {A.}~\bibnamefont {{Brazier}}},
  \bibinfo {author} {\bibfnamefont {P.~R.}\ \bibnamefont {{Brook}}}, \bibinfo
  {author} {\bibfnamefont {S.}~\bibnamefont {{Burke-Spolaor}}}, \bibinfo
  {author} {\bibfnamefont {R.}~\bibnamefont {{Case}}}, \bibinfo {author}
  {\bibfnamefont {J.~A.}\ \bibnamefont {{Casey-Clyde}}}, \bibinfo {author}
  {\bibfnamefont {M.}~\bibnamefont {{Charisi}}}, \bibinfo {author}
  {\bibfnamefont {S.}~\bibnamefont {{Chatterjee}}}, \bibinfo {author}
  {\bibfnamefont {T.}~\bibnamefont {{Cohen}}}, \bibinfo {author} {\bibfnamefont
  {J.~M.}\ \bibnamefont {{Cordes}}}, \bibinfo {author} {\bibfnamefont {N.~J.}\
  \bibnamefont {{Cornish}}}, \bibinfo {author} {\bibfnamefont {F.}~\bibnamefont
  {{Crawford}}}, \bibinfo {author} {\bibfnamefont {H.~T.}\ \bibnamefont
  {{Cromartie}}}, \bibinfo {author} {\bibfnamefont {K.}~\bibnamefont
  {{Crowter}}}, \bibinfo {author} {\bibfnamefont {M.~E.}\ \bibnamefont
  {{Decesar}}}, \bibinfo {author} {\bibfnamefont {P.~B.}\ \bibnamefont
  {{Demorest}}}, \bibinfo {author} {\bibfnamefont {M.~C.}\ \bibnamefont
  {{Digman}}}, \bibinfo {author} {\bibfnamefont {T.}~\bibnamefont {{Dolch}}},
  \bibinfo {author} {\bibfnamefont {B.}~\bibnamefont {{Drachler}}}, \bibinfo
  {author} {\bibfnamefont {E.~C.}\ \bibnamefont {{Ferrara}}}, \bibinfo {author}
  {\bibfnamefont {W.}~\bibnamefont {{Fiore}}}, \bibinfo {author} {\bibfnamefont
  {E.}~\bibnamefont {{Fonseca}}}, \bibinfo {author} {\bibfnamefont {G.~E.}\
  \bibnamefont {{Freedman}}}, \bibinfo {author} {\bibfnamefont
  {N.}~\bibnamefont {{Garver-Daniels}}}, \bibinfo {author} {\bibfnamefont
  {P.~A.}\ \bibnamefont {{Gentile}}}, \bibinfo {author} {\bibfnamefont
  {J.}~\bibnamefont {{Glaser}}}, \bibinfo {author} {\bibfnamefont {D.~C.}\
  \bibnamefont {{Good}}}, \bibinfo {author} {\bibfnamefont {K.}~\bibnamefont
  {{G{\"u}ltekin}}}, \bibinfo {author} {\bibfnamefont {J.~S.}\ \bibnamefont
  {{Hazboun}}}, \bibinfo {author} {\bibfnamefont {S.}~\bibnamefont
  {{Hourihane}}}, \bibinfo {author} {\bibfnamefont {R.~J.}\ \bibnamefont
  {{Jennings}}}, \bibinfo {author} {\bibfnamefont {A.~D.}\ \bibnamefont
  {{Johnson}}}, \bibinfo {author} {\bibfnamefont {M.~L.}\ \bibnamefont
  {{Jones}}}, \bibinfo {author} {\bibfnamefont {A.~R.}\ \bibnamefont
  {{Kaiser}}}, \bibinfo {author} {\bibfnamefont {D.~L.}\ \bibnamefont
  {{Kaplan}}}, \bibinfo {author} {\bibfnamefont {L.~Z.}\ \bibnamefont
  {{Kelley}}}, \bibinfo {author} {\bibfnamefont {M.}~\bibnamefont {{Kerr}}},
  \bibinfo {author} {\bibfnamefont {J.~S.}\ \bibnamefont {{Key}}}, \bibinfo
  {author} {\bibfnamefont {N.}~\bibnamefont {{Laal}}}, \bibinfo {author}
  {\bibfnamefont {M.~T.}\ \bibnamefont {{Lam}}}, \bibinfo {author}
  {\bibfnamefont {W.~G.}\ \bibnamefont {{Lamb}}}, \bibinfo {author}
  {\bibfnamefont {T.~J.~W.}\ \bibnamefont {{Lazio}}}, \bibinfo {author}
  {\bibfnamefont {N.}~\bibnamefont {{Lewandowska}}}, \bibinfo {author}
  {\bibfnamefont {T.}~\bibnamefont {{Liu}}}, \bibinfo {author} {\bibfnamefont
  {D.~R.}\ \bibnamefont {{Lorimer}}}, \bibinfo {author} {\bibfnamefont
  {J.}~\bibnamefont {{Luo}}}, \bibinfo {author} {\bibfnamefont {R.~S.}\
  \bibnamefont {{Lynch}}}, \bibinfo {author} {\bibfnamefont {C.-P.}\
  \bibnamefont {{Ma}}}, \bibinfo {author} {\bibfnamefont {D.~R.}\ \bibnamefont
  {{Madison}}}, \bibinfo {author} {\bibfnamefont {A.}~\bibnamefont {{McEwen}}},
  \bibinfo {author} {\bibfnamefont {J.~W.}\ \bibnamefont {{McKee}}}, \bibinfo
  {author} {\bibfnamefont {M.~A.}\ \bibnamefont {{McLaughlin}}}, \bibinfo
  {author} {\bibfnamefont {N.}~\bibnamefont {{McMann}}}, \bibinfo {author}
  {\bibfnamefont {B.~W.}\ \bibnamefont {{Meyers}}}, \bibinfo {author}
  {\bibfnamefont {P.~M.}\ \bibnamefont {{Meyers}}}, \bibinfo {author}
  {\bibfnamefont {C.~M.~F.}\ \bibnamefont {{Mingarelli}}}, \bibinfo {author}
  {\bibfnamefont {A.}~\bibnamefont {{Mitridate}}}, \bibinfo {author}
  {\bibfnamefont {C.}~\bibnamefont {{Ng}}}, \bibinfo {author} {\bibfnamefont
  {D.~J.}\ \bibnamefont {{Nice}}}, \bibinfo {author} {\bibfnamefont {S.~K.}\
  \bibnamefont {{Ocker}}}, \bibinfo {author} {\bibfnamefont {K.~D.}\
  \bibnamefont {{Olum}}}, \bibinfo {author} {\bibfnamefont {T.~T.}\
  \bibnamefont {{Pennucci}}}, \bibinfo {author} {\bibfnamefont {B.~B.~P.}\
  \bibnamefont {{Perera}}}, \bibinfo {author} {\bibfnamefont {P.}~\bibnamefont
  {{Petrov}}}, \bibinfo {author} {\bibfnamefont {N.~S.}\ \bibnamefont {{Pol}}},
  \bibinfo {author} {\bibfnamefont {H.~A.}\ \bibnamefont {{Radovan}}}, \bibinfo
  {author} {\bibfnamefont {S.~M.}\ \bibnamefont {{Ransom}}}, \bibinfo {author}
  {\bibfnamefont {P.~S.}\ \bibnamefont {{Ray}}}, \bibinfo {author}
  {\bibfnamefont {J.~D.}\ \bibnamefont {{Romano}}}, \bibinfo {author}
  {\bibfnamefont {S.~C.}\ \bibnamefont {{Sardesai}}}, \bibinfo {author}
  {\bibfnamefont {A.}~\bibnamefont {{Schmiedekamp}}}, \bibinfo {author}
  {\bibfnamefont {C.}~\bibnamefont {{Schmiedekamp}}}, \bibinfo {author}
  {\bibfnamefont {K.}~\bibnamefont {{Schmitz}}}, \bibinfo {author}
  {\bibfnamefont {B.~J.}\ \bibnamefont {{Shapiro-Albert}}}, \bibinfo {author}
  {\bibfnamefont {X.}~\bibnamefont {{Siemens}}}, \bibinfo {author}
  {\bibfnamefont {J.}~\bibnamefont {{Simon}}}, \bibinfo {author} {\bibfnamefont
  {M.~S.}\ \bibnamefont {{Siwek}}}, \bibinfo {author} {\bibfnamefont {I.~H.}\
  \bibnamefont {{Stairs}}}, \bibinfo {author} {\bibfnamefont {D.~R.}\
  \bibnamefont {{Stinebring}}}, \bibinfo {author} {\bibfnamefont
  {K.}~\bibnamefont {{Stovall}}}, \bibinfo {author} {\bibfnamefont
  {A.}~\bibnamefont {{Susobhanan}}}, \bibinfo {author} {\bibfnamefont {J.~K.}\
  \bibnamefont {{Swiggum}}}, \bibinfo {author} {\bibfnamefont {J.}~\bibnamefont
  {{Taylor}}}, \bibinfo {author} {\bibfnamefont {S.~R.}\ \bibnamefont
  {{Taylor}}}, \bibinfo {author} {\bibfnamefont {J.~E.}\ \bibnamefont
  {{Turner}}}, \bibinfo {author} {\bibfnamefont {C.}~\bibnamefont {{Unal}}},
  \bibinfo {author} {\bibfnamefont {M.}~\bibnamefont {{Vallisneri}}}, \bibinfo
  {author} {\bibfnamefont {R.}~\bibnamefont {{van Haasteren}}}, \bibinfo
  {author} {\bibfnamefont {S.~J.}\ \bibnamefont {{Vigeland}}}, \bibinfo
  {author} {\bibfnamefont {H.~M.}\ \bibnamefont {{Wahl}}}, \bibinfo {author}
  {\bibfnamefont {C.~A.}\ \bibnamefont {{Witt}}}, \bibinfo {author}
  {\bibfnamefont {O.}~\bibnamefont {{Young}}},\ and\ \bibinfo {author}
  {\bibnamefont {{Nanograv Collaboration}}},\ }\bibfield  {title} {\bibinfo
  {title} {{The NANOGrav 15 yr Data Set: Bayesian Limits on Gravitational Waves
  from Individual Supermassive Black Hole Binaries}},\ }\href
  {https://doi.org/10.3847/2041-8213/ace18a} {\bibfield  {journal} {\bibinfo
  {journal} {\apjl}\ }\textbf {\bibinfo {volume} {951}},\ \bibinfo {eid} {L50}
  (\bibinfo {year} {2023}{\natexlab{d}})},\ \Eprint
  {https://arxiv.org/abs/2306.16222} {arXiv:2306.16222 [astro-ph.HE]}
  \BibitemShut {NoStop}%
\bibitem [{\citenamefont {{Ennoggi}}\ \emph {et~al.}(2026)\citenamefont
  {{Ennoggi}}, \citenamefont {{Campanelli}}, \citenamefont {{Krolik}},
  \citenamefont {{Noble}}, \citenamefont {{Zlochower}},\ and\ \citenamefont
  {{de Simone}}}]{Ennoggi2026}%
  \BibitemOpen
  \bibfield  {author} {\bibinfo {author} {\bibfnamefont {L.}~\bibnamefont
  {{Ennoggi}}}, \bibinfo {author} {\bibfnamefont {M.}~\bibnamefont
  {{Campanelli}}}, \bibinfo {author} {\bibfnamefont {J.}~\bibnamefont
  {{Krolik}}}, \bibinfo {author} {\bibfnamefont {S.~C.}\ \bibnamefont
  {{Noble}}}, \bibinfo {author} {\bibfnamefont {Y.}~\bibnamefont
  {{Zlochower}}},\ and\ \bibinfo {author} {\bibfnamefont {M.~C.}\ \bibnamefont
  {{de Simone}}},\ }\bibfield  {title} {\bibinfo {title} {{Merger of Spinning,
  Accreting Supermassive Black Hole Binaries}},\ }\href
  {https://doi.org/10.1103/f74l-3c4y} {\bibfield  {journal} {\bibinfo
  {journal} {\prl}\ }\textbf {\bibinfo {volume} {136}},\ \bibinfo {eid}
  {111401} (\bibinfo {year} {2026})},\ \Eprint
  {https://arxiv.org/abs/2509.10319} {arXiv:2509.10319 [astro-ph.HE]}
  \BibitemShut {NoStop}%
\bibitem [{\citenamefont {{Harris}}\ \emph {et~al.}(2020)\citenamefont
  {{Harris}}, \citenamefont {{Millman}}, \citenamefont {{van der Walt}},
  \citenamefont {{Gommers}}, \citenamefont {{Virtanen}}, \citenamefont
  {{Cournapeau}}, \citenamefont {{Wieser}}, \citenamefont {{Taylor}},
  \citenamefont {{Berg}}, \citenamefont {{Smith}}, \citenamefont {{Kern}},
  \citenamefont {{Picus}}, \citenamefont {{Hoyer}}, \citenamefont {{van
  Kerkwijk}}, \citenamefont {{Brett}}, \citenamefont {{Haldane}}, \citenamefont
  {{del R{\'\i}o}}, \citenamefont {{Wiebe}}, \citenamefont {{Peterson}},
  \citenamefont {{G{\'e}rard-Marchant}}, \citenamefont {{Sheppard}},
  \citenamefont {{Reddy}}, \citenamefont {{Weckesser}}, \citenamefont
  {{Abbasi}}, \citenamefont {{Gohlke}},\ and\ \citenamefont
  {{Oliphant}}}]{numpy2020}%
  \BibitemOpen
  \bibfield  {author} {\bibinfo {author} {\bibfnamefont {C.~R.}\ \bibnamefont
  {{Harris}}}, \bibinfo {author} {\bibfnamefont {K.~J.}\ \bibnamefont
  {{Millman}}}, \bibinfo {author} {\bibfnamefont {S.~J.}\ \bibnamefont {{van
  der Walt}}}, \bibinfo {author} {\bibfnamefont {R.}~\bibnamefont {{Gommers}}},
  \bibinfo {author} {\bibfnamefont {P.}~\bibnamefont {{Virtanen}}}, \bibinfo
  {author} {\bibfnamefont {D.}~\bibnamefont {{Cournapeau}}}, \bibinfo {author}
  {\bibfnamefont {E.}~\bibnamefont {{Wieser}}}, \bibinfo {author}
  {\bibfnamefont {J.}~\bibnamefont {{Taylor}}}, \bibinfo {author}
  {\bibfnamefont {S.}~\bibnamefont {{Berg}}}, \bibinfo {author} {\bibfnamefont
  {N.~J.}\ \bibnamefont {{Smith}}}, \bibinfo {author} {\bibfnamefont
  {R.}~\bibnamefont {{Kern}}}, \bibinfo {author} {\bibfnamefont
  {M.}~\bibnamefont {{Picus}}}, \bibinfo {author} {\bibfnamefont
  {S.}~\bibnamefont {{Hoyer}}}, \bibinfo {author} {\bibfnamefont {M.~H.}\
  \bibnamefont {{van Kerkwijk}}}, \bibinfo {author} {\bibfnamefont
  {M.}~\bibnamefont {{Brett}}}, \bibinfo {author} {\bibfnamefont
  {A.}~\bibnamefont {{Haldane}}}, \bibinfo {author} {\bibfnamefont {J.~F.}\
  \bibnamefont {{del R{\'\i}o}}}, \bibinfo {author} {\bibfnamefont
  {M.}~\bibnamefont {{Wiebe}}}, \bibinfo {author} {\bibfnamefont
  {P.}~\bibnamefont {{Peterson}}}, \bibinfo {author} {\bibfnamefont
  {P.}~\bibnamefont {{G{\'e}rard-Marchant}}}, \bibinfo {author} {\bibfnamefont
  {K.}~\bibnamefont {{Sheppard}}}, \bibinfo {author} {\bibfnamefont
  {T.}~\bibnamefont {{Reddy}}}, \bibinfo {author} {\bibfnamefont
  {W.}~\bibnamefont {{Weckesser}}}, \bibinfo {author} {\bibfnamefont
  {H.}~\bibnamefont {{Abbasi}}}, \bibinfo {author} {\bibfnamefont
  {C.}~\bibnamefont {{Gohlke}}},\ and\ \bibinfo {author} {\bibfnamefont
  {T.~E.}\ \bibnamefont {{Oliphant}}},\ }\bibfield  {title} {\bibinfo {title}
  {{Array programming with NumPy}},\ }\href
  {https://doi.org/10.1038/s41586-020-2649-2} {\bibfield  {journal} {\bibinfo
  {journal} {\nat}\ }\textbf {\bibinfo {volume} {585}},\ \bibinfo {pages} {357}
  (\bibinfo {year} {2020})},\ \Eprint {https://arxiv.org/abs/2006.10256}
  {arXiv:2006.10256 [cs.MS]} \BibitemShut {NoStop}%
\bibitem [{\citenamefont {{Virtanen}}\ \emph {et~al.}(2020)\citenamefont
  {{Virtanen}}, \citenamefont {{Gommers}}, \citenamefont {{Oliphant}},
  \citenamefont {{Haberland}}, \citenamefont {{Reddy}}, \citenamefont
  {{Cournapeau}}, \citenamefont {{Burovski}}, \citenamefont {{Peterson}},
  \citenamefont {{Weckesser}}, \citenamefont {{Bright}}, \citenamefont {{van
  der Walt}}, \citenamefont {{Brett}}, \citenamefont {{Wilson}}, \citenamefont
  {{Millman}}, \citenamefont {{Mayorov}}, \citenamefont {{Nelson}},
  \citenamefont {{Jones}}, \citenamefont {{Kern}}, \citenamefont {{Larson}},
  \citenamefont {{Carey}}, \citenamefont {{Polat}}, \citenamefont {{Feng}},
  \citenamefont {{Moore}}, \citenamefont {{VanderPlas}}, \citenamefont
  {{Laxalde}}, \citenamefont {{Perktold}}, \citenamefont {{Cimrman}},
  \citenamefont {{Henriksen}}, \citenamefont {{Quintero}}, \citenamefont
  {{Harris}}, \citenamefont {{Archibald}}, \citenamefont {{Ribeiro}},
  \citenamefont {{Pedregosa}}, \citenamefont {{van Mulbregt}},\ and\
  \citenamefont {{SciPy 1. 0 Contributors}}}]{scipy2020}%
  \BibitemOpen
  \bibfield  {author} {\bibinfo {author} {\bibfnamefont {P.}~\bibnamefont
  {{Virtanen}}}, \bibinfo {author} {\bibfnamefont {R.}~\bibnamefont
  {{Gommers}}}, \bibinfo {author} {\bibfnamefont {T.~E.}\ \bibnamefont
  {{Oliphant}}}, \bibinfo {author} {\bibfnamefont {M.}~\bibnamefont
  {{Haberland}}}, \bibinfo {author} {\bibfnamefont {T.}~\bibnamefont
  {{Reddy}}}, \bibinfo {author} {\bibfnamefont {D.}~\bibnamefont
  {{Cournapeau}}}, \bibinfo {author} {\bibfnamefont {E.}~\bibnamefont
  {{Burovski}}}, \bibinfo {author} {\bibfnamefont {P.}~\bibnamefont
  {{Peterson}}}, \bibinfo {author} {\bibfnamefont {W.}~\bibnamefont
  {{Weckesser}}}, \bibinfo {author} {\bibfnamefont {J.}~\bibnamefont
  {{Bright}}}, \bibinfo {author} {\bibfnamefont {S.~J.}\ \bibnamefont {{van der
  Walt}}}, \bibinfo {author} {\bibfnamefont {M.}~\bibnamefont {{Brett}}},
  \bibinfo {author} {\bibfnamefont {J.}~\bibnamefont {{Wilson}}}, \bibinfo
  {author} {\bibfnamefont {K.~J.}\ \bibnamefont {{Millman}}}, \bibinfo {author}
  {\bibfnamefont {N.}~\bibnamefont {{Mayorov}}}, \bibinfo {author}
  {\bibfnamefont {A.~R.~J.}\ \bibnamefont {{Nelson}}}, \bibinfo {author}
  {\bibfnamefont {E.}~\bibnamefont {{Jones}}}, \bibinfo {author} {\bibfnamefont
  {R.}~\bibnamefont {{Kern}}}, \bibinfo {author} {\bibfnamefont
  {E.}~\bibnamefont {{Larson}}}, \bibinfo {author} {\bibfnamefont {C.~J.}\
  \bibnamefont {{Carey}}}, \bibinfo {author} {\bibfnamefont
  {{\.I}.}~\bibnamefont {{Polat}}}, \bibinfo {author} {\bibfnamefont
  {Y.}~\bibnamefont {{Feng}}}, \bibinfo {author} {\bibfnamefont {E.~W.}\
  \bibnamefont {{Moore}}}, \bibinfo {author} {\bibfnamefont {J.}~\bibnamefont
  {{VanderPlas}}}, \bibinfo {author} {\bibfnamefont {D.}~\bibnamefont
  {{Laxalde}}}, \bibinfo {author} {\bibfnamefont {J.}~\bibnamefont
  {{Perktold}}}, \bibinfo {author} {\bibfnamefont {R.}~\bibnamefont
  {{Cimrman}}}, \bibinfo {author} {\bibfnamefont {I.}~\bibnamefont
  {{Henriksen}}}, \bibinfo {author} {\bibfnamefont {E.~A.}\ \bibnamefont
  {{Quintero}}}, \bibinfo {author} {\bibfnamefont {C.~R.}\ \bibnamefont
  {{Harris}}}, \bibinfo {author} {\bibfnamefont {A.~M.}\ \bibnamefont
  {{Archibald}}}, \bibinfo {author} {\bibfnamefont {A.~H.}\ \bibnamefont
  {{Ribeiro}}}, \bibinfo {author} {\bibfnamefont {F.}~\bibnamefont
  {{Pedregosa}}}, \bibinfo {author} {\bibfnamefont {P.}~\bibnamefont {{van
  Mulbregt}}},\ and\ \bibinfo {author} {\bibnamefont {{SciPy 1. 0
  Contributors}}},\ }\bibfield  {title} {\bibinfo {title} {{SciPy 1.0:
  fundamental algorithms for scientific computing in Python}},\ }\href
  {https://doi.org/10.1038/s41592-019-0686-2} {\bibfield  {journal} {\bibinfo
  {journal} {Nature Methods}\ }\textbf {\bibinfo {volume} {17}},\ \bibinfo
  {pages} {261} (\bibinfo {year} {2020})},\ \Eprint
  {https://arxiv.org/abs/1907.10121} {arXiv:1907.10121 [cs.MS]} \BibitemShut
  {NoStop}%
\bibitem [{\citenamefont {{Hunter}}(2007)}]{matplotlib2007}%
  \BibitemOpen
  \bibfield  {author} {\bibinfo {author} {\bibfnamefont {J.~D.}\ \bibnamefont
  {{Hunter}}},\ }\bibfield  {title} {\bibinfo {title} {{Matplotlib: A 2D
  Graphics Environment}},\ }\href {https://doi.org/10.1109/MCSE.2007.55}
  {\bibfield  {journal} {\bibinfo  {journal} {Computing in Science \&
  Engineering}\ }\textbf {\bibinfo {volume} {9}},\ \bibinfo {pages} {90}
  (\bibinfo {year} {2007})}\BibitemShut {NoStop}%
\bibitem [{\citenamefont {{Streamlit}}(2024)}]{streamlit}%
  \BibitemOpen
  \bibfield  {author} {\bibinfo {author} {\bibnamefont {{Streamlit}}},\
  }\href@noop {} {\bibinfo {title} {{Streamlit: The fastest way to build and
  share data apps}}},\ \bibinfo {howpublished} {\url{https://streamlit.io}}
  (\bibinfo {year} {2024})\BibitemShut {NoStop}%
\bibitem [{\citenamefont {{Lyne}}\ \emph {et~al.}(1998)\citenamefont {{Lyne}},
  \citenamefont {{Manchester}}, \citenamefont {{Lorimer}}, \citenamefont
  {{Bailes}}, \citenamefont {{D'Amico}}, \citenamefont {{Tauris}},
  \citenamefont {{Johnston}}, \citenamefont {{Bell}},\ and\ \citenamefont
  {{Nicastro}}}]{Lyne1998}%
  \BibitemOpen
  \bibfield  {author} {\bibinfo {author} {\bibfnamefont {A.~G.}\ \bibnamefont
  {{Lyne}}}, \bibinfo {author} {\bibfnamefont {R.~N.}\ \bibnamefont
  {{Manchester}}}, \bibinfo {author} {\bibfnamefont {D.~R.}\ \bibnamefont
  {{Lorimer}}}, \bibinfo {author} {\bibfnamefont {M.}~\bibnamefont {{Bailes}}},
  \bibinfo {author} {\bibfnamefont {N.}~\bibnamefont {{D'Amico}}}, \bibinfo
  {author} {\bibfnamefont {T.~M.}\ \bibnamefont {{Tauris}}}, \bibinfo {author}
  {\bibfnamefont {S.}~\bibnamefont {{Johnston}}}, \bibinfo {author}
  {\bibfnamefont {J.~F.}\ \bibnamefont {{Bell}}},\ and\ \bibinfo {author}
  {\bibfnamefont {L.}~\bibnamefont {{Nicastro}}},\ }\bibfield  {title}
  {\bibinfo {title} {{The Parkes Southern Pulsar Survey -- II. Final results
  and population analysis}},\ }\href
  {https://doi.org/10.1046/j.1365-8711.1998.01144.x} {\bibfield  {journal}
  {\bibinfo  {journal} {\mnras}\ }\textbf {\bibinfo {volume} {295}},\ \bibinfo
  {pages} {743} (\bibinfo {year} {1998})}\BibitemShut {NoStop}%
\bibitem [{\citenamefont {{Levin}}\ \emph {et~al.}(2013)\citenamefont
  {{Levin}}, \citenamefont {{Bailes}}, \citenamefont {{Barsdell}},
  \citenamefont {{Bates}}, \citenamefont {{Bhat}}, \citenamefont {{Burgay}},
  \citenamefont {{Burke-Spolaor}}, \citenamefont {{Champion}}, \citenamefont
  {{Coster}}, \citenamefont {{D'Amico}}, \citenamefont {{Jameson}},
  \citenamefont {{Johnston}}, \citenamefont {{Keith}}, \citenamefont
  {{Kramer}}, \citenamefont {{Milia}}, \citenamefont {{Ng}}, \citenamefont
  {{Possenti}}, \citenamefont {{Stappers}}, \citenamefont {{Thornton}},\ and\
  \citenamefont {{van Straten}}}]{Levin2013}%
  \BibitemOpen
  \bibfield  {author} {\bibinfo {author} {\bibfnamefont {L.}~\bibnamefont
  {{Levin}}}, \bibinfo {author} {\bibfnamefont {M.}~\bibnamefont {{Bailes}}},
  \bibinfo {author} {\bibfnamefont {B.~R.}\ \bibnamefont {{Barsdell}}},
  \bibinfo {author} {\bibfnamefont {S.~D.}\ \bibnamefont {{Bates}}}, \bibinfo
  {author} {\bibfnamefont {N.~D.~R.}\ \bibnamefont {{Bhat}}}, \bibinfo {author}
  {\bibfnamefont {M.}~\bibnamefont {{Burgay}}}, \bibinfo {author}
  {\bibfnamefont {S.}~\bibnamefont {{Burke-Spolaor}}}, \bibinfo {author}
  {\bibfnamefont {D.~J.}\ \bibnamefont {{Champion}}}, \bibinfo {author}
  {\bibfnamefont {P.}~\bibnamefont {{Coster}}}, \bibinfo {author}
  {\bibfnamefont {N.}~\bibnamefont {{D'Amico}}}, \bibinfo {author}
  {\bibfnamefont {A.}~\bibnamefont {{Jameson}}}, \bibinfo {author}
  {\bibfnamefont {S.}~\bibnamefont {{Johnston}}}, \bibinfo {author}
  {\bibfnamefont {M.~J.}\ \bibnamefont {{Keith}}}, \bibinfo {author}
  {\bibfnamefont {M.}~\bibnamefont {{Kramer}}}, \bibinfo {author}
  {\bibfnamefont {S.}~\bibnamefont {{Milia}}}, \bibinfo {author} {\bibfnamefont
  {C.}~\bibnamefont {{Ng}}}, \bibinfo {author} {\bibfnamefont {A.}~\bibnamefont
  {{Possenti}}}, \bibinfo {author} {\bibfnamefont {B.}~\bibnamefont
  {{Stappers}}}, \bibinfo {author} {\bibfnamefont {D.}~\bibnamefont
  {{Thornton}}},\ and\ \bibinfo {author} {\bibfnamefont {W.}~\bibnamefont {{van
  Straten}}},\ }\bibfield  {title} {\bibinfo {title} {{The High Time Resolution
  Universe Pulsar Survey -VIII. The Galactic millisecond pulsar population}},\
  }\href {https://doi.org/10.1093/mnras/stt1103} {\bibfield  {journal}
  {\bibinfo  {journal} {\mnras}\ }\textbf {\bibinfo {volume} {434}},\ \bibinfo
  {pages} {1387} (\bibinfo {year} {2013})},\ \Eprint
  {https://arxiv.org/abs/1306.4190} {arXiv:1306.4190 [astro-ph.SR]}
  \BibitemShut {NoStop}%
\bibitem [{\citenamefont {{Shamohammadi}}\ \emph {et~al.}(2024)\citenamefont
  {{Shamohammadi}}, \citenamefont {{Bailes}}, \citenamefont {{Flynn}},
  \citenamefont {{Reardon}}, \citenamefont {{Shannon}}, \citenamefont
  {{Buchner}}, \citenamefont {{Cameron}}, \citenamefont {{Camilo}},
  \citenamefont {{Corongiu}}, \citenamefont {{Geyer}}, \citenamefont
  {{Kramer}}, \citenamefont {{Miles}},\ and\ \citenamefont
  {{Spiewak}}}]{Shamohammadi2024}%
  \BibitemOpen
  \bibfield  {author} {\bibinfo {author} {\bibfnamefont {M.}~\bibnamefont
  {{Shamohammadi}}}, \bibinfo {author} {\bibfnamefont {M.}~\bibnamefont
  {{Bailes}}}, \bibinfo {author} {\bibfnamefont {C.}~\bibnamefont {{Flynn}}},
  \bibinfo {author} {\bibfnamefont {D.~J.}\ \bibnamefont {{Reardon}}}, \bibinfo
  {author} {\bibfnamefont {R.~M.}\ \bibnamefont {{Shannon}}}, \bibinfo {author}
  {\bibfnamefont {S.}~\bibnamefont {{Buchner}}}, \bibinfo {author}
  {\bibfnamefont {A.~D.}\ \bibnamefont {{Cameron}}}, \bibinfo {author}
  {\bibfnamefont {F.}~\bibnamefont {{Camilo}}}, \bibinfo {author}
  {\bibfnamefont {A.}~\bibnamefont {{Corongiu}}}, \bibinfo {author}
  {\bibfnamefont {M.}~\bibnamefont {{Geyer}}}, \bibinfo {author} {\bibfnamefont
  {M.}~\bibnamefont {{Kramer}}}, \bibinfo {author} {\bibfnamefont
  {M.}~\bibnamefont {{Miles}}},\ and\ \bibinfo {author} {\bibfnamefont
  {R.}~\bibnamefont {{Spiewak}}},\ }\bibfield  {title} {\bibinfo {title}
  {{MeerKAT Pulsar Timing Array parallaxes and proper motions}},\ }\href
  {https://doi.org/10.1093/mnras/stae016} {\bibfield  {journal} {\bibinfo
  {journal} {\mnras}\ }\textbf {\bibinfo {volume} {530}},\ \bibinfo {pages}
  {287} (\bibinfo {year} {2024})},\ \Eprint {https://arxiv.org/abs/2401.06963}
  {arXiv:2401.06963 [astro-ph.HE]} \BibitemShut {NoStop}%
\bibitem [{\citenamefont {{McConnell}}\ and\ \citenamefont
  {{Ma}}(2013)}]{McConnellMa2013}%
  \BibitemOpen
  \bibfield  {author} {\bibinfo {author} {\bibfnamefont {N.~J.}\ \bibnamefont
  {{McConnell}}}\ and\ \bibinfo {author} {\bibfnamefont {C.-P.}\ \bibnamefont
  {{Ma}}},\ }\bibfield  {title} {\bibinfo {title} {{Revisiting the Scaling
  Relations of Black Hole Masses and Host Galaxy Properties}},\ }\href
  {https://doi.org/10.1088/0004-637X/764/2/184} {\bibfield  {journal} {\bibinfo
   {journal} {\apj}\ }\textbf {\bibinfo {volume} {764}},\ \bibinfo {eid} {184}
  (\bibinfo {year} {2013})},\ \Eprint {https://arxiv.org/abs/1211.2816}
  {arXiv:1211.2816 [astro-ph.CO]} \BibitemShut {NoStop}%
\end{thebibliography}%

\end{document}